\documentstyle[floats,aps,epsf,amsfonts,amsmath,amssymb]{revtex}
\tighten

\def\lesssim{\mathrel{\hbox{\rlap{\hbox{\lower4pt\hbox{$\sim$}}}\hbox{$<$}}}}
\def\gtrsim{\mathrel{\hbox{\rlap{\hbox{\lower4pt\hbox{$\sim$}}}\hbox{$>$}}}}
\def\alt{\mathrel{\hbox{\rlap{\hbox{\lower4pt\hbox{$\sim$}}}\hbox{$<$}}}}
\def\agt{\mathrel{\hbox{\rlap{\hbox{\lower4pt\hbox{$\sim$}}}\hbox{$>$}}}}
\newcommand {\be} {\begin{equation}} \newcommand {\ee}
{\end{equation}} \newcommand {\ban} {\begin{eqnarray}} \newcommand
{\ean} {\end{eqnarray}}

\begin{document}

\draft

\title{LISA Capture Sources: Approximate Waveforms, Signal-to-Noise Ratios,
and Parameter Estimation Accuracy}


\author{Leor Barack}
\address
{Department of Physics and Astronomy and
Center for Gravitational Wave Astronomy,
University of Texas at Brownsville,
Brownsville, Texas 78520}
\author{Curt Cutler}
\address
{Max-Planck-Institut fuer Gravitationsphysik,
Albert-Einstein-Institut, Am Muehlenberg 1,
D-14476 Golm bei Potsdam, Germany}



\date{\today}
\maketitle

\begin{abstract}

   Captures of stellar-mass compact objects (COs) by massive ($\sim 10^6
M_\odot$) black holes (MBHs) are potentially an important source
for LISA, the proposed space-based gravitational-wave (GW) detector.
The orbits of the inspiraling COs are highly complicated;
they can remain rather eccentric up until the final plunge, and
display extreme versions of relativistic perihelion precession and
Lense-Thirring precession of the orbital plane.  The amplitudes of the
strongest GW signals are expected to be roughly an order of magnitude
smaller than LISA's instrumental noise, but in principle (i.e., with
sufficient computing power) the GW signals can be disentangled from the
noise by matched filtering. The associated template waveforms are
not yet in hand, but theorists will very likely be able to provide
them before LISA launches. Here we introduce a family of approximate
(post-Newtonian) capture waveforms, given in (nearly) analytic form,
for use in advancing LISA studies until more accurate versions are
available. Our model waveforms include most of the key qualitative features
of true waveforms, and cover the full space of capture-event
parameters (including orbital eccentricity and the MBH's spin).
Here we use our approximate waveforms to (i) estimate the relative
contributions of different harmonics (of the orbital frequency) to the
total signal-to-noise ratio, and (ii) estimate the accuracy with which LISA
will be able to extract the physical parameters of the capture event
from the measured waveform.
For a typical source (a $10 M_\odot$ CO captured by a $10^6 M_\odot$ MBH
at a signal-to-noise ratio of 30), we find that LISA can
determine the MBH and CO masses to within a fractional
error of $\sim 10^{-4}$, measure $S/M^2$ (where $S$ and $M$
are the MBH's mass and spin) to within $\sim 10^{-4}$, and
determine the location to the source on the sky
to within $\sim 10^{-3}$ stradians.


\end{abstract}



\section{Introduction}\label{Sec:intro}

Captures of stellar-mass compact objects (COs) by massive
($\sim 10^6 M_\odot$)
black holes (MBHs) in galactic nuclei
represent an important potential source for
LISA, the proposed space-based gravitational-wave (GW) detector~\cite{Pre}.
The capture orbits, which can remain
rather eccentric right up to the final plunge,
display extreme versions of relativistic perihelion
precession and Lense-Thirring precession (i.e., precession of
the orbital plane due to the spin of the MBH), as
well as orbital decay.
Ryan~\cite{ryan_multipoles} has illustrated how the
measured waveforms can effectively map
out the spacetime geometry close to the MBH.

Rate estimates indicate that the strongest detectable sources will be
$D \sim 1$ Gpc from us~\cite{rates}, implying the
measured GW strain amplitude $h(t)$
will be roughly an order of magnitude smaller than the strain
amplitude $n(t)$ due
to detector noise. Nevertheless, because the capture waveform will be
``visible'' to LISA for $\sim 10^{5}$ cycles, in principle (i.e., with
infinite computing power) it should be possible
to dig these signals out of the noise using matched
filtering.
While until now, theorists have not been able to
calculate capture waveforms with the accuracy required for
matched filtering, great progress has been made on this problem in
recent years (see the recent review by Poisson~\cite{Poisson_03}
and references therein), and it seems very
likely that sufficiently-accurate
template waveforms {\it will} be available before
LISA's planned launch in $\sim 2011$.

Nevertheless, there are many theoretical issues related to captures that
must be addressed even {\it before} sufficiently-accurate templates are in hand,
as the final LISA mission specs are being driven substantially by the requirement
that LISA be sensitive enough to see captures (see the LISA Science Requirements
Document at the {\it LISA Sources and Data Analysis team} website
\cite{wg1_website}).
Among the most pressing questions  is how to
efficiently search through the data for capture signals.
Given the large dimensionality
of the space of capture waveforms, and the length and complexity of the
individual template signals, it is clear that
straightforward matched filtering---using a huge set of templates that cover
parameter space like a net---would require vastly more
computational power than is practical. Instead, we need to
develop suboptimal alternatives to coherent match filtering and to
estimate the sensitivity of LISA {\it when using these methods}.

This is the first in a series of papers designed to address such data
analysis problems, for LISA capture sources.
Here we introduce a family of relatively simple
inspiral waveforms, given in nearly analytic form [i.e., up to solutions
of ordinary differential equations (ODEs)], that should roughly
approximate the true waveforms and that include their key qualitative
features. Because our approximate waveforms are given in (nearly) analytic
form, we can
generate vast numbers of templates extremely fast---a feature
that we expect to be crucial in performing Monte Carlo
studies of search techniques.

Our capture waveforms are based on the lowest-order, quadrupolar
waveforms for eccentric-orbit binaries derived by Peters
and Matthews~\cite{pm},
but the orbits are corrected to include the effects of pericenter precession,
Lense-Thirring precession, and inspiral from radiation reaction [all calculated
using post-Newtonian (PN) formulae].
Our waveforms have the right dimensionality (except that in practice
we will neglect the spin of the CO) and the same qualitative features
as the true waveforms, except (i) they do not exhibit the extreme
``zoom-whirl'' behavior of true fully-relativistic waveforms just prior to plunge
\cite{Cutler-Kennefick-Poisson,Glampedakis-Kennefick}, and (ii) they
approximate as constant the angle $\lambda$ between the CO's orbital
angular momentum and the MBH's spin, whereas the correct evolution
would show a small secular change in $\lambda$. However, we do not
expect these two (missing) effects to be very important for our
purpose, since, again, the latter is small (at least for circular
orbits~\cite{scott1}), while extreme zoom-whirl behavior occurs
only for rather eccentric orbits extremely close to
plunge. [The number of ``whirls'' per ``zoom'' grows only
as $-\ln(p-p_s)$ as the orbit's semi-latus rectum $p$ approaches
the plunge value $p_s$~\cite{Cutler-Kennefick-Poisson,Glampedakis-Kennefick},
and so stretches of waveform exhibiting extreme zoom-whirl behavior will
likely contribute only a very small fraction of the total
signal-to-noise ratio (SNR).]

In this paper, we use our approximate waveforms to take a
first cut at estimating the accuracy with which LISA
should be able to extract the capture source's physical parameters, including
its distance and location on the sky, the masses of both
bodies, and the spin of the MBH.
We also illustrate in detail
the relative contributions of different harmonics (of the orbital frequency)
to the total SNR of the waveform. For these calculations,
besides using approximate waveforms, we also use
a low-frequency approximation to the LISA response
function and an approximate version of the Fisher-matrix inner product.
Since our methods are so approximate,
the results should be considered illustrative
rather than definitive.

The plan of this paper
is as follows. In Sec.\ \ref{Sec:work}, to put the present
work in context, we briefly summarize
the previous literature on this topic and indicate the several areas of active
theoretical research.  In Sec.\ \ref{Sec:waveform} we present our
approximate, (nearly-)analytic
waveforms.  Our scheme for incorporating the LISA response function
and the noise closely follows Cutler~\cite{cutler98}, but we give enough detail
that the reader could easily employ the response function detailed in Cornish and
Rubbo~\cite{Cornish} (which is more accurate at high frequencies).
Sec.\ \ref{Sec:Rev} provides a brief review
of signal analysis, partly to explain our conventions.
In Sec.\ \ref{Sec:SNR} we display plots showing the relative contributions
of different harmonics to the
total SNR.
We show that, even for relatively modest final eccentricity, the
higher harmonics contribute significantly.
Finally, in Sec.\ \ref{Sec:ParaAcc},
we present estimates of how accurately LISA can determine
the physical parameters of capture systems.
We emphasize that our treatment is highly modular, allowing for simple
improvements of the various approximations.

A few details of our analysis are left to the
Appendices. In Appendix A we derive a simple expression for the
contribution to the pericenter precession from the
spin of the MBH, and show its equivalence to the standard result.
Appendix B compares the magnitudes of the various PN
terms in our orbital evolution equations.
Since the near-plunge capture orbits are highly relativistic,
higher-order terms are comparable in magnitude to the lower-order ones,
as one would expect.
Estimates of the magnitudes of effects related to the CO's spin
are given in Appendix C. The spin of the CO may have
marginally important effects on the
templates (for rapidly rotating COs),
but for simplicity we leave these out of the rest of our analysis.
(They could be put back in rather easily.)

In later papers will turn to the problem that
is the main motivation for this work: designing a practical
algorithm for digging capture waveforms out of the LISA noise.
There we will
use our approximate waveforms
to estimate the scheme's sensitivity, compared to an
optimal search with infinite computing power.

Our index notation is the following. Indices for vectors and
tensors on parameter space are chosen from the beginning of the
Latin alphabet ($a,b,c,\ldots$).
Vectors and tensors on three-dimensional space have indices
chosen from the middle of the Latin alphabet ($i,j,k,\ldots$),
and run over $1,2,3$; their indices are raised and lowered with
the flat 3-metric, $\eta_{ij}$.
We use Greek indices ($\alpha,\beta,\ldots$), running only over
$I,II$, to label the two independent gravitational waveforms that
LISA effectively generates. (No four-dimensional, spacetime indices
occur in this paper.)

Throughout this paper we use units in which $G=c=1$.




\section{Summary of previous and ongoing theoretical work} \label{Sec:work}

Most, if not all, nucleated galaxies harbor MBHs
in their centers~\cite{Richstone_98,Kormendy_02}.
The MBH's gravity dominates the local stellar
dynamics within a cusp radius $r_c = M/\sigma_c^2$, where $M$ is the
MBH mass and $\sigma_c$ is the one-dimensional velocity dispersion
of stars inside the cusp. A typical MBH with $M=10^6 M_{\odot}$
would have $r_c\sim 1$ pc.
The total mass of stars inside
the cusp is typically of order the MBH mass~\cite{rates}.
Captures occur when two stars in the cusp undergo a close encounter,
sending one of them into the ``loss cone.'' These are orbits that
pass sufficiently close to the MBH that the timescale
on which the CO tends to spiral into the
MBH due to gravitational radiation reaction
is shorter than the timescale on which the CO is
scattered back out by other stars.

Because LISA's sensitivity band is centered at $f \sim 3 \times 10^{-3}$Hz, the
MBHs most ``visible'' to LISA are those with mass $M\sim 10^6 M_{\odot}$.
To avoid tidal disruption, while being close enough to the MBH to emit
GWs in the LISA frequency band, the captured star must be either
a white dwarf (WD), neutron star (NS), black hole (BH), or
a very low mass main-sequence star (LMMS)~\cite{rates}.

Early estimates of capture rates and LISA SNRs were made by Hils and
Bender~\cite{Hils_Bender_95}, who considered the capture of $1 M_{\odot}$
objects.
More recent rate estimates~\cite{rates} suggest that, while the total
capture rate is dominated by LMMSs and WDs, LISA's detection rate
should be dominated by
captures of $\sim 10 M_\odot$ BHs. This is partly because the BHs, being
more massive,
can be ``seen'' to
greater distance,
and partly because two-body relaxation enhances the density of
BHs nearer the MBH~\cite{sterl_notes}.
(Two-body stellar collisions tend to equalize kinetic energies,
causing heavier stars to sink to the center of the cusp.)

The first extended look at data analysis for capture sources was taken
by Finn and Thorne~\cite{finnthorne}.  They simplified the problem by
restricting to the case of circular, equatorial orbits, but
for this case they were able to calculate the correct relativistic
orbits and waveforms, and they showed how
the LISA SNR accumulates over
the last year of inspiral---during which typically
$\sim 10^5$ GW cycles
are emitted---for a range of CO masses and MBH masses and spins.
Their plots illustrate the salient fact that, typically, the entire
last year of inspiral contributes significantly to the SNR.
This is because, one year before plunge, the CO is already
quite close to the MBH. Indeed, we shall see below that considerable
signal-to-noise can accumulate even {\it before} the final year.

A more realistic treatment of the capture problem must
incorporate the facts that (i) capture orbits will generally be
non-equatorial (i.e., the CO's orbital angular momentum will
not be aligned with the spin of the MBH), and (ii) a fair fraction
of the inspiraling orbits will
remain moderately eccentric right up until
the final plunge. The latter fact may seem surprising, since
it is well known that gravitational radiation reaction tends to circularize
orbits rather efficiently.\footnote{Except very close to plunge, where
the very strong-field potential tends to decrease the rate of
circularization, and may even reverse the sign of $de/dt$---cf.\
\cite{Cutler-Kennefick-Poisson,GHK}.}
The point, however, is that
when the COs enter the loss cone, their orbits are initially
{\it extremely} eccentric: $1 - e_{\rm init} \sim 10^{-6}-10^{-3}$,
typically, while the initial pericenter distance is only
$r_{p,{\rm init}} \sim 8-100M$~\cite{Freitag_03a}.
Given the CO's initial trajectory, just after scattering
into the loss cone,
we would like to calculate the eccentricity at the last stable orbit,
$e_{LSO}$.
For non-spinning MBHs, at least, this is straightforward.
We find that $e_{LSO} > 0.1$ if
$r_{p,{\rm init}} \alt 20.0 M$, $e_{LSO} > 0.2$ for
$r_{p,{\rm init}} \alt 12.8 M$, and $e_{LSO} > 0.3$ for
$r_{p,{\rm init}} \alt 9.2 M$.
[These estimates were obtained as follows.
In the test particle limit, let $r_1$ and $r_2$ be the turning points
(pericenter and apocenter) of the radial motion, where $r$ is the
standard radial coordinate in Schwarzschild. Define
$p$ and $e$ by $r_1 (\equiv r_p) = p/(1+e)$ and $r_2 = p/(1-e)$.
Plunge occurs at $p/M = 6 + 2e_{LSO}$~\cite{Cutler-Kennefick-Poisson}.
Then $p_{\rm init}$ is given by
\be\label{pinit}
p_{\rm init}/M = 6 + 2\,e_{LSO} + \frac{1}{M}
\int_{e_{LSO}}^1{\frac{dp}{de} de} \, .
\ee
(Of course, the upper limit in the integral should
actually be slightly less than $1.0$---say, $e=0.99995$---but since the
integrand is smooth as $e\rightarrow 1.0$, it makes no practical
difference if we simply approximate the upper limit as $1.0$.)
The derivative $dp/de = \dot p/\dot e$ due to radiation reaction
was calculated numerically
by Cutler, Kennefick, and Poisson~\cite{Cutler-Kennefick-Poisson}
for orbits near the horizon.
We used the results from Fig.~1 of ~\cite{Cutler-Kennefick-Poisson}
to integrate (roughly, using a pencil and ruler)
$dp/de$ backwards (in time) from plunge to $p/M = 12$.
We then used the lowest-order post-Newtonian result~\cite{pm}
$dp/de = (12/19)(p/e)(1 + \frac{7}{8}e^2)/(1 + \frac{121}{304}e^2)$
to continue the integration backwards to $e=1.0$.]

Based on Freitag's Monte Carlo simulation of capture events in our Galaxy
(Fig.~1 of \cite{Freitag_03a}), we then estimate that roughly half
the captures of $\sim 10 M_{\odot}$ BHs (which, again, should dominate LISA's
detection rate) should have $e_{LSO} \agt 0.2$.
Note that a year or two before the final plunge the eccentricity of such
captures will, in fact, be significantly larger than $e_{LSO}$ (as
illustrated in Figs.\ \ref{fig:evo1},\ref{fig:evo2} below).

%

Capture sources are unlike some other LISA sources (e.g.,
galactic WD-WD binaries or MBH-MBH binaries at high redshift), in that
they may lie quite near the margin of detectability, given LISA's current
design specifications. (Put another way, a modest change in the height
or location of the noise floor may determine whether or not these sources
are detected.) Since detecting capture sources is very high priority for LISA,
it has been a high priority for LISA's Sources and Data Analysis team
[``Working Group 1'' (WG1)] to make as much progress as possible
studying capture sources before finalizing the LISA design~\cite{wg1_members}.
This motivates current research on several fronts, including work to
(i) improve estimates
for event rates and for the distribution of source parameters
(especially masses and initial pericenter distances); (ii) solve the
radiation reaction problem to determine the true orbit,
and construct the corresponding waveforms; (iii) investigate what science
can be done with these sources (both
astrophysics and tests of fundamental physics);
(iv) understand the limits on capture detection due to ``source confusion'',
i.e., the background ``noise'' caused by {\it other}, unresolved
capture sources; and (v) construct strategies
to dig the capture waveforms out of the instrumental and confusion noise.
The present work addresses issues (iii) and (iv) above, while later papers
will address problem (v).

Parameter estimation with LISA [clearly bearing on above issue (iii)]
has been looked at systematically for WD-WD binaries
by Peterseim {\it et al.}~\cite{peterseim96} and Cutler~\cite{cutler98},
and for mergers of MBH pairs by Cutler~\cite{cutler98}
and Vecchio~\cite{vecchio03}.  No comparable analysis has yet
been done for capture sources.
For captures, some initial estimates of parameter estimation
accuracy were made by Poisson~\cite{Poisson96} and
Ryan~\cite{ryan_multipoles} (the latter's main interest being to test
alternative gravitation theories).  However, both Poisson and Ryan
used extremely simplifying approximations: they both took the inspiral
orbits to be circular and equatorial a priori (effectively reducing
the number of unknown system parameters, while leaving uninvestigated
the significance of perihelion precession and Lense-Thirring
precession for parameter extraction), and they did not incorporate in
their signal models the amplitude and phase modulations that arise from
LISA's orbital motion (which LISA will use to determine the source
position).  By comparison, our treatment is far more realistic.  While
our results are also approximate, we believe they should at least give
correct order-of-magnitude estimates of LISA's parameter estimation
accuracy (while it seems doubtful that the earlier estimates can be
trusted even at that level).



\section{Our model inspiral waveform} \label{Sec:waveform}

We approximate the CO-MBH system as being, at any instant,
a Newtonian-orbit binary emitting a Peters-Matthews (i.e., lowest order,
quadrupolar) waveform.
We then use post-Newtonian (PN) equations to secularly evolve the parameters
of the orbit. In particular, we include orbital decay from radiation reaction,
pericenter precession, and Lense-Thirring precession of the orbital plane.
The modulation of the waveform's amplitude and phase due to
Lense-Thirring precession has been described (in the context of circular-orbit
binaries) by Apostolatos {\it et al.}~\cite{haris}.
The motion of the LISA detector introduces additional modulations; our
handling of these closely follows that of Cutler~\cite{cutler98}.
Cutler's treatment does not account for the decrease in LISA's sensitivity
at frequencies $f \agt 10$ mHz (where the GW wavelength is smaller than
the detector's armlength)---an effect recently accounted for by
Cornish and Rubbo~\cite{Cornish}. It would be straightforward to repeat
our analysis using the Cornish and Rubbo formalism. However, since most of
the SNR for astophysically-relevant capture sources will accumulate at frequencies
below $\sim 10\,$mHz, we expect our high-frequency approximation will have
only a modest impact on the results.

We emphasize that our treatment is highly modular.
E.g., while our choice of physical variables is a slight generalization of the
variables used in \cite{haris} and \cite{cutler98}, it
would be straightforward to re-write our waveforms using
a parameterization along the lines of Buonanno, Chen, and
Vallisneri~\cite{BCV} (hereafter, BCV), who found a
particularly convenient way of parametrizing circular-orbit binaries with
spin. (It seems the BCV parametrization should be readily extendible
to eccentric orbits.)
Similarly, it would be straightforward for us to treat the LISA noise
and response function along the lines of Cornish and
Rubbo~\cite{Cornish}.
We do not implement either of these treatments here because they
did not seem of critical importance, and because our work
was already well underway when they became available.

\subsection{Principal axes}
In this paper, we adopt a mixed notation for spatial vectors, sometimes
labelling them with spatial indices ($i,j,k,\ldots$), but sometimes
suppressing the indices and instead
using the standard $3-d$ vector notation: an over-arrow (as in $\vec A$)
to represent a vector, $\vec A \cdot \vec B$ to represent
a scalar (``dot'') product, and $\vec A \times \vec B$
to represent the vector (``cross'') product.
An over-hat (as in $\hat n$)
will indicate that a vector is normalized, i.e., has unit length. We trust our
meaning will always be clear, despite this mixed notation.

Let ${\hat n}$ be the unit vector pointing from the detector
to the source, and let ${ \hat L}(t)$ be the unit vector along
the CO's orbital angular momentum.
We find it convenient to work in a (time-varying) wave frame defined with
respect to ${ \hat n}$ and ${ \hat L}(t)$. We define unit vectors
${ \hat p}$ and ${ \hat q}$ by
\begin{eqnarray} \label{pq}
{ \hat p} &\equiv& ({ \hat n}\times { \hat L})/
                    |{ \hat n}\times { \hat L}|, \nonumber\\
{ \hat q} &\equiv& { \hat p} \times { \hat n},
\end{eqnarray}
based on which we then
define the two polarization basis tensors
\begin{eqnarray} \label{H}
H_{ij}^{+}(t)      & \equiv & \hat p_i \hat p_j - \hat q_i \hat q_j, \nonumber\\
H_{ij}^{\times}(t) & \equiv & \hat p_i \hat q_j + \hat q_i \hat p_j.
\end{eqnarray}
The general GW strain field at the detector can then be written as
\begin{equation} \label{hab}
h_{ij}(t)=
A^{+}(t)H_{ij}^{+}(t) + A^{\times}(t)H_{ij}^{\times}(t),
\end{equation}
where $A^{+}(t)$ and $A^{\times}(t)$ are the amplitudes of the two
polarizations.

\subsection{Peters-Mathews waveforms}

In the quadrupole approximation, the metric perturbation
far from the source is given (in the ``transverse/traceless''
gauge) by \cite{MTW}
\be\label{quad}
h_{ij} = (2/D)\bigl(P_{ik}P_{jl} -
\frac{1}{2}P_{ij}P_{kl}\bigr) \ddot I^{kl}
\ee
where 
$D$ is the distance to the source, the projection operator $P_{ij}$ is
given by
$P_{ij} \equiv \eta_{ij} - {\hat n}_i{\hat n}_j$, and
$\ddot I^{ij}$ is the second time derivative of the inertia tensor.
In this paper we work in the limit of small mass ratio, $\mu/M\ll 1$,
where $\mu$ and $M$ are the masses of the CO and MBH, respectively.
In this limit, the inertia tensor is just
$I^{ij}(t) = \mu\, r^i(t) r^j(t)$,
where $\vec r$ is the position vector of the CO with respect to the MBH.

Consider now a CO-MBH system described as a Newtonian binary, with semi-major
axis $a$, eccentricity $e$, and orbital frequency $\nu = (2\pi M)^{-1}
(M/a)^{3/2}$.
Let ${\hat e_1}$ and ${\hat e_2}$ be orthonormal vectors pointing
along the major and minor axes of the orbital ellipse, respectively.
Since the orbit is planar, $I^{ij}$ has only 3 independent components:
$I^{11}$, $I^{21}$, and $I^{22}$,
and as the motion is periodic, we can express $I^{ij}$ as a sum of harmonics
of the orbital frequency $\nu$:
$I^{ij}=\sum_n I^{ij}_n$.

We next denote
\ban
a_n &\equiv& {1\over 2}(\ddot I^{11}_n - \ddot I^{22}_n), \nonumber\\
b_n &\equiv& \ddot I^{12}_n, \nonumber\\
c_n &\equiv& {1\over 2}(\ddot I^{11}_n + \ddot I^{22}_n). \label{anbncn}
\ean
Peters and Matthews showed \cite{pm} that
\begin{eqnarray} \label{abc}
a_n &=& - n{\cal A}\bigl[J_{n-2}(ne)-2eJ_{n-1}(ne)+(2/n)J_n(ne)
+2eJ_{n+1}(ne)-J_{n+2}(ne)\bigr]\cos[n\Phi(t)],
\nonumber\\
b_n &=& - n{\cal A}(1-e^2)^{1/2}\bigl[J_{n-2}(ne)-2J_{n}(ne)
+J_{n+2}(ne)\bigr]\sin[n \Phi(t)],
\nonumber\\
c_n &=& 2{\cal A}J_n(ne)\cos[n\Phi(t)],
\end{eqnarray}
where
\begin{equation} \label{calA}
{\cal A}\equiv (2\pi \nu M)^{2/3}\mu ,
\end{equation}
$J_n$ are Bessel functions of the first kind,
and $\Phi(t)$ is the mean anomaly (measured from pericenter).
For a strictly Newtonian binary we have
\begin{equation} \label{Phi}
\Phi(t) = 2\pi\nu (t-t_0) +\Phi_0,
\end{equation}
where $\Phi_0$ is the mean anomaly at $t_0$.
Decomposing Eq.\ (\ref{hab}) into $n$-harmonic contributions and
using Eq.~(\ref{quad}), one then easily obtains explicit expressions
for the $n$-harmonic components of the two polarization
coefficients, $A^{+}\equiv \sum_n A^{+}_n$ and $A^{-}\equiv \sum_n A^{-}_n$.
They are
\begin{eqnarray} \label{A}
A^{+}_n &=&-[1+({ \hat L}\cdot{ \hat n})^2]\left[
a_n\cos(2\gamma)-b_n\sin(2\gamma)\right]
+[1-({ \hat L}\cdot{ \hat n})^2]c_n, \nonumber\\
A^{\times}_n&=& 2({ \hat L}\cdot{ \hat n})\left[
b_n \cos(2\gamma)+a_n \sin(2\gamma)\right],
\end{eqnarray}
where $\gamma$ is an azimuthal angle measuring the direction of
pericenter with respect to
$\hat x \equiv [-\hat n + \hat L (\hat L\cdot \hat n)]
/[1-(\hat L\cdot \hat n)^2]^{1/2}$.

\subsection{LISA's response function}

With its three arms, LISA functions as a pair of two-arm detectors,
outputting two orthogonal signals. Let $l_1^i, l_2^i, l_3^i$ be unit
vectors, each along one of LISA's three arms, and let $L$ be LISA's
average arm length. Let also $L_i(t)$ be the length of the $i$'th
arm when LISA measures an incident GW, and denote $\delta L_1(t)\equiv L_i(t)-L$.
We refer to the two-arm detector formed by arms $1$ and $2$ as
`detector I.' The strain amplitude in this detector is given by
\begin{equation}\label{hdef}
h_I(t)\equiv \left[\delta  L_1(t) - \delta L_2(t)\right]/L
= \frac{1}{2} \, h_{ij}(t)\bigl(l_1^i l_1^j - l_2^i l_2^j\bigl).
\end{equation}
The second, orthogonal signal is then given by \cite{cutler98}
\begin{equation}\label{hII}
h_{II}(t) =  3^{-1/2}\left[\delta L_1(t) + \delta L_2(t)
- 2\delta L_3(t)\right]/L
=\frac{1}{2\sqrt{3}} \, h_{ij}(t)\bigl(l_1^i l_1^j +
l_2^i l_2^j - 2 l_3^i l_3^j \bigl).
\end{equation}
For GW wavelengths much larger than the LISA arm length,
$h_I(t)$ and $h_{II}(t)$ coincide with the two ``Michelson variables''
\cite{AET}, describing
the responses of a pair of two-arm/$90^{\circ}$ detectors.
We can then write $h_I(t)$ and $h_{II}(t)$ as a sum over $n$-harmonic
contributions,
\be\label{sum}
h_{\alpha}(t) = \sum_n  h_{\alpha,n}(t), \quad\quad (\alpha=I,II),
\ee
where
\begin{equation} \label{halpha}
h_{\alpha,n}(t)=\frac{1}{D} \frac{\sqrt{3}}{2}
\left[F^{+}_{\alpha}(t)A_n^{+}(t) +
F^{\times}_{\alpha}(t)A_n^{\times}(t)\right].
\end{equation}
Here, $A_n^{+,\times}(t)$ are the two polarization coefficients
[given, in our model, by Eq.\ (\ref{A}) above], the factor
$\sqrt{3}/2$ accounts for the fact that the actual angle between
LISA arms is $60^{\circ}$ rather than $90^{\circ}$,
and $F_{\alpha}^{+,\times}$ are the ``antenna pattern'' functions,
reading~\cite{haris,300years}
\begin{mathletters} \label{F}
\begin{eqnarray}  \label{FI}
F^{+}_{I} &=&  \frac{1}{2}(1+\cos^2\theta)\cos(2\phi)\cos(2\psi)
-\cos\theta\sin(2\phi)\sin(2\psi), \nonumber\\
F^{\times}_{I} &=& \frac{1}{2}(1+\cos^2\theta)\cos(2\phi)\sin(2\psi)
+\cos\theta\sin(2\phi)\cos(2\psi),
\end{eqnarray}
\begin{eqnarray}  \label{FII}
F^{+}_{II} &=&  \frac{1}{2}(1+\cos^2\theta)\sin(2\phi)\cos(2\psi)
+\cos\theta\cos(2\phi)\sin(2\psi), \nonumber\\
F^{\times}_{II} &=& \frac{1}{2}(1+\cos^2\theta)\sin(2\phi)\sin(2\psi)
-\cos\theta\cos(2\phi)\cos(2\psi).
\end{eqnarray}
\end{mathletters}
In these expressions, $(\theta,\phi)$ is the source's sky location in a
detector-based coordinate system and $\psi$ is the ``polarization angle''
describing the orientation of the ``apparent ellipse'' drawn by the
projection of the orbit on the sky---see Fig.\ 1 in Ref.\ \cite{haris}
and the explicit relation (\ref{relations2}) given below.

It is more convenient to express the above response function in terms
of angles defined not in the rotating, detector-based system, but
rather in a fixed, ecliptic-based coordinate system.
The angles $\theta,\phi$ are related to $\theta_S,\phi_S$---the
source location in an ecliptic-based system---through
\begin{eqnarray} \label{relations1}
\cos\theta(t) &=& \frac{1}{2}\cos\theta_S-\frac{\sqrt{3}}{2}\sin\theta_S
\cos[\bar\phi_0+2\pi(t/T)-\phi_S], \nonumber\\
\phi(t) &=& \bar\alpha_0+2\pi(t/T) + \tan^{-1}\left[
\frac{\sqrt{3}\cos\theta_S+\sin\theta_S\cos[\bar\phi_0+2\pi(t/T)-\phi_S]}
{2\sin\theta_S\sin[\bar\phi_0+2\pi(t/T)-\phi_S]}\right],
\end{eqnarray}
where $T=1$ year and $\bar\phi_0,\bar\alpha_0$ are constant angles specifying,
respectively, the orbital and rotational phase of the detector at $t=0$. (See Cutler~\cite{cutler98}
for a complete definition of these angles; note, though, that the angle
$\bar\alpha_0=0$ in this paper is referred to as $\alpha_0=0$ in Cutler~\cite{cutler98}.)

Next, we express the polarization angle $\psi$ in terms of $\theta_S,\phi_S$
and $\theta_L,\phi_L$---the direction of the CO's orbital angular
momentum, ${\hat L}(t)$, in the ecliptic-based system. We have
\begin{eqnarray}  \label{relations2}
\tan\psi&=&\left\{\frac{1}{2}\cos\theta_L-\frac{\sqrt{3}}{2}
\sin\theta_L \cos[\bar\phi_0+2\pi(t/T)-\phi_L]-\cos\theta(t)\left[
\cos\theta_L\cos\theta_S+\sin\theta_L\sin\theta_S\cos(\phi_L-\phi_S)
\right]\right\}/  \nonumber\\
&&\left[
\frac{1}{2}\sin\theta_L\sin\theta_S\sin(\phi_L-\phi_S)
-\frac{\sqrt{3}}{2}\cos(\bar\phi_0+2\pi t/T)\left(\cos\theta_L\sin\theta_S\sin\phi_S
-\cos\theta_S\sin\theta_L\sin\phi_L\right)\right.
\nonumber\\&&\left.
-\frac{\sqrt{3}}{2}\sin(\bar\phi_0+2\pi t/T)\left(\cos\theta_S\sin\theta_L\cos\phi_L
-\cos\theta_L\sin\theta_S\cos\phi_S\right)\right].
\end{eqnarray}

For concreteness we shall hereafter take $\bar\phi_0=\bar\alpha_0=0$, but one
could specify any other value as appropriate.

Note that the angles $\theta_L,\phi_L$ are not constant, since
$\hat L$ precesses about the MBH's spin direction $\hat S$.
Let $\theta_K,\phi_K$ be the direction of $\vec S$ in the ecliptic-based
system (`K' standing for `Kerr');
let also $\lambda$ be the angle {\it between} $\hat L$ and $\hat S$, and
$\alpha(t)$ be an azimuthal angle (in the orbital plane) that measures the
precession of $\hat L$ {\it around} $\hat S$:
Specifically, let
\be \label{alpha}
\hat L = \hat S \, \cos\lambda +
\frac{\hat z - \hat S \cos\theta_K}{\sin\theta_K} \sin\lambda \cos\alpha
+ \frac{\hat S \times \hat z}{\sin\theta_K}  \, \sin\lambda \sin\alpha,
\ee
\noindent
where $\hat z$ is a unit vector normal to the ecliptic. Then the angles
$\theta_L(t),\phi_L(t)$ are given in terms of $\theta_K$, $\phi_K$,
$\lambda$, $\alpha(t)$ as
\begin{eqnarray}\label{relations3}
\cos\theta_L(t) &=& \cos\theta_K \cos\lambda
    +\sin\theta_K\sin\lambda\cos\alpha(t), \nonumber\\
\sin\theta_L(t)\cos\phi_L(t) &=&
\sin\theta_K\cos\phi_K\cos\lambda
-\cos\phi_K\cos\theta_K\sin\lambda\cos\alpha(t)
+\sin\phi_K\sin\lambda\sin\alpha(t),  \nonumber\\
\sin\theta_L(t)\sin\phi_L(t) &=&
\sin\theta_K\sin\phi_K\cos\lambda
-\sin\phi_K\cos\theta_K\sin\lambda\cos\alpha(t)
-\cos\phi_K\sin\lambda\sin\alpha(t) .
\end{eqnarray}

\subsection{The pericenter angle $\tilde \gamma$ }

As mentioned above, the angle $\gamma$ that appears in Eqs.\ (\ref{A})
measures the
direction of pericenter with respect to
$\hat x \equiv [-\hat n + \hat L (\hat L\cdot \hat n)]/
[1-(\hat L\cdot \hat n)^2]^{1/2}$. With this definition,
$\gamma$ is neither purely extrinsic nor purely intrinsic.
(In the terminology of BCV, ``intrinsic'' parameters describe the
system without reference to the location or orientation of the observer.)
We will find it convenient to introduce a
somewhat different convention for the
zero-point of this angle: We shall define $\tilde\gamma$ to be
the direction of pericenter
with respect to $\hat L \times \hat S$. Then $\tilde\gamma$
is a purely intrinsic quantity.

Clearly, $\gamma$  and $\tilde \gamma$ are related by
\be\label{beta}
\gamma = \tilde\gamma + \beta,
\ee
where $\beta$ is the angle from $\hat x \propto [\hat L(\hat L \cdot \hat n)
- \hat n] $ to $(\hat L \times \hat S)$.
It is straightforward to show that $\beta$ is given by
\begin{eqnarray}\label{sinbeta}
\sin\beta &=& \frac{\cos\lambda\, \hat L\cdot\hat n -\hat S\cdot \hat n }
{\sin\lambda\bigl[1 - (\hat L\cdot\hat n)^2\bigr]^{1/2}}, \nonumber \\
\cos\beta &=& \frac{\hat n \cdot (\hat S \times \hat L)}
{\sin\lambda\bigl[1 - (\hat L\cdot\hat n)^2\bigr]^{1/2} }.
\end{eqnarray}
To evaluate $\beta(t)$ in practice, we shall need the following
relations:
\be\label{SdotN}
{ \hat S}\cdot{ \hat n} = \cos\theta_S \cos\theta_K
+ \sin\theta_S \sin\theta_K \cos(\phi_S-\phi_K),
\ee
\be\label{ScrossLdotN}
\hat n \cdot (\hat S \times \hat L) =
\sin\theta_S \sin(\phi_K-\phi_S)\sin\lambda \cos\alpha
+ \frac{\hat S\cdot\hat n \cos\theta_K -\cos\theta_S}{\sin\theta_K}
\, \sin\lambda \sin\alpha,
\ee
and
\be\label{LdotN}
{ \hat L}\cdot{ \hat n} = { \hat S}\cdot{ \hat n}\cos\lambda
+ \frac{\cos\theta_S - \hat S\cdot\hat n \cos\theta_K}{\sin\theta_K}
\, \sin\lambda \cos\alpha + \frac{(\hat S \times \bar z)\cdot \hat n}
{\sin\theta_K}  \, \sin\lambda \sin\alpha,
\ee
or, equivalently,
\be\label{2LdotN}
{ \hat L}\cdot{ \hat n} = \cos\theta_S \cos\theta_L +
\sin\theta_S \sin\theta_L \cos(\phi_S-\phi_L).
\ee
Note that the time-variation of ${ \hat S}\cdot{ \hat n}$ is very small
in the extreme mass-ratio case considered here: this quantity is constant to
better than $\sim (\mu/M)(S/M^2)$ (see Appendix C).
In our model we shall approximate $\hat S$---and hence ${\hat S}\cdot{\hat n}$---as strictly constant.

\subsection{Parameter space}
The two-body system is described by 17 parameters. The spin of the
CO can be marginally relevant (see Appendix C), but in this paper
we shall ignore it, leaving us with 14 parameters.
We shall denote a vector in the 14-d parameter space by $\lambda^a$
($a=0,\ldots,13$).
We choose our parameters as follows:
\begin{eqnarray} \label{lambda}
\lambda^a &\equiv& (\lambda^0,\ldots,\lambda^{13}) =\nonumber\\
&&
\left[t_0\,\ln\mu,\,\ln M,\,S/M^2,\,e_0,\,\tilde\gamma_0,\,\Phi_0,\,
\mu_S\equiv\cos\theta_S,\,\phi_S,\,\cos\lambda,\,\alpha_0,
\mu_K\equiv\cos\theta_K,\,\phi_K,\,\ln(\mu/D)\right].
\end{eqnarray}
Here, $t_0$ is a time parameter that allows us to specify ``when''
the inspiral occurs---we shall generally choose $t_0$ to be the instant of
time when the (radial) orbital frequency sweeps through
some fiducial value $\nu_0$ (typically, we shall choose $\nu_0$ of
order $1\,$mHz),
$\mu$ and $M$ are the masses of the CO and MBH, respectively, and $S$ is the
magnitude of the MBH's spin angular momentum (so $0 \le S/M^2 \le 1$).
The parameters $e_0$, $\tilde\gamma_0$, and $\Phi_0$ describe,
respectively, the eccentricity, the direction of the pericenter within
the orbital plane, and the mean anomaly---all at time $t_0$.
More specifically, we take $\tilde\gamma_0$ to be the angle (in the plane
of the orbit) from $\hat L \times \hat S$ to pericenter, and, as usual,
$\Phi_0$ to be the mean anomaly with respect to pericenter passage.
The parameter $\alpha_0\equiv\alpha(t=t_0)$ [where $\alpha(t)$
is defined in Eq.~(\ref{alpha})] describes the direction
of $\hat L$ around $\hat S$ at $t_0$.
The angles $(\theta_S,\phi_S)$ are the direction to the source, in
ecliptic-based coordinates; $(\theta_K,\,\phi_K)$ represent the direction
$\hat S$ of the MBH's spin (approximated as constant) in ecliptic-based
coordinates; and $\lambda$ is the angle between $\hat L$ and $\hat S$ (also
approximated as constant\footnote
{In reality, radiation reaction will impose a small time variation in
$\lambda$; however, this variation is known to be very small
(See Ref.\ \cite{scott1}) and we shall ignore
it here. When a model of the time-variation of $\lambda$ is eventually
at hand, it would be trivial to generalize our treatment to
incorporate it: one would just need an equation for $d\lambda/dt$,
and in the parameter list $\lambda$ would be replaced by
$\lambda_0$---the value of $\lambda$ at time $t_0$.}).
Finally, $D$ is the distance to the source.

\begin{figure}[htb]
\centerline{\epsfysize 6cm \epsfbox{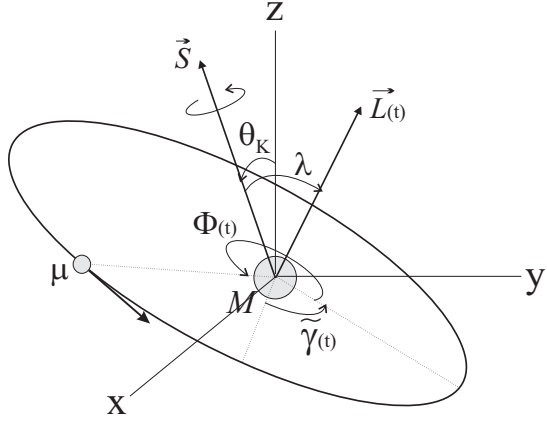}}
\caption{\protect\footnotesize
The MBH-CO system: setup and notation.
$M$ and $\mu$ are the masses of the MBH and the CO, respectively.
The axes labeled $x{-}y{-}z$ represent a Cartesian system {\em based on
ecliptic coordinates} (the Earth's motion around
the Sun is in the x--y plane). The spin $\vec S$ of the MBH is parametrized
by its magnitude $S$ and the two angular coordinates $\theta_K,\phi_K$,
defined (in the standard manner) based on the system $x{-}y{-}z$.
$\vec L(t)$ represents the (time-varying) orbital angular momentum;
its direction is parametrized by the (constant) angle $\lambda$ between
$\vec L$ and $\vec S$, and by an azimuthal angle $\alpha(t)$ (not
shown in the figure).
The angle $\tilde\gamma(t)$ is the (intrinsic) direction of pericenter,
as measured with respect to $\vec L\times\vec S$. Finally,
$\Phi(t)$ denotes the mean anomaly of the orbit, i.e., the average
orbital phase with respect to the direction of pericenter.}
\label{fig1}
\end{figure}

The various parameters and their meaning are summarized in Table
\ref{tableI}. Fig.\ \ref{fig1} illustrates the various angles involved
in our parameterization.

\begin{table}[thb]
\centerline{$\begin{array}{c|c|l}\hline\hline
\lambda^0 & t_0      & \text{$t_0$ is time where orbital frequency
    sweeps through fiducial value (e.g., 1mHz)}   \\
\lambda^1 & \ln\mu        & \text {($\ln$ of) CO's mass}\\
\lambda^2 & \ln M         & \text {($\ln$ of) MBH's mass}\\
\lambda^3 & S/M^2         & \text{magnitude of (specific) spin
    angular momentum of MBH} \\
\lambda^4 & e_0           & \text{$e(t_0)$, where $e(t)$ is the
    orbital eccentricity} \\
\lambda^5 & \tilde\gamma_0 &  \text{$\tilde\gamma(t_0)$,
    where $\tilde\gamma(t)$ is the angle (in orbital plane)
    between $\hat L\times\hat S$ and pericenter}      \\
\lambda^6 & \Phi_0        & \text{$\Phi(t_0)$, where $\Phi(t)$ is
    the mean anomaly}\\
\lambda^7 & \mu_S\equiv\cos\theta_S  & \text{(cosine of) the source direction's
    polar angle }  \\
\lambda^8 & \phi_S        & \text{azimuthal direction to source}  \\
\lambda^9 & \cos\lambda   & \hat L\cdot\hat S(={\rm const}) \\
\lambda^{10} & \alpha_0     & \text{$\alpha(t_0)$, where $\alpha(t)$
    is the azimuthal direction of $\hat L$ (in the orbital plane)}   \\
\lambda^{11} & \mu_K\equiv\cos\theta_K & \text{(cosine of) the polar angle
   of MBH's spin}  \\
\lambda^{12} & \phi_K       &  \text{azimuthal direction of MBH's
    spin}  \\
\lambda^{13} & \ln(\mu/D)          & \text{($\ln$ of) CO's mass divided by distance to source}\\
\hline\hline
\end{array}$}
\caption{\protect\footnotesize
Summary of physical parameters and their meaning.
The angles ($\theta_S$,$\phi_S$) and ($\theta_K$,$\phi_K$)
are associated with a spherical coordinate system attached to the ecliptic.
$\hat L$ and $\hat S$ are unit vectors in the directions
of the orbital angular momentum and the MBH's spin, respectively.
For further details see figure \ref{fig1} and the description
in the text.}\label{tableI}
\end{table}

Note for simplicity we are treating the background spacetime as
Minkowski space, not Robertson-Walker. To correct this, for a source
at redshift $z$, requires only
the simple translation: $M\rightarrow M(1+z)$, $\mu\rightarrow \mu(1+z)$,
$S\rightarrow S(1+z)^2$, $D\rightarrow D_L$, where $D_L$ is the
``luminosity distance''~\cite{markovic}.

The parameters can be divided into ``intrinsic'' and ``extrinsic''
parameters, following BCV.
Extrinsic parameters refer to the observer's position or orientation, or to
the zero-of-time on the observer's watch.
There are seven extrinsic parameters: the four parameters
$t_0$,  $\mu_S$, $\phi_S$, and $D$ correspond to the
freedom to translate the same binary in space and time, and
the three parameters $\mu_K$, $\phi_K$, and $\alpha_0$ are
basically Euler angles that specify the orientation of the
orbit with respect to the observer (at $t_0$).
The intrinsic parameters are the ones that control the detailed
dynamical evolution of the system, without reference to the
observer's location or orientation.
In our parametrization, the seven intrinsic parameters are
$\ln\mu$, $\ln M$, $S/M^2$, $\cos\lambda$, $e_0$, $\tilde\gamma_0$,
and $\Phi_0$.
BCV observed (in the context of circular-orbit binaries with spin) that
extrinsic parameters are generally much ``cheaper'' to search over
than intrinsic parameters. We shall make good use of this important
observation in further papers.

\subsection{Orbital evolution equations}
\label{EvEq}

We evolve $\Phi(t)$, $\nu(t)$, $\tilde\gamma(t)$, $e(t)$,
and $\alpha(t)$ using the following PN formulae:
\begin{eqnarray}
\frac{d\Phi}{dt} &=& 2\pi\nu, \label{Phidot} \\
%
\frac{d\nu}{dt} &=&
\frac{96}{10\pi}(\mu/M^3)(2\pi M\nu)^{11/3}(1-e^2)^{-9/2}
\bigl\{
\left[1+(73/24)e^2+(37/96)e^4\right](1-e^2) \nonumber \\
&&+ (2\pi M\nu)^{2/3}\left[(1273/336)-(2561/224)e^2-(3885/128)e^4
-(13147/5376)e^6 \right] \nonumber \\
&&- (2\pi M\nu)(S/M^2)\cos\lambda (1-e^2)^{-1/2}\bigl[(73/12)
+ (1211/24)e^2 \nonumber \\
&&+(3143/96)e^4 +(65/64)e^6 \bigr]
\bigr\}, \label{nudot} \\
\frac{d\tilde\gamma}{dt} &=& 6\pi\nu(2\pi\nu M)^{2/3} (1-e^2)^{-1}
\left[1+\frac{1}{4}(2\pi\nu M)^{2/3} (1-e^2)^{-1}(26-15e^2)\right] \nonumber \\
&&-12\pi\nu\cos\lambda (S/M^2) (2\pi M\nu)(1-e^2)^{-3/2},
\label{Gamdot} \\
\frac{de}{dt}  &=& -\frac{e}{15}(\mu/M^2) (1-e^2)^{-7/2} (2\pi M\nu)^{8/3}
\bigl[(304+121e^2)(1-e^2)\bigl(1 + 12 (2\pi M\nu)^{2/3}\bigr) \, \nonumber \\
&&- \frac{1}{56}(2\pi M\nu)^{2/3}\bigl( (8)(16705) + (12)(9082)e^2 - 25211e^4 \bigr)\bigr]\,
\nonumber \\
&&+ e (\mu/M^2)(S/M^2)\cos\lambda\,(2\pi M\nu)^{11/3}(1-e^2)^{-4}
\, \bigl[(1364/5) + (5032/15)e^2 + (263/10)e^4\bigr] ,
\label{edot} \\
\frac{d\alpha}{dt} &=& 4\pi\nu (S/M^2) (2\pi M\nu)(1-e^2)^{-3/2}.
\label{alphadot}
\end{eqnarray}
Equations (\ref{nudot}), (\ref{Gamdot}), and (\ref{edot}) are from
Junker and Sch\"afer~\cite{JunkerSchaefer}, except (i) the second line
of Eq.\ (\ref{Gamdot}) is from Brumberg~\cite{Brumberg}
(cf.\ our Appendix A),
and the last term in Eq.\ (\ref{nudot})---the term $\propto S/M^2$---is from
Ryan~\cite{ryan96}.  Eq.~(\ref{alphadot}) is from Barker and
O'Connell~\cite{Barker}.
The dissipative terms $d\nu/dt$ and
$de/dt$ are given accurately through 3.5PN order (i.e., one order higher
than 2.5PN order, where radiation reaction first becomes manifest).\footnote{
The currently undetermined term in the 3.5PN expressions
[see \cite{Blanchet02}, in particular Eqs.\ (12)--(14) therein] does
not show up in our calculation, since here we are ignoring terms that
are higher-order in the mass ratio $\mu/M$.}
The non-dissipative equations, for $d\tilde\gamma/dt$ and $d\alpha/dt$,
are accurate through 2PN order.\footnote{In fact, the equations for
$d\tilde\gamma/dt$ and $d\alpha/dt$ are missing terms proportional
to $(S/M^2)^2$, which, according to usual ``order counting'' are
classified as 2PN. However, this usual counting is misleading when
the central object is a spinning BH: Because BHs are ultracompact, their spins
are smaller than suggested by the usual counting, and the missing terms
$\propto (S/M^2)^2$ have, in fact, the same magnitude as 3PN terms.
Similarly, the terms  $\propto (S/M^2)$ in Eqs.~(\ref{Gamdot}) and
(\ref{alphadot}) can be viewed as effectively 1.5PN terms.}

In solving the above time-evolution equations, the initial values (at time
$t_0$) of $\Phi$, $\nu$, $\tilde\gamma$, $e$, and $\alpha$ are just the parameters
$\Phi_0$, $\nu_0$, $\tilde\gamma_0$, $e_0$, and $\alpha_0$.

We emphasize again that our treatment is highly modular:
The PN expressions in Eqs.\ (\ref{nudot})--(\ref{alphadot})
could be replaced with improved ones as soon as higher-order PN
expressions are available. Also, one might wish to improve these
evolution equations using values from look-up tables, or results
from numerical studies of the orbital evolution in Kerr (such
as in \cite{scott1}).

\subsection{Doppler phase modulation}

Doppler phase modulation
due to LISA's orbital motion becomes important
for integration times longer than a few weeks.
We incorporate this effect by shifting the phase
$\Phi(t)$, according to
\begin{equation}\label{phid1}
\Phi(t)\to \Phi(t)+\Phi^D(t),
\end{equation}
where
\begin{equation}\label{phid2}
\Phi^D(t)=2\pi \nu(t) R \sin\theta_S \cos[2\pi(t/T)-\phi_S].
\end{equation}
Here $R \equiv 1\, {\rm AU} = 499.00478$ sec.

\subsection{Putting the pieces together}

The algorithm for constructing our approximate waveform is then:
Fix some fiducial frequency $\nu_0$ and choose waveform parameters
$(t_0,\,\ln\mu,\,\ln M,\,S/M^2,\,e_0,\,\tilde\gamma_0,\,\Phi_0,\,\cos\theta_S,\,
\phi_S,\,\cos\lambda,\,\alpha_0,\cos\theta_K,\,\phi_K,D)$.
Solve the ODEs (\ref{Phidot})--(\ref{alphadot})
for $\Phi(t)$, $\nu(t)$, $\tilde\gamma(t)$, $e(t)$, $\alpha(t)$.
Calculate $\theta_L(t),\phi_L(t)$ using Eqs.~(\ref{relations3}) and then
obtain $\psi(t)$ from (\ref{relations2}).
Calculate $\gamma(t)$ from $\tilde\gamma(t)$ using Eqs.~(\ref{beta}) and
(\ref{sinbeta}). Use $e(t)$ and $\nu(t)$ to
calculate $a_n(t), b_n(t), c_n(t)$ in Eqs.~(\ref{abc}),
remembering to include the Doppler modulation via
$n\Phi(t)\to n[\Phi(t)+\Phi^D(t)]$, a la Eqs.~(\ref{phid1}) and (\ref{phid2}).
Calculate the amplitude coefficients $A_n^{+,\times}$ and the antenna
pattern functions $F_{\alpha}^{+,\times}$ using Eqs.~(\ref{A}) and
(\ref{F}), respectively.
Then finally calculate $h_\alpha(t)$ (for $\alpha = I, II$) using Eqs.\
(\ref{halpha}) and (\ref{sum}).

Note that, in our treatment, pericenter precession and Lense-Thirring
precession have no effect on the intrinsic signal (the signal in a frame
that rotates with the system), since we always use the Peters-Matthews
``lowest-order'' waveforms. The effect of these motions is simply to
rotate the binary system with respect to the detector.
This relative rotation modulates the polarization angle $\psi$
[which appears in the response functions $F_{\alpha}^{+,\times}(t)$]
since $\psi$ depends on $\hat L$, and it affects the amplitudes
$A^{+,\times}$ since the latter depend on $\hat L \cdot \hat n$ and $\gamma$.

\begin{figure}[htb]
\centerline{\epsfysize 7cm \epsfbox{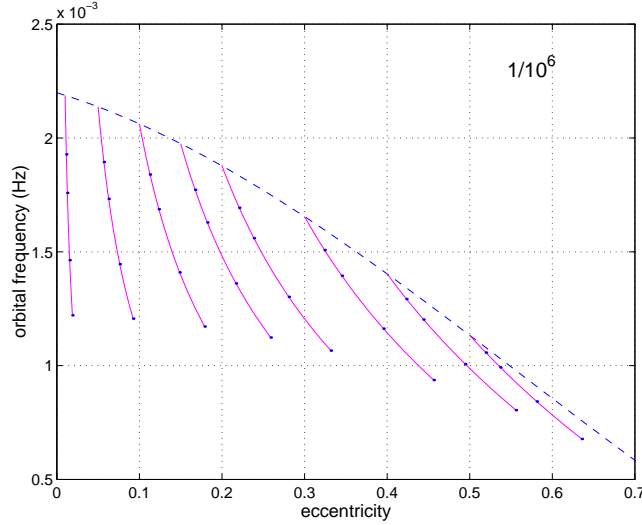}}
\caption{\label{fig:evo1}
Evolution of orbits in our model,
for a system composed of a $1 M_{\odot}$ CO inspiralling into a
$10^6 M_{\odot}$ (non-spinning) MBH. The dashed line represents the last
stable orbit (LSO). Each of the solid lines shows the $\nu-e$
trajectory of a system with given initial data (the orbit evolves in
time ``from bottom to top''). The four dots plotted along each trajectory
indicate, from bottom to top, the state of the system 10, 5, 2, and 1
years before the LSO. 
}
\end{figure}

\begin{figure}[htb]
\centerline{\epsfysize 7cm \epsfbox{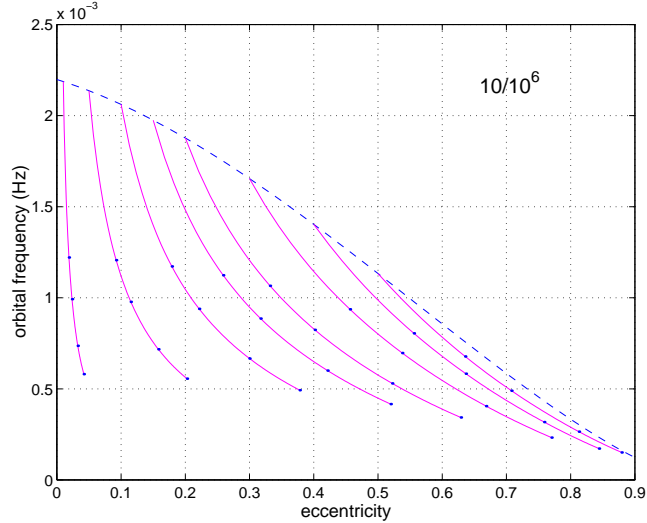}}
\caption{\label{fig:evo2}
Same as in Fig.\ \ref{fig:evo1}, for a $10 M_{\odot}$ CO inspiralling
into a $10^6 M_{\odot}$ MBH.
}
\end{figure}

Figures \ref{fig:orb1}--\ref{fig:wf2} show some sample
orbits and waveforms, obtained from employing the above algorithm.
Figures \ref{fig:evo1} and \ref{fig:evo2} demonstrate
the evolution of orbits in the eccentricity-frequency plane.

\section{Formalism of Signal Analysis} \label{Sec:Rev}


This section briefly reviews the basic formulae of signal
analysis, with application to LISA. We follow closely the treatment
of Cutler~\cite{cutler98}. In particular, our analysis is strictly
valid only in the low-frequency regime, where
the light-travel-time up and down one arm is much less than
the gravitational wave period. (However, we expect it to
be a reasonable approximation at higher frequencies as well).

As discussed above, LISA functions as a pair of two-arm, Michelson detectors,
which we label I and II. The output from these two detectors can be
represented by the vector $s_\alpha(t)$ (with $\alpha = I, II$).
In what follows it will be convenient to work with the Fourier
transform of the signal; the convention we use is
\begin{equation}
\label{fourierT}
{\tilde s_\alpha}(f) \equiv \int_{-\infty}^{\infty}\,
e^{2\pi i f t}s_\alpha(t)\, dt  \;.
\end{equation}

Now, the output $s_\alpha(t)$ is the sum of incident gravitational waves
$h_\alpha(t)$ and instrumental noise $n_\alpha(t)$.
For simplicity we assume that (i) the noise is stationary and Gaussian,
(ii) the noise in detectors I and II is uncorrelated, and
(iii) the noise spectral density $S_h(f)$ is the same in the two detectors.
`Gaussianity' means that each Fourier component ${\tilde n_\alpha}(f)$
has a Gaussian probability distribution.
The combination of assumptions (ii) and (iii) is expressed by the
relation
\begin{equation}
\label{pr}
\langle {\tilde n_\alpha}(f) \, {\tilde n_\beta}(f^\prime)^* \rangle ={1 \over 2}
\delta(f - f^\prime) S_n(f) \delta_{\alpha \beta},
\end{equation}
where `$\langle \ \rangle$' denotes the ``expectation value'' and
$S_n(f)$ is the (single-sided) {\it noise spectral
density} for each detector.

Under the assumptions of stationarity and Gaussianity, we have
a natural definition of the inner product on the vector space of signals
\cite{cutler_flanagan}:
\begin{equation}\label{inner}
\left( {\bf p} \,|\, {\bf q} \right)
\equiv 2\sum_{\alpha} \int_0^{\infty}\left[ \tilde p_\alpha^*(f)
\tilde q_\alpha(f) + \tilde p_\alpha(f) \tilde q_\alpha^*(f)\right]
/S_n(f)\,df,
\end{equation}
where $p_{\alpha}(t)$ and $q_\alpha(t)$ are any two signals.
One can show, based on this definition, that the
inner product of pure noise $\bf n$ with any signal is a random
variable with zero mean and unit variance. In particular,
the probability for the noise to have some realization ${\bf n}_0$
is just
\begin{equation}
\label{pn0}
p({\bf n} = {\bf n}_0) \, \propto \, e^{- \left( {\bf n}_0\, |\,
{\bf n}_0 \right) /2}.
\end{equation}
Thus, if the actual incident waveform is ${\bf h}$, the probability of
measuring a signal ${\bf s}$ in the detector output is proportional to
$e^{-\left( {\bf s-h} \, | \, {\bf s-h}\right)/2}$.  Correspondingly,
given a measured signal ${\bf s}$, the gravitational waveform ${\bf h}$
that ``best fits'' the data is the one that minimizes the quantity
$\left( {\bf s-h} \, | \, {\bf s-h} \right)$.

The SNR for an incident waveform ${\bf h}$ filtered by a perfectly matched
template ${\bf T}={\bf h}$ is
\begin{equation}
\label{SNR-def}
{\rm SNR}[ {\bf h}] \equiv
{{({\bf h}|{\bf h})}\over { {\rm rms}\,({\bf h}|{\bf n})}},
\end{equation}
where ${\rm rms}\,({\bf h}|{\bf n})$ is the rms value for an ensemble
of realizations of the detector noise, ${\bf n}$.
From the definition (\ref{inner}) it follows \cite{cutler_flanagan}
that for any two signals $p_\alpha(t)$ and $q_\alpha(t)$, the expectation
value of $({\bf p}|{\bf n}) ({\bf q}|{\bf n})$ is just $({\bf p}|{\bf q})$.
In particular, we have ${\rm rms}\,({\bf h}|{\bf n})=({\bf h}|{\bf h})^{1/2}$,
and hence the SNR of the detection is approximately given by just
\begin{equation}
\label{SNR}
{\rm SNR}[ {\bf h}] =({\bf h}|{\bf h})^{1/2}.
\end{equation}

For a given incident gravitational wave, different realizations
of the noise will give rise to somewhat different best-fit
parameters.  However, for large SNR, the best-fit parameters will have a
Gaussian distribution centered on the correct values.
Specifically, let ${\tilde \lambda}^a$ be the ``true'' values of the
physical parameters,
and let ${\tilde \lambda}^a + \delta \lambda^a$ be the best
fit parameters in the presence of some realization of the noise.  Then
for large SNR, the parameter-estimation errors $\delta \lambda^a$ have
the Gaussian probability distribution
\begin{equation}
\label{gauss}
p(\delta \lambda^a)=\,{\cal N} \, e^{-{1\over 2}\Gamma_{ab}\delta \lambda^a
\delta \lambda^b}.
\end{equation}
Here $\Gamma_{ab}$ is the so-called {\it Fisher information matrix}, defined by
\begin{equation}
\label{sig}
\Gamma_{ab} \equiv \bigg( {\partial {\bf h} \over \partial \lambda^a}\, \bigg| \,
{\partial {\bf h} \over \partial \lambda^b }\bigg),
\end{equation}
and ${\cal N} = \sqrt{ {\rm det}({\bf \Gamma} / 2 \pi) }$ is the
appropriate normalization factor.  For large SNR, the
variance-covariance matrix is given by
\begin{equation}
\label{bardx}
\left< {\delta \lambda^a} {\delta \lambda^b}
 \right>  = (\Gamma^{-1})^{ab} + {\cal O}({\rm SNR})^{-1} .
\end{equation}
We define $\Delta \lambda^a \equiv \left< {\delta \lambda^a} {\delta \lambda^a}
 \right>^{1/2}$ (the repeated index is {\it not} being summed here).
The uncertainty in the source's angular position, $\Delta \Omega_S$
(a solid angle), is given by \cite{cutler98}\footnote{
Note Eq.\ (3.7) of \cite{cutler98} is erroneous. However, the results
quoted in that paper are based on the correct expression, Eq.\ (\ref{deloms})
here.}
\begin{equation}\label{deloms}
\Delta \Omega_S  = 2 \pi\,\sqrt{(\Delta {\mu}_S \, \Delta {\phi}_S)^2
- \left< \delta {\mu}_S \, \delta {\phi}_S \right>^2} \, .
\end{equation}
The second term in brackets in Eq.~(\ref{deloms}) accounts for the
fact that errors in $\mu_S$ and $\phi_S$ will in general be correlated,
so that the ``error box'' on the sky is elliptical in general, not circular.
The $2\pi$ factor on the right-hand side (RHS) of Eq.~(\ref{deloms})
is our convention; with this definition,
the probability that the source lies {\it outside} an
(appropriately shaped) error ellipse enclosing solid angle
$\Delta \Omega$ is $e^{-\Delta \Omega/\Delta \Omega_S}$.
In the same way, the error ellipse for the spin direction,
$\Delta \Omega_K$, is given by
\begin{equation}\label{delomk}
\Delta \Omega_K  = 2 \pi\,\sqrt{(\Delta {\mu}_K \, \Delta {\phi}_K)^2
- \left< \delta {\mu}_K \, \delta {\phi}_K \right>^2} .
\end{equation}

The actual inner product, Eq.~(\ref{inner}), is formulated in the
frequency domain. For a white noise [i.e., $S_n(f) =$ constant],
the inner product is equivalent to $2 S_n^{-1}\sum_{\alpha}
\int_{-\infty}^{\infty} \, p_\alpha(t) q_\alpha(t) dt$,
by Parseval's theorem. Motivated by this formula, we shall adopt
the following approximate version of the inner product in calculating
the Fisher matrix:
First, we define the ``noise-weighted'' waveform
\be\label{replace}
\hat h_{\alpha}(t) \equiv \sum_n  h_{\alpha,n}(t)/S^{1/2}_h
\bigl(f_n(t)\bigr),
\ee
where we take
\be\label{fn}
f_n(t) = n\nu(t) + \dot{\tilde\gamma}(t)/\pi.
\ee
Then we approximate the covariance matrix, Eq.~(\ref{sig}), as
\be\label{inner_approx}
\Gamma_{ab} = 2\sum_{\alpha}\int_0^T{\partial_a \hat h_{\alpha}(t) \partial_b \hat h_{\alpha}(t) dt} \, .
\ee
That is, we simply re-weight each harmonic by the square root of
the inverse spectral density of the noise, and thereafter treat the
noise as if it were white.

The decision to set
$f_n(t) = n[\nu(t) + (2/n)(\dot{\tilde\gamma}(t)/2\pi)]$ is something of a
compromise: The radial orbital frequency is $\nu$, the azimuthal orbital
frequency is
$\nu(t) + (\dot{\tilde\gamma}/2\pi)$, and the fully correct waveforms will
contain harmonics of both. Our ``compromise'' approximates the
signal as harmonics of $\nu + (2/n)(\dot{\tilde\gamma}/2\pi)$, which
lies between the radial and azimuthal frequencies\footnote{except for the
$n=1$ harmonic, which, however, contributes very little to the GW signal
and to the overall SNR---cf.\ the plots in Figs.~(\ref{fig:SN2})--(\ref{fig:SN4})
below.}
and is the ``correct'' choice for circular motion (in which case
the $n=2$ harmonic dominates the GW output, and only the azimuthal piece enters
the waveform).

\section{SNR estimates} \label{Sec:SNR}

Our analysis of the SNR build-up follows, basically, that of Finn
and Thorne~\cite{finnthorne}. The main advance here is, of course,
the fact that we consider realistically eccentric orbits, whereas
Finn and Thorne confined their analysis to circular orbits.
Unlike in the circular-orbit case, where the contribution from
the $n=2$ harmonic always dominates the SNR, eccentric orbits
have their emitted power (and contribution to SNR) distributed
among higher $n$-harmonics. One of the goals of this section is
to explore this mode distribution, for realistic values of
the orbital eccentricity.

\subsection{LISA noise model}

LISA's noise has three components: instrumental noise, confusion noise
from short-period galactic binaries, and confusion noise from
extragalactic binaries. Our treatment of these three sources follows
Hughes~\cite{Hughes02}, though we use somewhat different estimates for the
sizes of the three pieces.

For LISA's instrumental noise, $S^{\rm inst}_h(f)$, we use the
following analytic fit by Finn and
Thorne~\cite{finnthorne}\footnote{To obtain this
expression, we used the expression given at the beginning of page 8 of
Ref.\ \cite{finnthorne}, where for $[h^{SA}_{SN5,1yr}(f)]^2$ we used the
expression given in footnote [44] therein, and for $\Delta f$ we have
taken $1/1yr$.}, based on the noise
budget given in the LISA Pre-Phase A Report~\cite{Pre}:
\be\label{noise_Sum}
S^{\rm inst}_h(f) = 9.18 \times 10^{-52}f^{-4} + 1.59 \times 10^{-41}
+ 9.18 \times 10^{-38}f^{2}\;\;
{\rm Hz}^{-1},
\ee
where the frequency $f$ is to be given in Hz.

Next we turn to WD confusion noise. Any isotropic background
of indistinguishable GW sources represents (for the
purpose of analyzing {\it other} sources) a noise source with
spectral density~\cite{stochUL}
\be\label{ShOm}
S^{\rm conf}_h(f) = \, \frac{3}{5\pi} f^{-3} \rho_c \Omega_{GW}(f),
\ee
where $\rho_c \equiv 3 H^2_0/(8\pi)$ is the critical energy density needed
to close the universe (assuming it is matter-dominated) and
$\Omega_{GW} \equiv (\rho_c)^{-1} d \rho_{GW}/d(\ln f)$ is the energy density in gravitational
waves (expressed as a fraction of the closure density) per logarithmic
frequency interval.\footnote{Note the RHS of our Eq.~(\ref{ShOm}) is a factor
$\frac{3}{4}$ as large as the RHS in Eq.~(3.4) in\cite{stochUL};
this difference arises simply because the angle between any two
LISA arms
is $\pi/3$ (instead of the $\pi/2$ for LIGO's arms),
and $\sin^2(\pi/3) = 3/4$.}
For the extragalactic WD background, Farmer and Phinney~\cite{FarmerPhinney}
estimate that, for $f$ near $1\,$mHz,
$\Omega_{GW}(f) = 3.6 \times 10^{-12} (f/{10^{-3} {\rm Hz}})^{2/3}$
[at $H_0 = 70 {\rm km/(sec\cdot Mpc)}$], so
\be\label{extragal}
S_h^{\rm ex.\ gal} =
4.2 \times 10^{-47} \left(\frac{f}{1{\rm Hz}}\right)^{-7/3} {\rm Hz}^{-1}.
\ee
Note Eq.~(\ref{extragal}) is not a good fit to $S_h^{\rm ex.\ gal}$ for
$f\gtrsim 10^{-2}$Hz, where mergers cause the spectrum to decrease
more sharply.
However, at such high frequencies, instrumental noise
dominates the total noise in any case, so for our purposes
the extrapolation of Eq.~(\ref{extragal}) to high frequencies is
harmless.

A recent calculation of the galactic confusion background by
Nelemans {\it et al}.~\cite{Nelemans_2001c} yields an $\Omega^{gal}_{GW}$ that is
$5.0\times 10^1$ times larger than $\Omega^{ex.\ gal}_{GW}$
(near 1 mHz)~\cite{FarmerPhinney}; therefore\footnote
{Note our prefactor $2.1\times 10^{-45}$ is
a factor $\sim 25$ lower than the prefactor cited in Hughes~\cite{Hughes02},
based on his private communication with S. Phinney.
This large discrepancy seems to be the product of the following two factors.
First, it was based on the estimate of $\Omega^{gal}_{GW}$ by
Webbink and Han~\cite{Webbink_Han_98}, which is $\sim 3$ times larger than
the result of Nelemans {\it et al}.~\cite{Nelemans_2001c}. Second, it contained
a factor $20/3$ error due to a misunderstanding of Phinney's normalization
convention.}
\begin{equation}
S^{\rm gal}_h(f) = 2.1\times10^{-45}\,\left(\frac{f}{1{\rm Hz}}\right)^{-7/3}
{\rm Hz}^{-1}.
\label{eq:wdgal}
\end{equation}
This is larger than instrumental noise in the range $\sim 10^{-4}$--$10^{-2}$
Hz. However, at frequencies $f \agt 3 \times 10^{-3}\,$Hz, galactic sources
are sufficiently sparse, in frequency space, that one expects to be able
to ``fit them out'' of the data. An estimate of the resulting noise is~\cite{Hughes02}
\begin{equation}\label{inst+gal}
S_h^{\rm inst + gal}(f) = {\rm min}
\left\{S_h^{\rm inst}(f)/\exp(-\kappa T_{\rm mission}^{-1} dN/df), \;\;
S_h^{\rm inst}(f) + S_h^{\rm gal}(f)\right\}\, .
\end{equation}
Here $dN/df$ is the number density of galactic white
dwarf binaries per unit GW frequency, $T_{\rm mission}$ is
the LISA mission lifetime (so $\Delta f = 1/T_{\rm mission}$ is the bin size
of the discretely Fourier transformed data), and $\kappa$ is
the average number of frequency bins that are ``lost'' (for the purpose
of analyzing other sources) when each galactic binary is fitted out
($\kappa$ is larger than one because LISA's motion effectively smears
the signal from each binary over several frequency bins).
The factor $\exp(-\kappa T_{\rm mission}^{-1} dN/df)$ is therefore the
fraction of ``uncorrupted'' bins, where instrumental noise still dominates.
For $dN/df$ we adopt the estimate~\cite{Hughes02}
\begin{equation}\label{eq:dNdf}
{dN\over df} = 2\times10^{-3}\,{\rm Hz}^{-1}\left(1\,{\rm Hz}\over
f\right)^{11/3},
\end{equation}
and take $\kappa T_{\rm mission}^{-1} = 1.5/{\rm yr}$ (corresponding to
$T_{\rm mission} \approx 3\,$yr and $\kappa \approx 4.5$~\cite{Cornish_confusion}).
To obtain the {\it total} LISA noise, we
just add to Eq.~(\ref{inst+gal})
the contribution from the
extragalatic confusion background, Eq.~(\ref{extragal}):
\begin{equation}\label{Shtot}
S_h(f) = S_h^{\rm inst + gal}(f) + S_h^{\rm ex.\ gal}(f) \, .
\end{equation}


\subsection{SNR estimates for inspiral orbits}

For a Keplerian orbit, the source is strictly periodic and hence the
GWs are at harmonics $f_n \equiv n\nu$ of the orbital frequency.
However, as discussed in Sec.\ \ref{Sec:Rev}, to partially compensate for the
fact that the radial and azimuthal periods are different, we set
$f_n(t)$ equal to $n\nu(t) + \dot{\tilde\gamma}(t)/\pi$.
Assuming the contributions from the various harmonics to be
approximately orthogonal, we may approximate the SNR from a single synthetic
2-arm Michelson detector (denoted here by $\rm SNR_I$) as
\be\label{sn2}
\langle {\rm SNR_I}^2\rangle_{\rm SA} =
\Sigma_n\int\frac{h^2_{c,n}(f_n)}{(20/3) f_n S_h (f_n)}\, d(\ln f),
\ee
where ``$\langle\ldots\rangle_{\rm SA}$'' means ``sky average'', i.e.,
average over all source directions (the factor $20/3$ in the denominator
results from this averaging).
The {\it characteristic amplitude} $h_{c,n}$ is given, following
Finn and Thorne\footnote{
In their SNR estimates, Finn and Thorne~\cite{finnthorne} tend to consider
the quantity $h'_{c,m}$ (in their notation), the ``modified'' characteristic
amplitude, introduced to account for the reduction in the
GW signal near the plunge, where the available bandwidth becomes very small.
Here we rather consider $h_{c,m}$ itself: This quantity has the
convenient characteristic that when integrated against the frequency
[through Eq.\ (\ref{sn2})] it yields the SNR (squared). An estimate
of the SNR based on the plots given in \cite{finnthorne}, which show the
modified amplitudes $h'_{c,m}$ rather than $h_{c,m}$ itself, actually takes
into account {\em twice} the
effect of the final plunge: The fact that the frequency changes rapidly near
the plunge is already accounted for in the definition of $h_{c,m}$, just
above Eq.\ (2.2) therein.},
by
\be\label{hcn}
h_{c,n} = (\pi D)^{-1}\sqrt{2\dot E_n/\dot f_n}\, ,
\ee
where $\dot E_n$ is the power radiated to infinity by GWs
at frequency $f_n$. To lowest order, this is
\be\label{dotEn}
\dot E_n = \frac{32}{5} \mu^2M^{4/3}(2\pi \nu)^{10/3} g(n,e),
\ee
where $g(n,e)$ is given by \cite{pm}
\ban\label{gne}
g(n,e) &=& \frac{n^4}{32}\bigl\{\bigl[J_{n-2}(ne) - 2e\,J_{n-1}(ne)
+\frac{2}{n}\,J_{n}(ne) + 2e\,J_{n+1}(ne) - J_{n+2}(ne)\bigr]^2
\nonumber \\
&& + (1-e^2)[J_{n-2}(ne) - 2\,J_{n}(ne) + J_{n+2}(ne)]^2
+\frac{4}{3n^2}[J_{n}(ne)]^2 \bigr\}.
\ean
In this section, our major motivation is to investigate the effect
of non-zero eccentricity. For this reason, we ignore the effect of
the MBH's spin---effectively assuming the MBH is Schwarzschild.

The curves in Figs.~\ref{fig:SN1}--\ref{fig:SN4}
show the buildup of SNR with time, for each harmonic.
As is customary in the LISA literature, our plots actually give
${\rm SNR_I}$---the SNR from a single 2-arm Michelson; the actual
LISA SNR buildup will be a factor $\sim \sqrt{2}$ times larger.
The curves are derived as follows.
We use our PN Eqs.~(\ref{edot}) and (\ref{nudot}) (with $S$ set to zero)
to evolve $e(t)$ and $\nu(t)$ forward in time, up
to the point when the CO plunges over the top of the
effective potential barrier. For a point particle in Schwarzschild, the
plunge occurs at
$a_{\rm min} = M (6 + 2e)(1-e^2)^{-1}$~\cite{Cutler-Kennefick-Poisson},
so we set
\be\label{numax}
\nu_{\rm max} = (2\pi M)^{-1}[(1-e^2)/(6 + 2e)]^{3/2} \, .
\ee
We denote by $e_f$ the ``final'' value of $e$, i.e., the value
of $e$ when $\nu$ reaches the plunge frequency $\nu_{\rm max}$.
Then, for each harmonic $n$ we use our solution $\left\{\nu(t),e(t)\right\}$
along with Eqs.~({\ref{hcn}), ({\ref{dotEn}), and ({\ref{gne})
to determine $h_{c,n}(f_n)$.

The upper ``signal'' curves in Figs.~(\ref{fig:SN2})--(\ref{fig:SN4}) show
$h_{c,n}(t)$ for each $n$; we ``cut off'' each curve at
$f_{n,max} = n\nu_{max} + \pi^{-1}\dot{\tilde\gamma}(\nu_{max},e_f)$.
Marks along each curve indicate (from right to left) one, two,
five, and ten years before the final plunge.
The lower ``noise'' curve depicts $h_n(f) \equiv [(20/3) f S_h(f)]^{1/2}$,
the rms noise amplitude per logarithmic frequency interval
[the factor of $20/3$ comes from sky-averaging;
Finn and Thorne~\cite{finnthorne} define $S^{SA}_h(f) = (20/3) S_h(f)$].
For comparison, the instrumental contribution,
$h_n^{\rm inst}(f)\equiv[(20/3) f_n S_h^{\rm inst}(f_n)]^{1/2}$, is also plotted.
With these conventions, the contribution to the SNR from each harmonic is
\be\label{snn}
({\rm SNR_I})_n^2 = \int(h_{c,n}/h_n)^2 \, d(\ln f),
\ee
so using the curves one can ``integrate by eye''
to estimate the contribution to the SNR from each harmonic, and
to detect {\it when} (i.e., how long before the final plunge) most of
the contribution is accumulated.

\begin{figure}
\centerline{\epsfysize 9cm \epsfbox{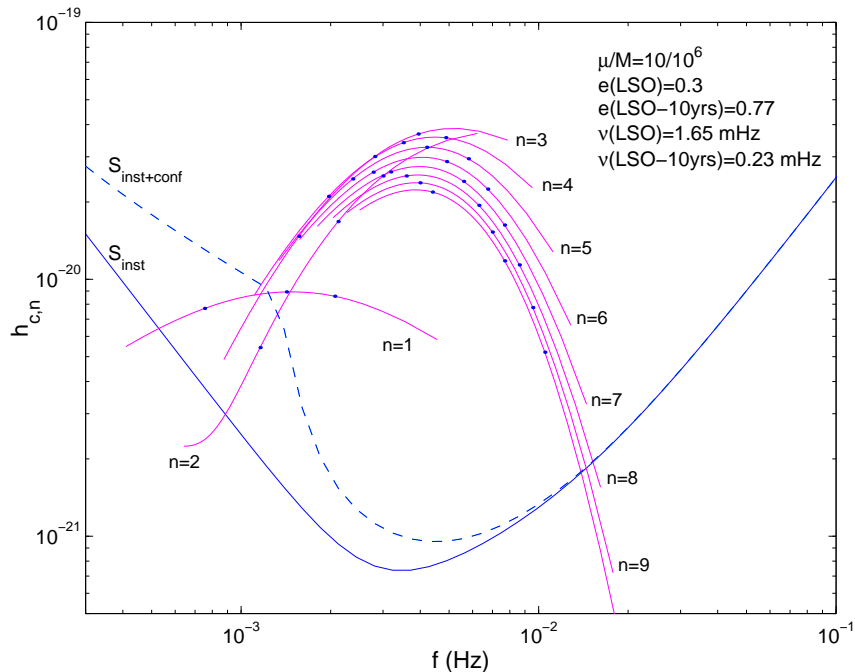}}
\caption{\label{fig:SN1}
GW signal from a $10M_{\odot}$ CO spiralling into a (non-spinning)
$10^6M_{\odot}$ MBH at $D=1\,$Gpc:
case where eccentricity at the last stable orbit
(LSO) is $e_{\rm LSO}=0.3$.
The curve labeled `$S_{\rm inst}$' shows LISA's sky-averaged
instrumental noise level, $h^{\rm inst}_n(f)$.
The dashed line is an estimate of LISA's overall noise level,
$h_n(f)$,
including the effect of stochastic-background ``confusion'' due to WD
binaries (both galactic and extra-galactic).
The convex curves show the amplitudes $h_{c,n}$ of the first 10
$n$-harmonics
of the GW signal, over the last 10 years of evolution prior to the final
plunge.
Along each of these curves we marked 3 dots, indicating (from left to right)
the GW amplitude 5, 2, and 1 years before the plunge.
The orbital eccentricity 10, 5, 2, and 1 years before plunge is
0.77, 0.67, 0.54, and 0.46, respectively.
The orbital frequency 10, 5, 2, and 1 years before plunge is
0.23, 0.41, 0.70, and 0.94 mHz, respectively. The frequency at the LSO is
1.65 mHz.
}
\end{figure}

\begin{figure}
\input{epsf}
\centerline{\epsfysize 9cm \epsfbox{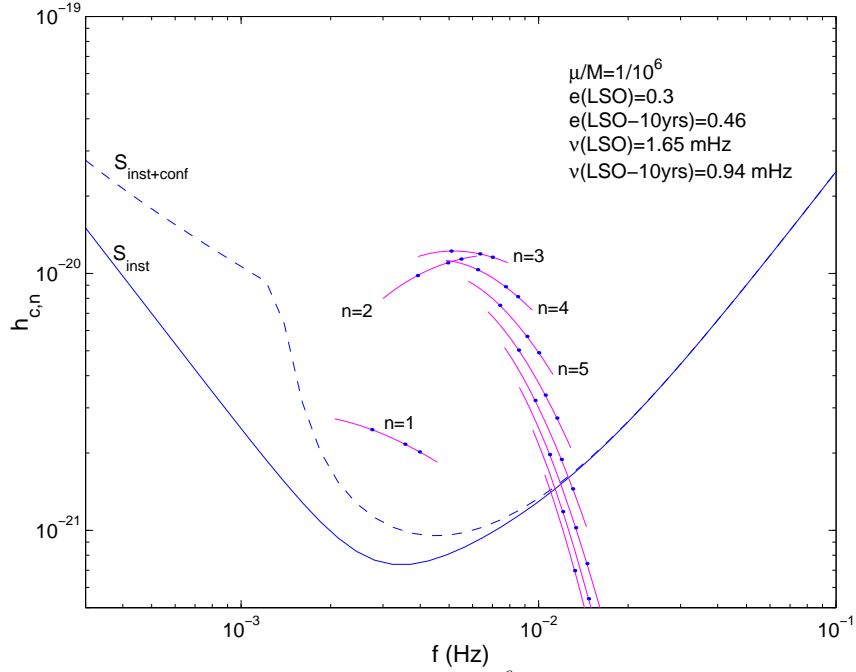}}
\caption{\label{fig:SN2}
Same as in Fig.\ (\ref{fig:SN1}), but for inspiral of a $1M_{\odot}$ CO
into a $10^6M_{\odot}$ MBH.
The orbital eccentricity 10, 5, 2, and 1 years before plunge is
0.46, 0.40, 0.35, and 0.32, respectively.
The orbital frequency $\nu$ 10, 5, 2, and 1 years before plunge is
0.94, 1.16, 1.39, and 1.51 mHz, respectively. The frequency at the LSO is
1.65 mHz.
}
\end{figure}

\begin{figure}
\input{epsf}
\centerline{\epsfysize 9cm \epsfbox{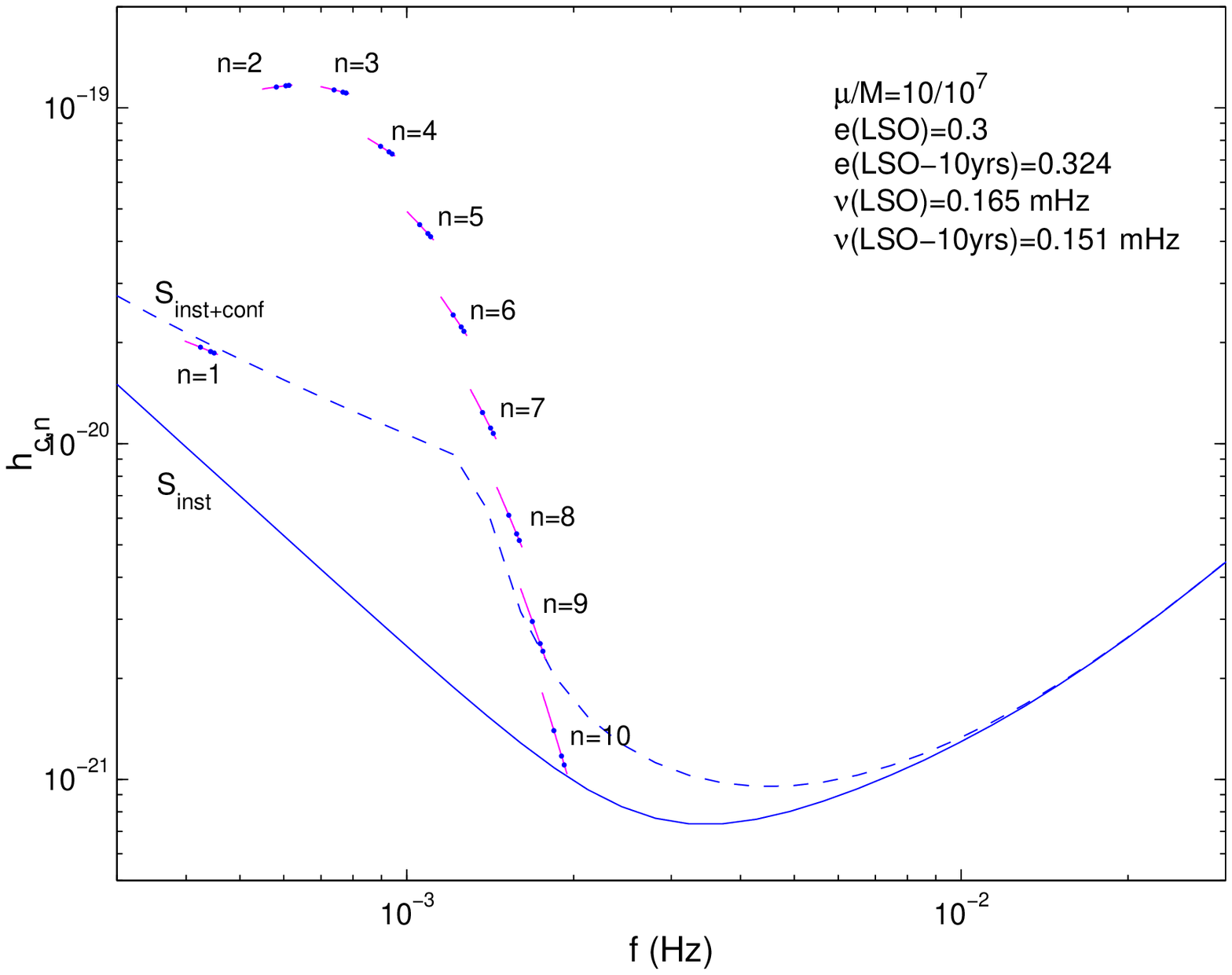}}
\caption{\label{fig:SN3}
Same as in Fig.\ (\ref{fig:SN1}), but for inspiral of a $10M_{\odot}$ CO
into a $10^7M_{\odot}$ MBH.
The orbital eccentricity 10, 5, 2, and 1 years before plunge is
0.324, 0.313, 0.305, and 0.303, respectively.
The orbital frequency 10, 5, 2, and 1 years before plunge is
0.151, 0.158, 0.162, and 0.164 mHz, respectively. The frequency at
the LSO is 0.165 mHz.
}
\end{figure}

\begin{figure}
\input{epsf}
\centerline{\epsfysize 9cm \epsfbox{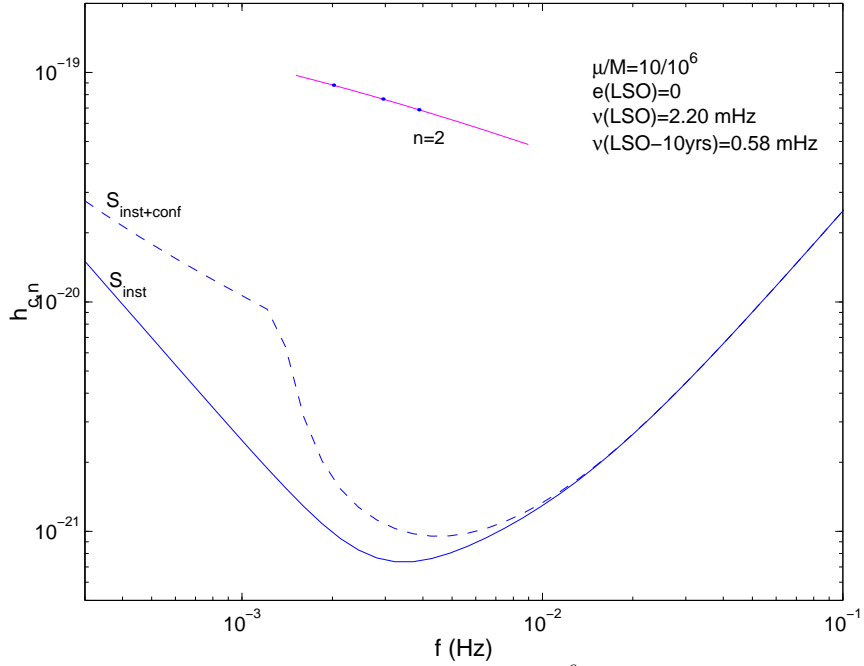}}
\caption{\label{fig:SN0}
GW signal from a $10M_{\odot}$ CO spiralling into a (non-spinning)
$10^6M_{\odot}$ MBH: case of a circular orbit (compare with Fig.\ 4
of Finn and Thorne). Notation is the same as in Fig.\ (\ref{fig:SN1}).
Note that in our model, a CO in a circular orbit emits GW only into
the $n=2$ harmonic.
The orbital frequency 10, 5, 2, and 1 years before plunge is
0.58, 0.74, 0.99, and 1.22 mHz, respectively. The frequency at the
LSO is 2.20 mHz.
}
\end{figure}

\begin{figure}
\input{epsf}
\centerline{\epsfysize 9cm \epsfbox{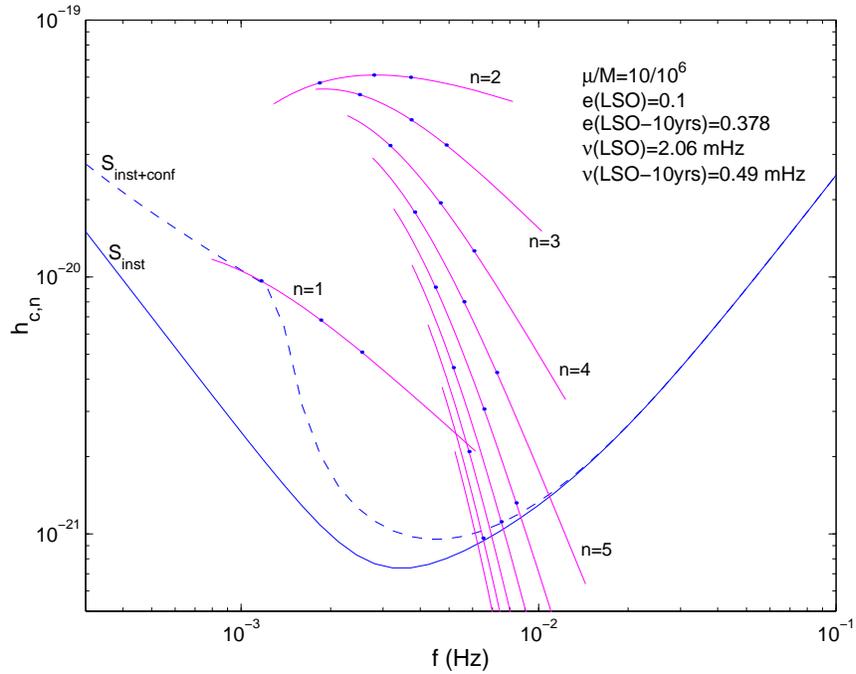}}
\caption{\label{fig:SN5}
Same as in Fig.\ (\ref{fig:SN1}), except that the LSO eccentricity
is taken to be $0.1$.
The orbital eccentricity 10, 5, 2, and 1 years before plunge is
0.38, 0.30, 0.22, and 0.18, respectively.
The orbital frequency 10, 5, 2, and 1 years before plunge is
0.49, 0.67, 0.94, and 1.17 mHz, respectively. The frequency at the
LSO is 2.06 mHz.
}
\end{figure}

\begin{figure}
\input{epsf}
\centerline{\epsfysize 9cm \epsfbox{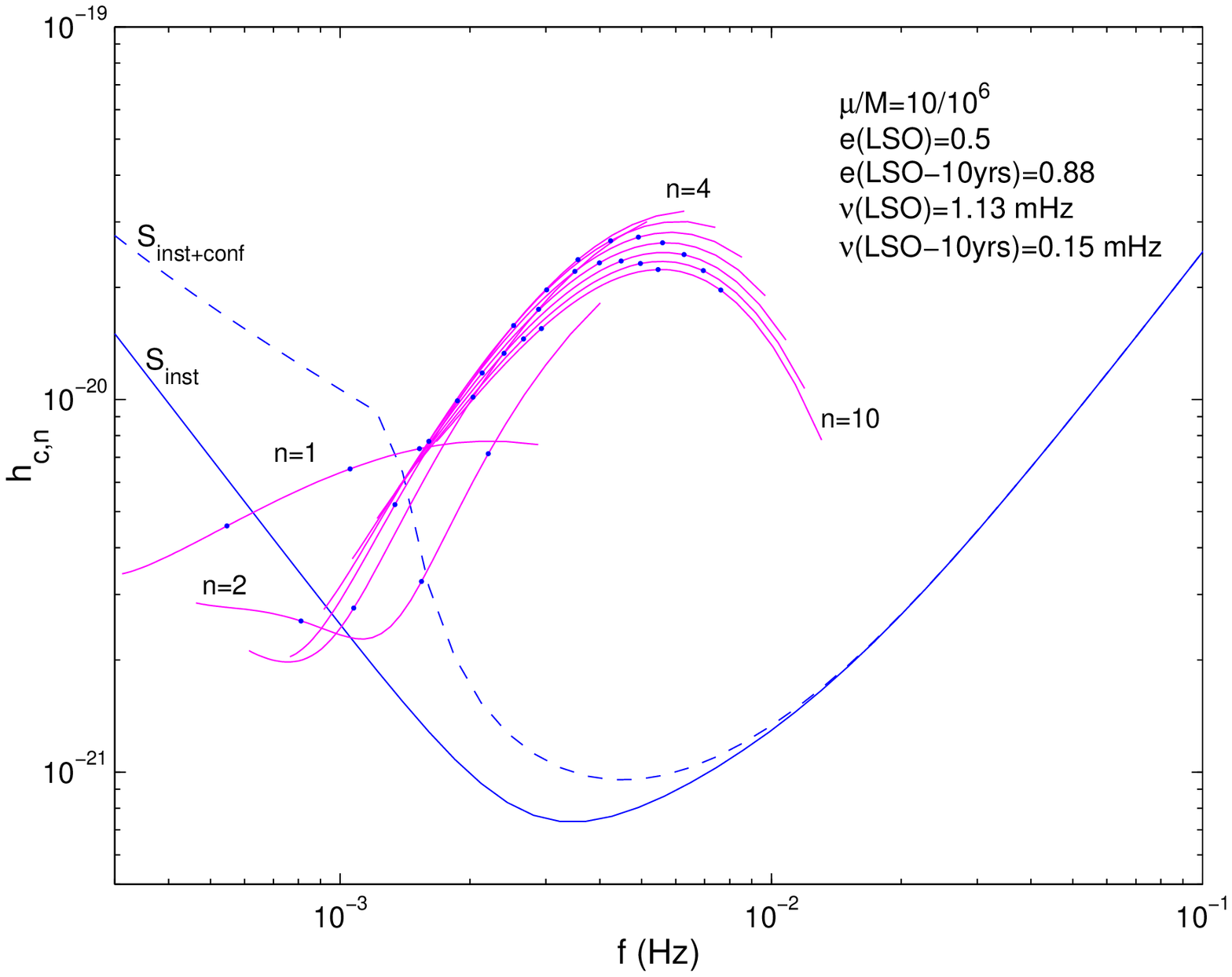}}
\caption{\label{fig:SN4}
GW signal from a $10M_{\odot}$ CO spiralling into a
$10^6M_{\odot}$ MBH: case where eccentricity at the last stable orbit
is $e_{\rm LSO}=0.5$. Notation is the same as in Fig.\ (\ref{fig:SN1}).
The orbital eccentricity 10, 5, 2, and 1 years before plunge is
0.88, 0.81, 0.71, and 0.64, respectively.
The orbital frequency 10, 5, 2, and 1 years before plunge is
0.151, 0.265, 0.490, and 0.678 mHz, respectively. The frequency at
the LSO is 1.13 mHz.
}
\end{figure}

Here are some points to pay attention to when examining the SNR
plots in Figs.\ \ref{fig:SN1}--\ref{fig:SN4}:

\begin{itemize}

\item Systems with MBH's mass of $\sim 10^6$ are ideally ``located''
in the LISA band. Systems with $M=10^7$ radiate at frequencies
where WD confusion noise would likely obscure the capture signal.

\item Systems with $M \approx 10^6$ at $D = 1\ $Gpc
are detectable with $\rm SNR_I$ of $\sim 5 (\mu/1 M_{\odot})$, assuming
1 year of signal integration. Combining both synthetic Michelsons,
and for two years of integration, LISA's SNR is $\sim 10 (\mu/1 M_{\odot})$.

\item As expected, the higher the orbital eccentricity, the more
SNR is contributed by high $n$-harmonics. At $e_{\rm LSO}=0.3$, the
contribution from $n=3,4$ is equally important to that of $n=2$.
At $e_{\rm LSO}=0.5$, the dominant contribution comes at $n=4$, and
one needs to sum the contributions of at least a dozen
modes in order to properly estimate the overall SNR.
\item A related point: For $10^6 M_{\odot}$ MBHs  and $\sim 1 M_{\odot}$ COs,
the last year prior to plunge contributes only a small fraction of
the potential SNR. Even for $\sim 10 M_{\odot}$ COs, the contribution between
10 and 1 years prior to plunge can easily exceed the contribution from the
last year.

\end{itemize}

The last effect may be further visualized by looking at the total amount
of energy radiated in GW up to a time $t$ prior to plunge, as a function
of $t$. This is demonstrated in Fig.\ \ref{fig:Eradiated}
for a mass ratio of $(10M_{\odot})/(10^6M_{\odot})$ and for a variety
of LSO eccentricities. For this plot, we used the leading-order
expression given in \cite{pm} for the total power radiated from all
$n$-harmonics:
\begin{equation} \label{power}
\dot{E}(t)=\frac{32}{5}(\mu/M)^2[2\pi\nu(t) M]^{10/3}[1-e(t)^2]^{-7/2}
\left[1+(73/24)e^2(t)+(37/96)e^4(t)\right].
\end{equation}
The percentage of energy radiated up to time $t$, out of the total
energy radiated during the capture, is then calculated through
\begin{equation} \label{Epercentage}
\%E_{\rm rad}(t)=
\left(1-\frac{\int_{t}^{t_{\rm LSO}}\dot{E}(t')dt'}
{\int_{-\infty}^{t_{\rm LSO}}\dot{E}(t')dt'}
\right)\times 100.
\end{equation}
The results are striking:
For a $10^6 M_{\odot}$ MBH and a $10 M_{\odot}$ CO, about
half the total GW energy is released earlier than 10 years
before the final plunge; for a $1 M_{\odot}$ CO, half
the energy is emitted already 100 years before plunge.
Most of this energy is released in the LISA band, in short ``spurts''
as the CO passes close to the MBH.
This provides a vivid demonstration of the potential ``threat'' imposed
by self-confusion. A systematic analysis of this issue
will be presented elsewhere.

\begin{figure}
\input{epsf}
\centerline{\epsfysize 8cm \epsfbox{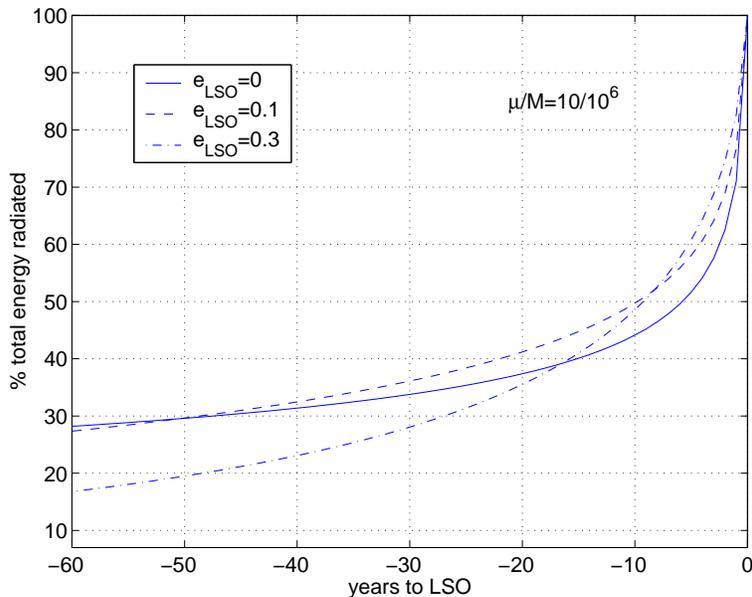}}
\caption{\label{fig:Eradiated}
Percentage of energy radiated (out of the total energy radiated
in GW to infinity during the entire capture process) as a function of the time
left to the final plunge. The CO and MBH masses are $10 M_{\odot}$
and $10^6 M_{\odot}$, respectively. This plot was generated
using Eq.\ (\ref{Epercentage}) with $\dot {E}$ approximated using
Eq.\ (\ref{power}). Note how large the time scale is for energy
to be emitted in GW by the CO-MBH system, compared to LISA mission
time ($\sim 2$ years).
}
\end{figure}

Finally, recall that all of our results regarding SNR values
should be regarded as ballpark estimates rather than definite,
due to the many approximations made here.
Here is a summary of the various approximations involved:
(i) We incorporate an approximate model
of LISA's instrumental noise, which is inaccurate at the
high-frequency end;
(ii) We model the source using PN evolution and quadrupole
emission---both approximations being worse, again, at high frequencies;
(iii) We assume a non-rotating MBH;
(iv) We assume the contributions from the various $n$-modes
to the SNR are uncoupled; and
(v) We assume that an ideal, coherent matched filtering search is
carried out over the entire observation period.
Nevertheless, we believe our results amply illustrate the
importance of higher harmonics and of searches with multi-year
integration times.

\subsection{SNR for a low-mass main-sequence star at Sgr A*}

Freitag~\cite{Freitag_03a,FreitagNew} recently pointed out that LISA might be
able to detect a few $\sim 0.1 M_{\odot}$ main-sequence stars
captured by the MBH at the center of our own galaxy.
The strongest such source will still have $\sim 10^6\, $years to go
before plunge, and so will currently be at the low end of the
LISA frequency range, but because it is so close to us, Freitag estimated
that it could still yield a ${\rm SNR_I}$ as high as $\sim 100$.
We have re-examined this estimate, using the example
from Fig.\ 1 in Freitag~\cite{FreitagNew}, in which $\mu= 0.06 M_{\odot}$,
$M=2.6\cdot 10^6 M_{\odot}$, $D=8$ kpc, and $e=0.8$ at
ten million years prior to plunge. The results are shown in
Fig.~\ref{fig:SNFreitag}.
Since for the galactic source considered here, there is practically no
frequency evolution during the observation time,
our convention in Fig.~\ref{fig:SNFreitag}
differs somewhat from those in Figs.\ \ref{fig:SN1}--\ref{fig:SN4}.
As in Figs.~\ref{fig:SN1}--\ref{fig:SN4}, the different curves
correspond to the contributions from different harmonics, but here
we plot $h_{c,n}$ times the factor $ [(f^{-1}_n df_n/dt)\times 2{\rm yr}]$, so
the height of the marked point above the noise
curve now gives SNR accumulated over 2 years of observation.
The marked points correspond
to 2-yr observations carried out
$10^7$, $5\cdot 10^6$, $2\cdot 10^6$, $10^6$, and $10^5$ years
before plunge, respectively. (Actually, there is no ``plunge,'' as
the low-mass star is tidally disrupted at least $\sim 50\ $yrs before
plunge would occur, but we can ignore that for this analysis.)

For a signal coming $10^6$ years before plunge, we find
a 2-yr $\rm SNR_I$ of $\sim 11$ (the square root of the sum of
the squares of the contributions from all modes).
A two-year observation only $10^5$ years before plunge would yield
$\rm SNR_I \approx 55$. These $\rm SNR_I$ values are a factor $\sim 4$
times smaller than the values obtainable from Fig.~1
of Frietag~\cite{FreitagNew}
(when the 1-yr results he gives there are
scaled up to two years), but still support the idea that
this is a potentially observable source.
As Freitag points out~\cite{FreitagNew},
tidal heating could possibly (under pessimistic assumptions)
disrupt the low-mass star sometime between $10^6$ and $10^5$ years
prior to plunge, so the higher $\rm SNR_I$ value must
be treated with some caution.

\begin{figure}
\input{epsf}
\centerline{\epsfysize 9cm \epsfbox{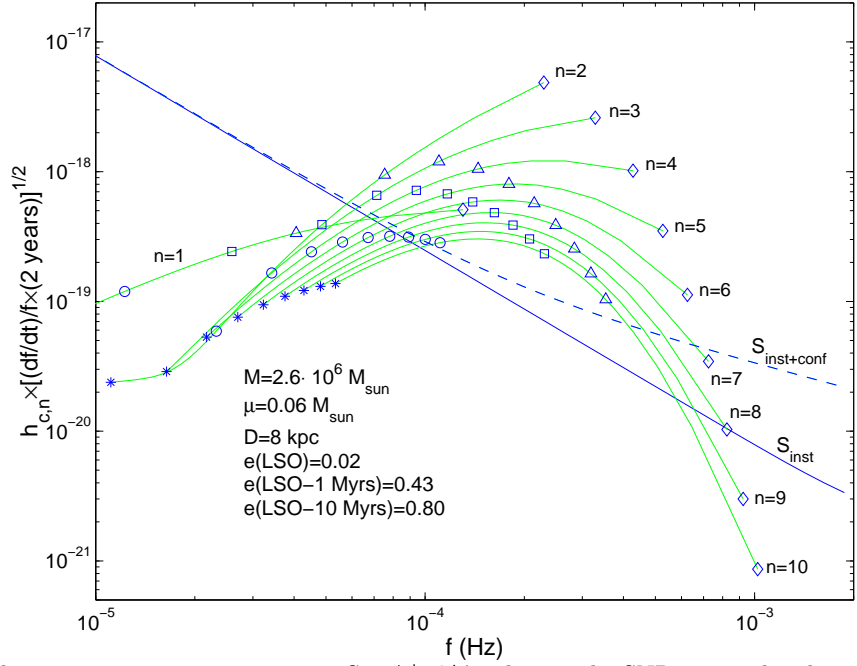}}
\caption{\label{fig:SNFreitag}
SNR for a low-mass main-sequence star at Sgr A*.
`$\diamondsuit$' indicates the SNR accumulated over 2 years
of observation when the system is $10^5$ years before plunge.
`$\triangle$', `$\Box$', `$\circ$', and `$\ast$' similarly correspond
to $10^6$, $2\cdot 10^6$, $5\cdot 10^6$, and $10^7$ years before
plunge, respectively.
}
\end{figure}


\section{Parameter Estimation} \label{Sec:ParaAcc}

\subsection{Numerical implementation}

For our parameter-accuracy estimates we wrote a simple numerical
code, based on the following prescription.
\begin{itemize}

\item Pick a specific point $\lambda^a$ in parameter space.
We found it convenient to first prescribe a value for the eccentricity
at the LSO, $e_{\rm LSO}$; then get the associated LSO frequency
using, for simplicity, the Schwarzschild value
\cite{Cutler-Kennefick-Poisson}
\be\label{nuLSO}
\nu_{\rm LSO} = (2\pi M)^{-1}[(1-e_{LSO}^2)/(6 + 2e_{LSO})]^{3/2};
\ee
and finally obtain $e_0$ and $\nu_0$ by integrating the evolution
equations of Sec.\ \ref{EvEq} one year back in time.
The parameter $t_0$ is set to be in the middle of the integration
time. Namely, the integration is carried out from
$t_{\rm init}\equiv t_0-(1/2)$yr to $t_{\rm LSO}\equiv t_0+(1/2)$yr.

\item For that point, calculate the $n$-modes $h_{I,n}(t)$ and
$h_{II,n}(t)$ of Eq.\ (\ref{halpha}), for the last year of inspiral,
as described in Sec.\ \ref{Sec:waveform}.
The number of modes one has to take into account varies with
the prescribed LSO eccentricity. We determined this number by
requiring that the relative error in the final Fisher matrix components due to
omission of higher-$n$ modes is not greater than $\sim 10^{-6}$.
At $e_{\rm LSO}=0.5$ this meant summing over $\sim 20$ modes.
Note that at the $n$'th mode, the 1-year long function $h_{\alpha,n}$
contains a huge number of wave cycles---roughly $\sim 10^5 n$.
The time resolution has been set such that each wave-cycle is sampled
at least $10$ times. At $n=20$, this meant a time resolution
of about 5 seconds.

\item For each relevant $n$, use Eq.\ (\ref{Shtot}) to calculate
LISA'a noise $S_n(f_n)$. Then sum over modes using Eq.\ (\ref{replace})
to obtain the ``noise-weighted'' waveforms $\hat h_{I}(t)$ and
$\hat h_{II}(t)$.

\item Calculate the SNR through Eq.\ (\ref{SNR}), using the time-domain
approximation for the inner product:
\be\label{SNR_approx}
{\rm SNR}^2= 2\sum_{\alpha=I,II}\int_{t_{\rm init}}^{t_{\rm LSO}}
\hat h_{\alpha}^2(t)dt.
\ee

\item Calculate the $13\times 2$ derivatives
$\partial_a \hat h_{\alpha}\equiv \partial\hat h_{\alpha}/\partial\lambda^a$.
We take these derivatives numerically, through
$\left[\hat h_{\alpha}(\lambda^a\to \lambda^a+\delta\lambda^a/2)-
\hat h_{\alpha}(\lambda^a\to \lambda^a-\delta\lambda^a/2)\right]/
\delta\lambda^a$ (for each $\lambda^a=\lambda^0,\ldots,\lambda^{13}$).
Namely, for each derivative we calculate the waveform twice, with
the relevant parameter shifted by $\pm\delta\lambda^a/2$.
The shift $\delta\lambda^a$ is set such that the resulting relative
error in the Fisher-matrix elements is $\lesssim 10^{-6}$.
[The dependence of the waveform in some of the parameters is such
the derivatives can, in principle, be taken analytically. For some
other parameters (like $e_0$ or $\Phi_0$) the dependence is less
explicit, and one is required to take the derivatives numerically.
We found it convenient (and accurate enough) to take all derivatives
numerically.]

\item Calculate all elements of the Fisher matrix $\Gamma_{ab}$,
using Eq.\ (\ref{inner_approx}).

\item Use Eq.\ (\ref{bardx}) to calculate the measurement error for
each of the parameters:
\be\label{delta lambda}
\Delta\lambda^a= \sqrt{(\Gamma^{-1})^{aa}}
\ee
(no summation over $a$ on the RHS). Obtain $\Delta\Omega_S$
and $\Delta\Omega_K$ using Eqs.\ (\ref{deloms}) and (\ref{delomk}).
To invert the Fisher matrix, we used a numerical subroutine based on
{\bf dgaussj(\,)} of Ref.\ \cite{NR}.

\item Finally, multiply each of the various $\Delta\lambda^a$ by SNR/30.
This, in effect, normalizes the distance to the source such
that the SNR becomes 30 (roughly the SNR output from a $3 M_{\odot}$
CO captured by a $10^6 M_{\odot}$ at $D=1$ Gpc).

\end{itemize}

\subsection{Results for 1-yr integrations}
Although in Sec.~V we stressed the importance
of the signal-to-noise built up in the last several years prior to plunge,
in this section, due to limitations of computer memory and speed,
we restrict ourselves to waveforms coming from the last year of
inspiral. In essence, in this section we pretend that LISA
is ``off-line'' prior to $t_{LSO} - 1\,$yr, where
$t_{LSO}$ is the instant of plunge.

We presents our results from inverting the Fisher matrix
for MBH mass of $10^6 M_{\odot}$, CO masses of $1 M_{\odot}$ and
$10 M_{\odot}$, and a range of values of the MBH spin and the orbital
eccentricity at the LSO.
The rest of the parameters have been set as follows:
$t_0=t_{LSO}-(1/2)$yr (middle of integration),
$\tilde\gamma_0=0$,
$\Phi_0=0$,
$\theta_S=\pi/4$,
$\phi_S=0$,
$\lambda=\pi/6$,
$\alpha_0=0$,
$\theta_K=\pi/8$, and
$\phi_K=0$.
Again, the angles $\bar\phi_0$ and $\bar\alpha_0$ specifying
LISA's position and orientation at $t_0$ are
set to zero.
Tables \ref{tableII} and \ref{tableIII} give the results.
We expect that the measurement accuracies for intrinsic
parameters will not depend very sensitively on
the source's position and orientation,
but that the measurement accuracies for extrinsic parameters (e.g., LISA's
angular resolution) will depend rather more sensitively on
the actual values of these angles.
(This was found to be the case in LISA measurements of MBH-MBH
coalescences~\cite{vecchio03}, and we have also been able to verify
it, to a limited extent, in the LISA capture case.)

\begin{table}[thb]
\centerline{$\begin{array}{lcccccccccc}\hline\hline
{S/M^2} & 0.1 & 0.1 & 0.1
   & 0.5 & 0.5 & 0.5
   & 1 & 1 & 1 &1\\
{e_{\rm LSO}} & 0.1 & 0.3 & 0.5
   & 0.1 & 0.3 & 0.5
   & 0.1 & 0.3 & 0.4 & 0.5 \\
\hline\hline
\Delta(\ln  M)   & {2.3e{-}3} & {1.1e{-}3} & {2.0e{-}3}
                 & {6.9e{-}4} & {1.3e{-}3} & {1.8e{-}3}
                 & {9.2e{-}4} & {1.3e{-}3} & {2.0e{-}3} & {6.5e{-}3} \\ \hline
\Delta (S/M^2)   & {2.5e{-}4} & {1.3e{-}4} & {3.0e{-}4}
                 & {3.6e{-}4} & {8.5e{-}4} & {1.7e{-}3}
                 & {1.0e{-}3} & {2.1e{-}3} & {6.3e{-}3} & {1.3e{-}2} \\ \hline
\Delta(\ln\mu)   & {1.0e{-}3} & {3.2e{-}4} & {1.7e{-}3}
                 & {3.9e{-}4} & {2.2e{-}4} & {4.0e{-}3}
                 & {3.4e{-}4} & {6.0e{-}4} & {7.4e{-}3} & {5.4e{-}2} \\ \hline
\Delta (e_{0})   & {7.1e{-}4} & {2.7e{-}4} & {5.0e{-}4}
                 & {7.6e{-}4} & {4.4e{-}4} & {8.2e{-}4}
                 & {1.2e{-}3} & {8.6e{-}4} & {3.0e{-}3} & {9.7e{-}3} \\ \hline
\Delta (\cos\lambda) & {6.2e{-}2} & {3.1e{-}2} & {6.5e{-}2}
                 & {3.6e{-}3} & {7.9e{-}3} & {1.4e{-}2}
                 & {2.5e{-}3} & {4.8e{-}3} & {1.3e{-}2} & {2.3e{-}2} \\ \hline
\Delta (\Omega_s) & {1.3e{-}3} & {9.4e{-}4} & {3.4e{-}4}
                 & {1.4e{-}3} & {9.8e{-}4} & {3.7e{-}4}
                 & {1.6e{-}3} & {9.4e{-}4} & {7.8e{-}4} & {4.2e{-}4} \\ \hline
\Delta (\Omega_K) & {5.9e{-}2} & {6.4e{-}2} & {8.7e{-}2}
                 & {5.4e{-}2} & {5.1e{-}2} & {4.9e{-}2}
                 & {5.3e{-}2} & {5.0e{-}2} & {5.1e{-2}} & {5.3e{-}2} \\ \hline
\Delta (\tilde\gamma_0) & {5.6e{-}1} & {9.7e{-}1} & {3.1e{-}1}
                 & {5.4e{-}1} & {8.9e{-}1} & {3.1e{-}1}
                 & {4.6e{+}0} & {3.5e{-}1} & {5.1e{-}1} & {3.8e{-}1} \\ \hline
\Delta (\Phi_0)  & {4.1e{-}1} & {8.8e{-}1} & {7.9e{-}2}
                 & {4.2e{-}1} & {9.3e{-}1} & {8.1e{-}2}
                 & {5.8e{+}0} & {1.8e{-}1} & {5.6e{-}1} & {3.0e{-}1} \\ \hline
\Delta (\alpha_0) & {6.2e{-}1} & {5.7e{-}1} & {5.4e{-}1}
                 & {6.2e{-}1} & {5.8e{-}1} & {5.5e{-}1}
                 & {9.7e{-}1} & {5.9e{-}1} & {5.7e{-}1} & {5.6e{-}1} \\ \hline
\Delta [\ln(\mu/D)]  & {2.2e{-}1} & {4.0e{-}2} & {7.1e{-}2}
                 & {2.2e{-}1} & {3.8e{-}2} & {6.4e{-}2}
                 & {3.7e{-}2} & {8.2e{-}2} & {3.9e{-}2} & {4.2e{-}2} \\ \hline
\Delta(t_0)\nu_0 & {8.0e{-}2} & {1.5e{-}1} & {1.4e{-}2}
                 & {8.4e{-}2} & {1.6e{-}1} & {1.5e{-}2}
                 & {1.0e{+}0} & {3.6e{-}2} & {9.5e{-2}} & {4.8e{-}2} \\
\hline\hline
\end{array}$}
\caption{\protect\footnotesize
Parameter accuracy estimates for inspiral of a $1 M_\odot$
CO onto a $10^6 M_\odot$ MBH at SNR=30 (based on data collected
during the last year of inspiral).
Shown are results for various values of the MBH's
spin magnitude $S$ and the final eccentricity $e_{\rm LSO}$.
The rest of the parameters are set as follows:
$t_0=t_{LSO}- (1/2)$yr (middle of integration),
$\tilde\gamma_0=0$,
$\Phi_0=0$,
$\theta_S=\pi/4$,
$\phi_S=0$,
$\lambda=\pi/6$,
$\alpha_0=0$,
$\theta_K=\pi/8$,
$\phi_K=0$.
}
\label{tableII}
\end{table}

\begin{table}[thb]
\centerline{$\begin{array}{lccccccccc}\hline\hline
{S/M^2} & 0.1 & 0.1 & 0.1
   & 0.5 & 0.5 & 0.5
   & 1 & 1 & 1 \\
{e_{\rm LSO}} & 0.1 & 0.3 & 0.5
   & 0.1 & 0.3 & 0.5
   & 0.1 & 0.3 & 0.5 \\
\hline\hline
\Delta(\ln  M)   & {2.6e{-}4} & {5.6e{-}4} & {5.3e{-}5}
                 & {2.7e{-}4} & {9.2e{-}4} & {7.7e{-}5}
                 & {2.8e{-}4} & {2.5e{-}4} & {1.5e{-}4} \\ \hline
\Delta (S/M^2)   & {3.6e{-}5} & {7.9e{-}5} & {4.5e{-}5}
                 & {1.3e{-}4} & {6.3e{-}4} & {5.1e{-}5}
                 & {2.6e{-}4} & {3.7e{-}4} & {2.6e{-}4} \\ \hline
\Delta(\ln\mu)   & {6.8e{-}5} & {1.5e{-}4} & {7.4e{-}5}
                 & {6.8e{-}5} & {9.2e{-}5} & {1.0e{-}4}
                 & {6.1e{-}5} & {9.1e{-}5} & {1.0e{-}3} \\ \hline
\Delta (e_{0})   & {6.3e{-}5} & {1.3e{-}4} & {2.9e{-}5}
                 & {8.5e{-}5} & {2.8e{-}4} & {3.2e{-}5}
                 & {1.2e{-}4} & {1.1e{-}4} & {1.6e{-}4} \\ \hline
\Delta (\cos\lambda) & {6.0e{-}3} & {1.7e{-}2} & {1.3e{-}3}
                     & {1.3e{-}3} & {5.8e{-}3} & {2.4e{-}4}
                     & {6.5e{-}4} & {8.4e{-}4} & {4.7e{-}4} \\ \hline
\Delta (\Omega_s) & {1.8e{-}3} & {1.7e{-}3} & {7.9e{-}4}
                  & {2.0e{-}3} & {1.7e{-}3} & {7.6e{-}4}
                  & {2.1e{-}3} & {1.1e{-}3} & {6.7e{-}4} \\ \hline
\Delta (\Omega_K)& {5.6e{-}2} & {5.3e{-}2} & {4.7e{-}2}
                 & {5.5e{-}2} & {5.1e{-}2} & {4.7e{-}2}
                 & {5.6e{-}2} & {5.1e{-}2} & {4.8e{-}2} \\ \hline
\Delta (\tilde\gamma_0) & {4.0e{-}1} & {6.3e{-}1} & {3.8e{-}1}
                 & {1.0e{+}0} & {6.1e{-}1} & {3.9e{-}1}
                 & {9.3e{-}1} & {3.4e{-}1} & {3.9e{-}1} \\ \hline
\Delta (\Phi_0)  & {2.6e{-}1} & {6.7e{-}1} & {2.2e{-}1}
                 & {1.4e{+}0} & {7.5e{-}1} & {2.7e{-}1}
                 & {1.5e{+}0} & {1.7e{-}1} & {3.3e{-}1} \\ \hline
\Delta (\alpha_0) & {6.2e{-}1} & {5.8e{-}1} & {5.5e{-}1}
                 & {6.3e{-}1} & {5.9e{-}1} & {5.6e{-}1}
                 & {6.4e{-}1} & {5.9e{-}1} & {5.9e{-}1} \\ \hline
\Delta [\ln(\mu/D)]  & {8.7e{-}2} & {3.8e{-}2} & {3.7e{-}2}
                 & {3.8e{-}2} & {3.7e{-}2} & {3.7e{-}2}
                 & {3.8e{-}2} & {7.0e{-}2} & {3.7e{-}2} \\ \hline
\Delta(t_0)\nu_0 & {4.5e{-}2} & {1.1e{-}1} & {3.3e{-}2}
                 & {2.3e{-}1} & {1.3e{-}1} & {4.4e{-}2}
                 & {2.5e{-}1} & {3.2e{-}2} & {5.5-2} \\
\hline\hline
\end{array}$}
\caption{\protect\footnotesize
Parameter extraction accuracy for inspiral of a $10 M_\odot$
CO onto a $10^6 M_\odot$ MBH at SNR=30 (based on data collected
during the last year of inspiral).
Shown are results for various values of the MBH's
spin magnitude $S$ and the final eccentricity $e_{\rm LSO}$.
The rest of the parameters are set as follows:
$t_0=t_{LSO}-(1/2)$yr (middle of integration),
$\tilde\gamma_0=0$,
$\Phi_0=0$,
$\theta_S=\pi/4$,
$\phi_S=0$,
$\lambda=\pi/6$,
$\alpha_0=0$,
$\theta_K=\pi/8$,
$\phi_K=0$.
}
\label{tableIII}
\end{table}

Summarizing the results in Table \ref{tableIII}, for the case of a $10 M_\odot$
BH spiraling into a $10^6 M_\odot$ MBH, the measurement accuracies for
the seven intrinsic parameters are (very roughly) $\Delta(\ln  M)
\sim 2\times 10^{-4}$, $\Delta(\ln\mu) \sim 10^{-4}$, $\Delta (S/M^2) \sim 10^{-4}$,
$\Delta (\cos\lambda) \sim 10^{-3}$, $\Delta e_0 \sim 10^{-4}$, and
$\Delta \Phi_0 \sim \Delta \tilde\gamma_0 \sim 0.5$.

Comparing Tables \ref{tableII} and \ref{tableIII} ($1 M_\odot$ vs.\
$10 M_\odot$ CO), we see $\Delta(\ln  M)$, $\Delta(\ln\mu)$,
$\Delta (S/M^2)$, $\Delta (\cos\lambda)$, and $\Delta e_0$ are all typically
about an order of magnitude smaller for the  $10 M_\odot$ CO case (again, for
fixed SNR and fixed 1-yr integration time). This general trend of better
accuracy for higher CO mass is not
hard to understand. First, since $\dot \nu \propto \mu$, it is clear
that $\Gamma_{\ln\mu\, \ln\mu}$
should scale roughly like $\mu^2$, so $\Delta(\ln\mu) \equiv
[(\Gamma^{-1})^{\ln\mu\, \ln\mu}]^{1/2}$ should scale roughly as $\mu^{-1}$.
Similarly, the derivative
$\partial_{\ln M}h_{\alpha}(t)$ has size of order
$h_{\alpha}(t)\partial_{\ln M}\Phi(t)$ (during most of the integration time),
and it is clear from Eqs.~(\ref{Phidot})--(\ref{nudot}) that the typical size of
$\partial_{\ln M}\Phi(t)$ scales roughly like $\mu$. The magnitudes of
$\partial_{S/M^2}h_{\alpha}(t)$,  $\partial_{{\rm cos}\lambda}h_{\alpha}(t)$,
and $\partial_{e_0}h_{\alpha}(t)$ also
scale roughly linearly with $\mu$, for the same reason.
Hence, it is reasonable to expect that errors in these five intrinsic
variables should scale roughly like $\mu^{-1}$.

We next turn to the extrinsic parameters. Our few examples suggest
that LISA's angular resolution for capture sources is
$\Delta \Omega_s \sim 10^{-3}\,$ radians, while the MBH spin
direction can be determined to within
$\Delta \Omega_K \sim 5\times 10^{-2}$.
We find $\Delta [\ln(\mu/D)] \sim 5\times 10^{-2}$, typically.
Since $\Delta (\ln \mu) \ll \Delta [\ln(\mu/D)]$, it is clear
that $\Delta (\ln D) \sim 5\times 10^{-2}$ too.
[We may verify the last statement by examining the [$\ln\mu,\ln(\mu/D)$]
minor of $\Gamma^{-1}$: We have
$$
(\Gamma^{-1})^{\ln D,\ln D} = (\Gamma^{-1})^{\ln(\mu/D),\ln(\mu/D)}
+ (\Gamma^{-1})^{\ln\mu,\ln\mu}  -2(\Gamma^{-1})^{\ln\mu,\ln(\mu/D)},
$$
and since typical values are
$(\Gamma^{-1})^{\ln(\mu/D),\ln(\mu/D)}\sim 10^{-4}$,
$(\Gamma^{-1})^{\ln\mu,\ln\mu}\sim 10^{-9}$, and
$(\Gamma^{-1})^{\ln\mu,\ln(\mu/D)}\sim 10^{-9}$,
we find that indeed
$\Delta (\ln D) \approx \Delta [\ln(\mu/D)]$.]
Finally, as a check, it is easy to see that $\Delta [\ln(\mu/D)]$
must be greater than SNR$^{-1} = 0.033$ (since the signal amplitude is
linear in $\mu/D$), which is
indeed satisfied in every column of our tables.


\subsection{Comparison with other results in the literature}

Our angular resolution results can be compared to
results by Cutler and Vecchio~\cite{cutler_vecchio}
on LISA's angular resolution for monochromatic sources.
For a monochromatic
source with $f_{gw} = 3$ mHz and SNR=30, LISA's angular resolution is
typically $\Delta \Omega_s \sim 5 \times 10^{-4}\,$ (estimated by
interpolating between Figs.~2 and 3 in \cite{cutler_vecchio}, after
rescaling those figures to SNR=30), which is only a factor $\sim 2$ smaller
than our result for capture sources. Since capture sources have
twice as many unknown parameters as monochromatic sources
($14$ versus $7$), it is clear that LISA's angular resolution must be
worse for the former (at the same SNR), but the ``good news'' is that
this degradation appears to be quite modest,
based on our limited sample.

Our results on the mass and spin determination accuracy can be compared
to previous results by Ryan~\cite{ryan_multipoles} and
Poisson~\cite{Poisson96}.
Ryan's  waveforms are based on PN
evolution equations (similar to ours), while Poisson's are based on a
Taylor expansion of the waveform phase near plunge, with
expansion coefficients
obtained from numerical solution of the Teukolsky equation.
Both these authors consider only circular, equatorial orbits
(so $e_0 = 0$ and $\cos\lambda = 1.0$, a priori).
Both simplify the calculation further by
ignoring the waveform modulation caused by LISA's motion
(so they effectively pretend LISA is fixed at the
center of our solar system), and by restricting attention to the
waveform generated by just a single pair of LISA's arms.
Thus, their waveforms are determined by only $5$ parameters: an overall
amplitude and phase, the two masses, and $S/M^2$.
(Clearly, these simplifications were intended to make the
Fisher matrix calculation essentially identical to the
corresponding calculation for LIGO measurements of
binary black hole coalescences.)

The fact that Ryan and Poisson effectively ``toss out'' most of
the unknown parameters obviously tends to decrease the calculated
error bars for the included parameters.
On the other hand, their highly simplified waveforms obviously carry much less
information than the true waveforms, which
tends to have the opposite effect. A priori, it would seem difficult to guess
whether the net effect of their approximations is to underestimate
or overestimate $\Delta(\ln  M)$,  $\Delta(\ln\mu)$, and
$\Delta (S/M^2)$. Therefore, unfortunately, their work does not seem to
provide a useful check on ours. Nevertheless, Ryan's and Poisson's papers
were an interesting first-cut at the parameter estimation problem, and
it seems interesting to compare our results to theirs.

For a $10 M_\odot$ CO and
$10^6 M_\odot$ MBH, Ryan~\cite{ryan_multipoles} obtains (at SNR=30):
$\Delta(\ln  M) = 1.8 \times 10^{-4}$, $\Delta(\ln\mu) = 1.9 \times 10^{-5}$, and
$\Delta (S/M^2) = 4.9 \times 10^{-4}$.  For the same masses and SNR,
Poisson~\cite{Poisson96} states the results
$\Delta(\ln  M) = 6.7 \times 10^{-5}$, $\Delta (S/M^2) = 1.7 \times 10^{-3}$,
and $\Delta(\ln\eta) =  1.8\times 10^{-3}$, where $\eta \equiv \mu/M$.
Since both Ryan and Poisson consider only the case $e_0 =0$ and
evaluate the Fisher matrix at the point $S/M^2 = 0$, their results are
most usefully compared to those in column 1 of our Tables (i.e.,
$e_0 = 0.1$ and $S/M^2 = 0.1$).
Our estimates of mass-determination accuracy are within roughly
an order of magnitude of those quoted by Ryan and Poisson, and in
fact lie between them.
[Poisson does not quote a result for $\Delta(\ln\mu)$, but
we can still compare our results directly to his by using
\be\label{Delta_eta}
(\Gamma^{-1})^{\ln\eta , \ln\eta} = (\Gamma^{-1})^{\ln\mu, \ln\mu}
+ (\Gamma^{-1})^{\ln M, \ln M}  -2(\Gamma^{-1})^{\ln\mu, \ln M} \, ,
\ee
which gives $\Delta \eta = 3.0\times 10^{-4}$ for our case.]
Our $\Delta (S/M^2)$ is $\sim 10$ times
smaller than Ryan's and $\sim 50$ times smaller than Poisson's.
We guess this is because in Ryan's and Poisson's waveforms
the MBH spin affects only the orbital phase (whereas in the true waveforms
the spin also controls the Lense-Thirring precession rate), leading them
to overestimate the covariance of the spin with the two mass parameters, and
hence to overestimate $\Delta (S/M^2)$.



\subsection{Parameter extraction for a low-mass main-sequence star at Sgr A*.}

Finally, it's interesting to consider the parameter extraction accuracy
for captures of LMMSs at the center of the
Milky Way---the type of source whose anticipated SNR we discussed
at the end of Sec.\ \ref{Sec:SNR}. In Table \ref{tableIV} we consider
a few possible low-frequency orbits of a $0.06 M_{\odot}$ CO around
the $2.6\cdot10^6 M_{\odot}$ MBH at the known distance and sky location
of Sgr A*. We take D=7.9 kpc \cite{D} and $\theta_S=1.66749$; the value
for $\phi_S$ is picked arbitrarily, since the orbital location of the
detector during the observation is unknown. The columns with eccentricity
$0.43$ and frequency $0.0035$ mHz correspond to the orbit whose SNR
output has been discussed above (in Fig.\ \ref{fig:SNFreitag}; also
\cite{FreitagNew}), and assume the observation is made $10^6$ years prior
to plunge (just before tidal forces play an important role). In other
columns we consider other possible values of the eccentricity, frequency,
and MBH's spin. In calculating the Fisher matrix, we assumed a data
integration time of 2 years.

\begin{table}[thb]
\centerline{$\begin{array}{lccccccccc}\hline\hline
{S/M^2} & 0.1 & 0.1 & 0.1
   & 0.5 & 0.5 & 0.5
   & 1 & 1 & 1 \\
{e} & 0.10 & 0.43 & 0.80
   & 0.10 & 0.43 & 0.80
   & 0.10 & 0.43 & 0.80 \\
\nu_0 ({\rm mHz}) & 0.044 & 0.035  & 0.013
                & 0.044 & 0.035  & 0.013
                & 0.044 & 0.035  & 0.013 \\
{\rm SNR}       & 19.8  &  29.9   & 26.4
                & 19.0  &  28.3   & 25.7
                & 18.8  &  28.4   & 29.7 \\
\hline\hline
\Delta(\ln  M)   & {3.7e{-}3} & {3.6e{-}2} & {2.1e{-}1}
                 & {5.9e{-}3} & {3.8e{-}2} & {2.1e{-}1}
                 & {1.0e{-}2} & {3.8e{-}2} & {2.1e{-}1} \\ \hline
\Delta (S/M^2)   & {2.5e{-}3} & {1.9e{-}3} & {6.9e{-}3}
                 & {3.2e{-}3} & {2.5e{-}3} & {5.1e{-}3}
                 & {9.6e{-}3} & {6.8e{-}3} & {9.9e{-}3} \\ \hline
\Delta(\ln\mu)   & {6.1e{+}3} & {4.0e{+}3} & {4.5e{+}3}
                 & {6.4e{+}3} & {4.3e{+}3} & {4.6e{+}3}
                 & {6.5e{+}3} & {4.2e{+}3} & {4.1e{+}3} \\ \hline
\Delta (e_{0})   & {1.2e{-}2} & {2.4e{-}2} & {3.3e{-}2}
                 & {1.3e{-}2} & {2.5e{-}2} & {3.3e{-}2}
                 & {1.3e{-}2} & {2.5e{-}2} & {3.2e{-}2} \\ \hline
\Delta (\cos\lambda) & {3.2e{-}2} & {1.8e{-}2} & {2.0e{-}2}
                 & {2.6e{-}2} & {1.7e{-}2} & {1.8e{-}2}
                 & {2.6e{-}2} & {1.7e{-}2} & {1.6e{-}2} \\  \hline
\Delta (\Omega_s) & {3.1e{-}2} & {9.4e{-}3} & {1.1e{-}2}
                 & {2.0e{-}2} & {7.6e{-}3} & {7.0e{-}3}
                 & {2.0e{-}2} & {7.9e{-}3} & {7.1e{-}3} \\ \hline
\Delta (\Omega_K) & {3.1e{-}2} & {1.1e{-}2} & {1.0e{-}2}
                 & {1.9e{-}2} & {7.5e{-}3} & {6.2e{-}3}
                 & {2.0e{-}2} & {8.0e{-}3} & {7.5e{-}3} \\
{}               & {(1.1e{-}2)} & {(5.1e{-}3)} & {(6.4e{-}3)}
                 & {(1.2e{-}2)} & {(4.6e{-}3)} & {(4.2e{-}3)}
                 & {(1.2e{-}2)} & {(4.9e{-}3)} & {(4.8e{-}3)} \\ \hline
\Delta (\tilde\gamma_0) & {1.6e{+}1} & {3.6e{-}1} & {2.5e{-}1}
                 & {1.5e{+}1} & {3.5e{-}1} & {2.3e{-}1}
                 & {1.3e{+}1} & {3.1e{-}1} & {2.1e{-}1} \\ \hline
\Delta (\Phi_0)  & {2.0e{+}2} & {3.6e{+}0} & {1.7e{+}0}
                 & {2.0e{+}2} & {3.8e{+}0} & {1.7e{+}0}
                 & {2.1e{+}2} & {3.8e{+}0} & {1.6e{+}0} \\ \hline
\Delta (\alpha_0) & {1.8e{-}1} & {7.6e{-}2} & {6.5e{-}2}
                 & {7.4e{-}1} & {7.2e{-}2} & {6.3e{-}2}
                 & {1.5e{+}0} & {7.8e{-}2} & {7.4e{-}2} \\ \hline
\Delta [\ln(\mu/D)]  & {8.0e{-}2} & {7.8e{-}2} & {2.6e{-}1}
                 & {7.1e{-}2} & {7.8e{-}2} & {2.6e{-}1}
                 & {7.1e{-}2} & {7.7e{-}2} & {2.5e{-}1} \\ \hline
\Delta(t_0)\nu_0 & {3.1e{+}1} & {6.0e{-}1} & {2.9e{-}1}
                 & {3.2e{+}1} & {6.3e{-}1} & {2.9e{-}1}
                 & {3.2e{+}1} & {6.3e{-}1} & {2.8e{-}1} \\
\hline\hline
\end{array}$}
\caption{\protect\footnotesize
Parameter extraction accuracy for a low-mass main-sequence star
at Sgr A*. We assume $M=2.6\cdot 10^6 M_\odot$, $\mu=0.06 M_\odot$,
and data integration lasting {\bf 2 years}.
We also assume the star is observed a million years before the
(theoretical) plunge, just before tidal effects become important.
Each column of the table refers to a different choice of the MBH's
spin, orbital eccentricity $e$ and frequency $\nu$ at the
time of observation.
The other parameters are set as follows:
$\tilde\gamma_0=0$,
$\Phi_0=0$,
$\theta_S=1.66749$ (true value for Sgr A*),
$\phi_S=0$,
$\lambda=\pi/6$,
$\alpha_0=0$,
$\theta_K=\pi/8$,
$\phi_K=0$.
Most of the values given in the table result from inverting the full,
$14\times 14$-d Fisher matrix. The values for $\Delta \Omega_K$ obtained by
inverting the $11\times 11$ minor that excludes the
CO's mass $\mu$ and the two sky-location
coordinates $\theta_S$ and $\phi_S$ (whose precise values are known for Sgr A*)
are given in parentheses.
(For all other parameters, using the known sky position did not
signficantly improve measurement accuracy.)
}
\label{tableIV}
\end{table}

Most of the values in the table result from inverting the full, 14-d
Fisher matrix. We find that the sky location of the source can be
determined to within $\sim 0.01$ steradians. Unfortunately, the
distance $D$ to the source is entirely degenerate with the CO's
mass, and the latter cannot be determined by itself from the GW signal.
[This is as expected, since the accumulated effect of radiation reaction
on the waveform phase, after observation time $T_{obs} \sim 1\ $yr, is
merely
$\Delta\phi \sim  \pi \dot f T^2_{obs} \sim \pi f T^2_{obs}/(10^6 {\rm yr})
\sim 10^{-2}$, for $f\sim 10^{-4}\ $ Hz.]
Hence, the distance to the galactic center cannot be determined
by the GW signal alone.
However, in case where a detected GW signal appears to come from a capture
at Sgr A* (as its sky position is consistent with the galactic center
and its $\dot f$ is extremely small),
one could use the value of $D$ known from astronomical measurements,
in order to determine the CO's mass $\mu$.
(Recent studies of stellar dynamics and RR Lyrae stars have specified
$D$ to within $~4\%$ \cite{D}).
The relative error in
$\mu$ will then be approximately $\Delta[\ln(\mu/D)]$, or roughly $10\%$.

Once a source is confirmed to be at the galactic center,
we may eliminate the sky location $\Omega_S$ (known exactly for
Sgr A*) from the search, in order to improve the accuracy in determining
the other parameters. We may also eliminate the CO's mass $\mu$ from the
parameter list (while keeping $\mu/D$)---again, because
radiation reaction is negligible for this source.

To estimate the effect of this extra information on parameter extraction,
we inverted the 11-d minor of the Fisher matrix obtained by excluding
the rows and columns associated with $\mu_S$, $\phi_S$, and $\ln\mu$.
We found that this produces only a negligible improvement in
measurement accuracy for the rest of the parameters, except for
the MBH spin's direction;
$\Delta \Omega_K$ decreases by a factor $\sim 1.5$--$3$.
These improved values of $\Delta\Omega_K$ are given in the table
as well, set off by parentheses.

The more interesting results of Table \ref{tableIV} concern the
intrinsic parameters of the galactic hole. We estimate that the mass of
the central MBH (as well as the orbital eccentricity) could be
determined to within relative error $\sim 4\times 10^{-3}$ to
$2\times 10^{-1}$, i.e., comparable to or better than what is achieved today
using astrometric methods.
More impressive is the determination of the MBH's spin from the GW waveform
to within $\sim 0.005$ (in units of $M^2$).
It is hard for us to imagine an alternative method
that would allow such an accurate measurement of the spin of the
MBH at the Milky Way's center.

Finally, we point out that two approximations we have made---using
PN equations of motion and a low-frequency approximation to LISA's
response---should be really quite accurate for LMMS captures in our galaxy,
since these inspirals are viewed at a very early stage
(the orbits considered in Table \ref{tableIV} have their
pericenters at $\gtrsim 20M$), and most of the SNR comes from
frequencies in the range $10^{-4}$--$10^{-3}\ $Hz.

\section{Conclusions, Caveats, and Future Work} \label{Sec:summary}

Realistic capture orbits will be nonequatorial
and somewhat eccentric, in general. We have made a first cut
at answering some LISA data analysis questions, for such
realistic cases.

The figures in Sec.~V illustrate how the
LISA SNR builds up over time for eccentric orbits
(assuming there, for simplicity, that the MBH is nonspinning).
These figures show that, for $\sim 10 M_{\odot}$ COs captured by
$\sim 10^6 M_\odot$ MBHs, the last few years
of inspiral can all contribute significantly to the SNR;
for $\sim 1 M_{\odot}$ COs, the last few decades can be
significant. Clearly, these long waveforms will increase the
computational burden on matched-filtering detection schemes, and will
exacerbate the self-confusion problem, since eccentric-orbit
inspirals can deliver signicant GW energy into the LISA band many
years before they become individually detectable.

The tables in Sec.~VI represent our attempt to estimate
LISA's parameter estimation accuracy, for captures.
For a typical source (a $10 M_\odot$ CO captured by a $10^6 M_\odot$ MBH
at SNR of 30)
we find that $\Delta(\ln M), \Delta(\ln \mu), \Delta(S/M^2)$,
and $\Delta e_0$ will all be roughly $\sim 10^{-4}$, while
$\Delta\Omega_S \sim 10^{-3}$
and $\Delta\Omega_K \sim 5\times 10^{-2}$. Due to computational
limitations, those results are based on measurements from only the
final year prior to plunge. We naturally expect the measurement
errors to decrease when one considers waveforms lasting the
entire length of the LISA mission. In this sense, the numbers
above represent rough upper limits to LISA's measurement accuracy.
On the other hand, these estimates are based on a small
sample of hand-picked points in
parameter space. In the future we will improve these estimates by
doing several-year integrations and a full Monte Carlo sampling of
parameter space. Of course, our PN waveforms are probably not
very accurate for COs very close to the plunge, but still the above
estimates are the best ones available.

LISA has a reasonable chance of detecting LMMSs
captured by the MBH at the center of our own galaxy.
In this case, the sources will be detected $\sim 10^6\ $yrs before the
final plunge, when our PN waveforms and our low-frequency approximation
to LISA's response function should both be more reliable.
We find that, in a two-year integration, LISA could determine the
magnitude and direction of our
MBH's spin to within
$\Delta (S/M^2) \sim 5 \times 10^{-3}$
and $\Delta\Omega_K \sim 10^{-2}$, and measure the
mass of the infalling star to within $\sim 10\%$.

\begin{acknowledgments}
C.C.'s work was partly supported by NASA grants
NAG5-4093 and NAG5-12834.
L.B.'s work was supported by a Marie Curie Fellowship of the European
Community program IHP-MCIF-99-1 under contract number HPMF-CT-2000-00851;
by NSF Grant NSF-PHY-0140326 (`Kudu'); and by a grant from
NASA-URC-Brownsville (`Center for Gravitational Wave Astronomy').

This paper is part of larger effort by LISA's Working Group I (WG-I)
to address scientific issues related to capture sources.
Regular progress reports were given at WG-I telecons and posted
on the WG-I website. We benefited from interactions with
the rest of WG-I members, and in particular wish to thank Kip Thorne,
Sterl Phinney, Scott Hughes, and Michele Vallisneri.
We also thank Yanbei Chen for helpful discussions and
especially thank Jonathon Gair for helping check our code by
comparing some results to those from a similar code he has
written. Lastly, L.B. wishes to thank Carlos Lousto,
Manuela Campanelli, and the rest of CGWA members, for discussions
and continuous support.
\end{acknowledgments}

\appendix

\section{Lense-Thirring component of the pericenter precession}


At 1PN order, the piece of $\dot{\gamma}\equiv d\gamma/dt$
proportional to $S$ is~\cite{Brumberg}
\ban\label{gammadotS}
\dot{\gamma}|_{\propto S} &=&
\frac{{\hat L}\cdot{\hat n}}{1-({\hat L}\cdot{\hat n})^2}
({\hat L}\cdot{\hat n}\cos\lambda -{\hat S}
\cdot{\hat n})\dot{\alpha}
- 8\pi\nu (S/M^2)\cos\lambda (2\pi M\nu)(1-e^2)^{-3/2},
\ean
where $\dot{\alpha}\equiv d\alpha/dt$ is given in Eq.\ (\ref{alphadot}).
[To derive this equataion, use Eq.~(4.4.45) of Brumberg~\cite{Brumberg},
together with his Eqs.~(4.4.41) and (4.4.33), and along with the definitions
in Eqs. (1.1.6)--(1.10.10) therein.]

The expression (\ref{gammadotS}) is somewhat complicated (and worse,
badly behaved as $\hat L$ approaches $\hat n$) due to the usual
convention of defining $\gamma$ as the
angle from $\hat x \equiv
[\hat L(\hat L \cdot \hat n) - \hat n]/(1 - (\hat L \cdot \hat n)^2)^{1/2} $
to pericenter. In the terminology of BCV \cite{BCV}, $\gamma$ is neither wholly an
intrinsic variable, nor an extrinsic one; it is of a mixed type.
As explained in Sec.\ \ref{Sec:waveform}, it is preferrable to introduce a pericenter
angle that is defined purely intrinsically (i.e., without reference
to the observer). Accordingly, we define
$\tilde\gamma$ as the
angle from $\hat L \times \hat S$ to the direction of pericenter.
Then $\gamma$ and $\tilde\gamma$ are related by
\be\label{beta2}
\gamma = \tilde\gamma + \beta,
\ee
where $\beta$ is given by Eq.~(\ref{sinbeta}).
Not only is $\tilde\gamma$ wholly intrinsic, but we claim it also
obeys a simpler (and better behaved) evolution equation, Eq.~(\ref{Gamdot}):
\be\label{GamdotS}
\dot{\tilde\gamma}|_{\propto S} = -12\pi\nu\cos\lambda (S/M^2)
(2\pi M\nu)(1-e^2)^{-3/2}.
\ee

The purpose of this Appendix is to prove this claim; i.e., we show
that Brumberg's Eq.~(\ref{gammadotS})
is indeed equivalent to our Eqs.~(\ref{GamdotS}) and (\ref{beta2}).

Since it is obvious from the definitions that
\be
\gamma_0 = \tilde\gamma_0 + \beta_0 \, ,
\ee
what we need to prove is that
\be\label{dotbeta}
\dot\beta = \dot\gamma - \dot{\tilde\gamma} \, .
\ee
From Eqs.~(\ref{gammadotS}), (\ref{GamdotS}), and (\ref{alphadot}),
and noting that
$\dot\gamma$ and $\dot{\tilde\gamma}$ differ only in their piece
$\propto S$, the RHS of Eq.~(\ref{dotbeta}) becomes
\ban \label{rhs1}
\dot\gamma-\dot{\tilde\gamma}&=&
\frac{{\hat L}\cdot{\hat n}}{1-({\hat L}\cdot{\hat n})^2}
({\hat L}\cdot{\hat n}\cos\lambda -{\hat S}\cdot{\hat n})\, \dot\alpha
+\cos\lambda\,  \dot\alpha\, \nonumber\\
&=& \frac{\cos\lambda - ({\hat L}\cdot{\hat n})({\hat S}\cdot{\hat n})}
{1- ({\hat L}\cdot{\hat n})^2} \, \dot\alpha
\ean
The left-hand side (LHS) of Eq.~(\ref{dotbeta}) can be written as
\ban\label{lhs}
\dot\beta = (\cos\beta)^{-1}d(\sin\beta)/dt,
\ean
which, substituting for $\sin\beta$ and $\cos\beta$ from Eq.~(\ref{sinbeta}),
yields
\ban\label{lhs2}
\dot\beta =
\frac
{\bigl[1 - (\hat L\cdot\hat n)^2\bigr]^{1/2}}
{\hat n \cdot (\hat S \times \hat L)}
\left\{\frac{\cos\lambda \,(\dot{\hat L}\cdot{\hat n})}
{\bigl[1 - (\hat L\cdot\hat n)^2\bigr]^{1/2}}
+\frac{(\cos\lambda \hat{L}\cdot\hat{n}-\hat{S}\cdot\hat{n})
(\hat{L}\cdot\hat{n})(\dot{\hat{L}}\cdot{\hat n})}
{\bigl[1 - (\hat L\cdot\hat n)^2\bigr]^{3/2}}
\right\}
.
\ean
Using now the evolution equation for $\hat L$,
$\dot{\hat L} = \dot\alpha \, \hat S \times \hat L$,
the last expression reduces to the RHS of Eq.\ (\ref{rhs1}).
\noindent This proves Eq.\ (\ref{dotbeta}), and
hence Eq.~(\ref{GamdotS}) is proved.

We note Eq.~(\ref{GamdotS}) can also be seen as a special case of
Eq.~(\ref{gammadotS}), obtained by setting $\hat n$ equal
to $\hat S$ in the latter.
Finally, we also note that the term
\be\label{Thomas}
\frac{{\hat L}\cdot{\hat n}}{1-({\hat L}\cdot{\hat n})^2}
({\hat L}\cdot{\hat n}\cos\lambda -{\hat S}\cdot{\hat n})\, \dot\alpha
\ee
in Eq.~(\ref{gammadotS}) for $\dot\gamma$
is precisely the time derivative of what has been referred to in Ref.\ \cite{haris}
(in the context of quasi-circular orbits) as the ``Thomas precession phase''.
[The term (\ref{Thomas}) is precisely the RHS of Eq.~(28) in~\cite{haris}.]
We see  here that this ``Thomas precession'' term is really just the
``${\hat{L}}\times{\hat{S}}$ piece'' of the pericenter precession
[while the remainder of Eq.\ (\ref{gammadotS}) is the
``${\hat{L}}\cdot{\hat{S}}$ piece''].

\section{Relative magnitudes of PN contributions}

Our goal here is to gauge the suitability of the PN formulae
we use in evolving the CO's orbit within our model
[Eqs.\ (\ref{Phidot})--(\ref{edot})]. The part of the inspiral
relevant for detection by LISA takes place entirely within the
highly relativistic region right next to the horizon, a regime
where the validity of the PN expansion is normally to be
suspected. 
Accordingly, we do not expect our PN expressions to show any
convergence; we do wish to make sure, though, that truncating
the PN expansion as in Eqs.\ (\ref{Phidot})--(\ref{edot}) does
not lead to a pathological, unphysical evolution within the range
of parameters relevant for our analysis.

In Fig.\ (\ref{fig:PN}) we compare the magnitudes of the contributions
from the various PN expansion terms to each of the time derivatives
$\dot{\nu}$, $\dot{e}$, and $\tilde{\dot\gamma}$.
We consider the ``worse case'', in which the CO has just reached
the last stable orbit (LSO) and is about to plunge into the MBH.
For simplicity, we take $(2\pi M\nu)_{\rm LSO}=
\left[(1-e^2)/(6+2e)\right]$, which is the value for a non-rotating MBH.

As expected, the various PN terms are of comparable magnitudes.
Nevertheless, our evolution equations do not develop pathologies,
so long as the LSO eccentricity is not too high. In particular,
the orbital frequency increases monotonically (no ``outspiral''!)
and the eccentricity decreases monotonically (the orbit circularizes)
throughout the entire evolution.

\begin{figure}[htb]
\input{epsf}
\centerline{\epsfysize 7cm \epsfbox{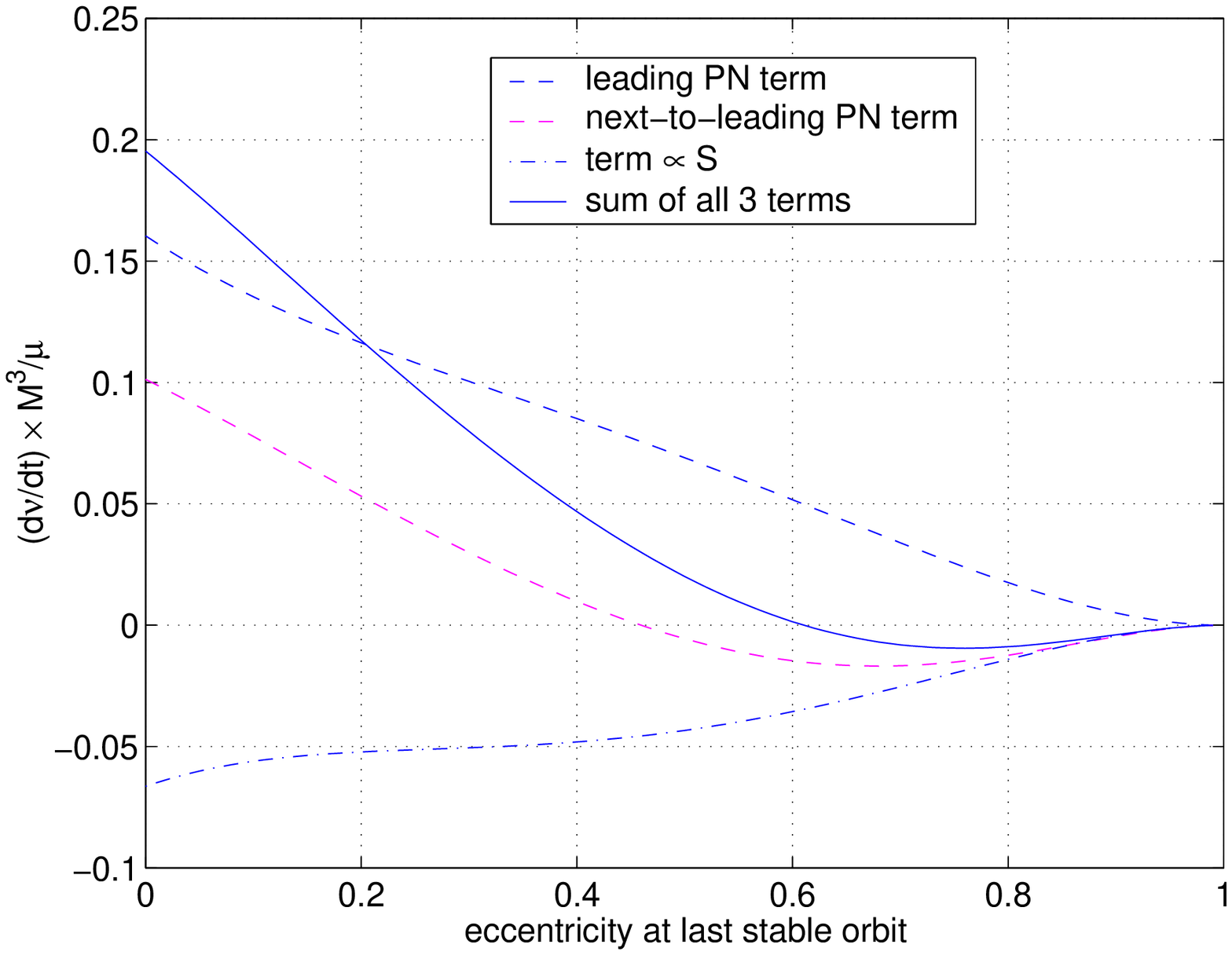}
\epsfysize 7cm \epsfbox{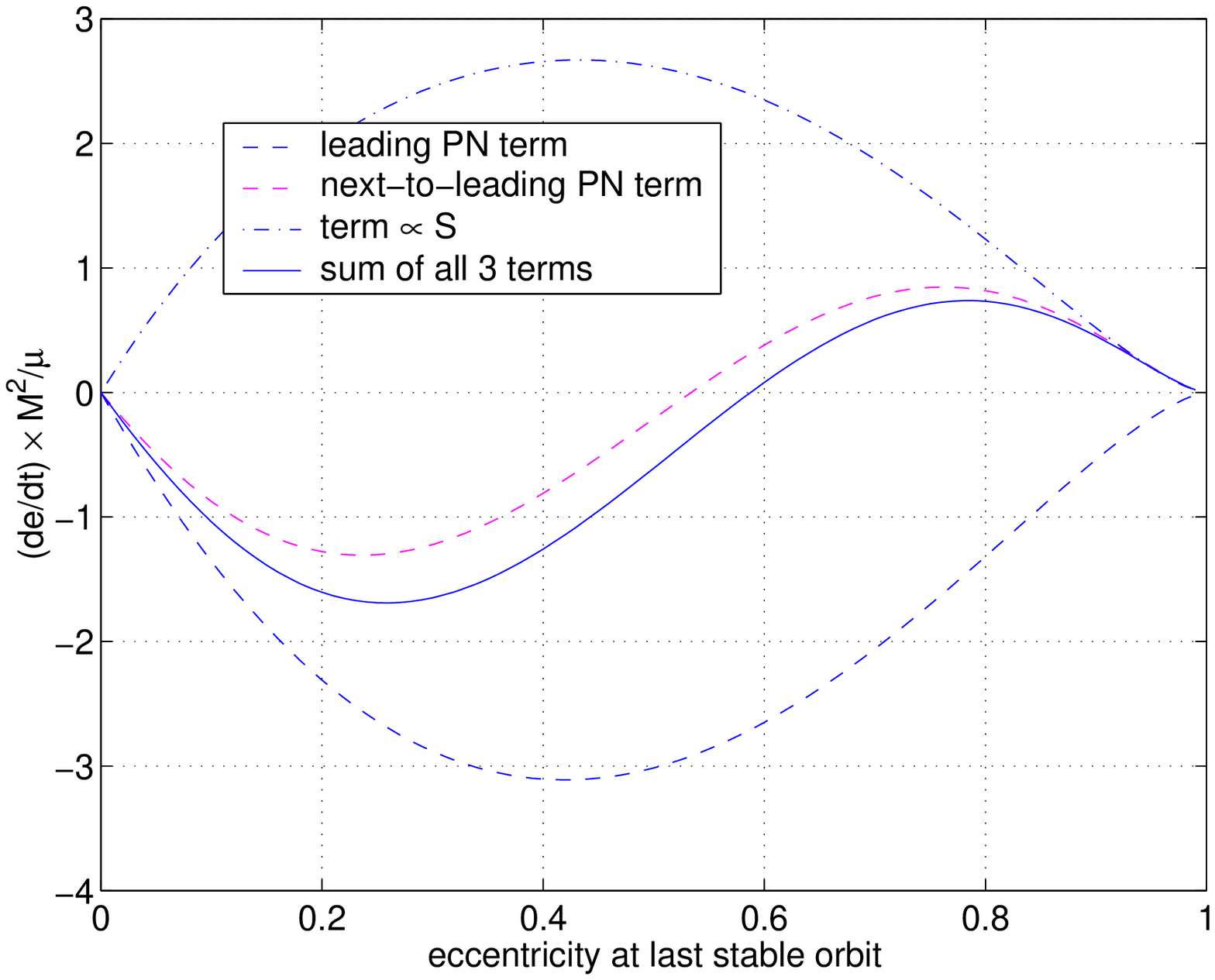}}
\centerline{\epsfysize 7cm \epsfbox{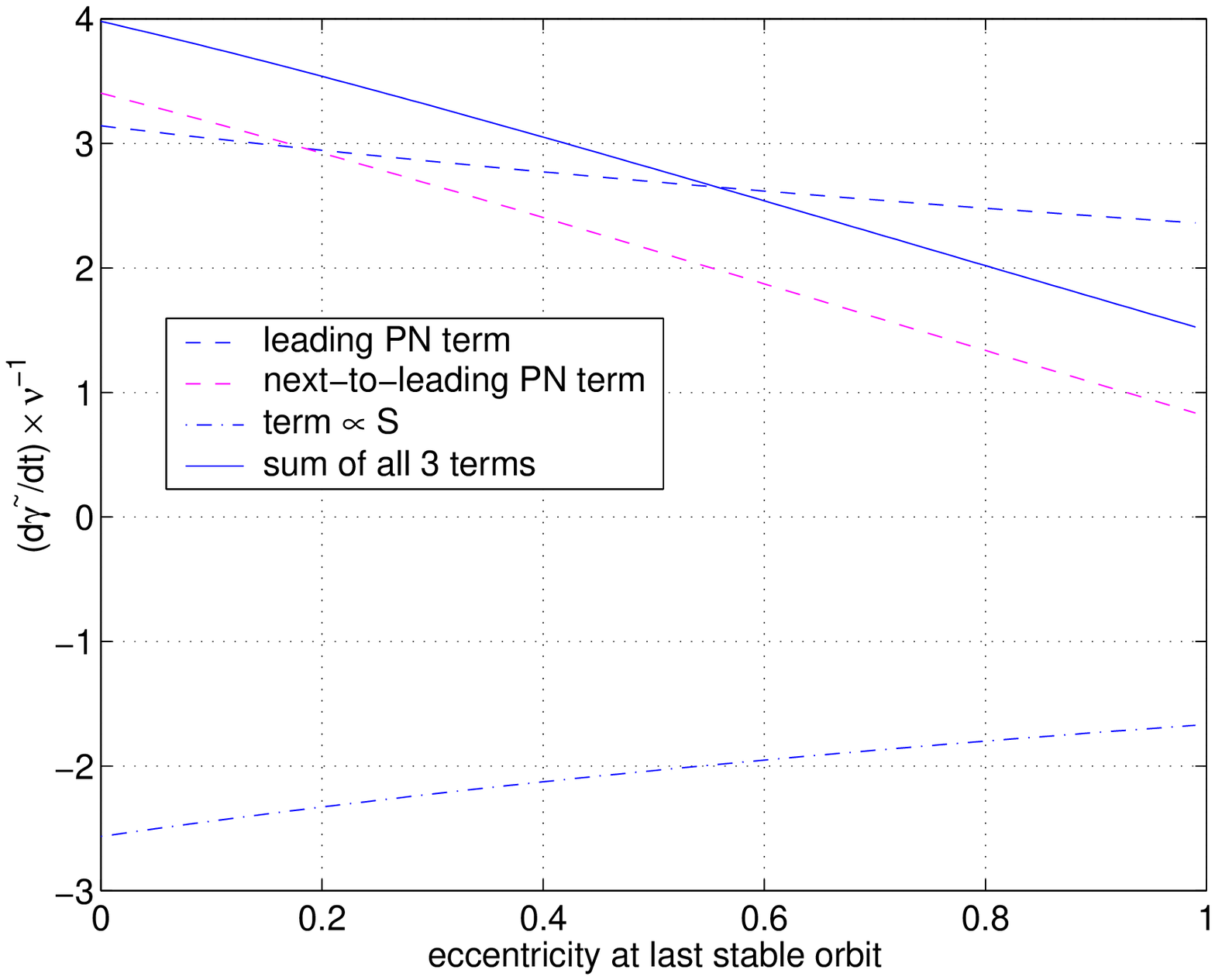}}
\caption{\label{fig:PN}
Relative magnitudes of the various PN terms.
The plots compare the contributions
to $\dot\nu$ (upper left panel), $\dot e$ (upper right panel), and
$\dot\gamma$ (bottom panel) from the various terms in the
PN expressions (\ref{nudot})--(\ref{edot}), when the CO is at its last stable
orbit. The various terms are of comparable magnitude, as expected
in this ``worse case''. Note, though, that so long as
$e_{\rm LSO}\lesssim 0.6$, the evolution of the orbital parameters at
this order is free of potential pathologies [like ``outspiral''
($\dot\nu<0$), or a negative pericenter advance ($\dot{\tilde\gamma}<0$)].
}
\end{figure}

\section{Estimating the effect of the CO's spin}

In this appendix we refer to the spin of the MBH as $\vec S_1$, and
denote by $\vec S_2$ the spin of the CO (in the rest of the paper
the MBH's spin is denoted as simply $\vec S$).
We present simple estimates of the effects of $\vec S_2$ on the
orbital evolution, showing that, at any instant, these are smaller than the
effects of $\vec S_1$ by an amount of order $(\mu/M)(\chi_2/\chi_1)$,
where $\chi_1 \equiv |\vec S_1|/M^2$ and $\chi_2 \equiv |\vec S_2|/\mu^2$.
This implies that over the course of the last $\sim 1\,$yr of inspiral,
the presence of the CO's spin modifies the accumulated orbital phase
and precession angles by a few radians at most.

Clearly $\chi_1 < 1$. Theoretical upper limits on $\chi_2$ are:
$\chi_2 < 1$ when the CO is a BH; $\chi_2 \alt 2$ for a $1.4 M_{\odot}$ NS;
and $\chi_2 \alt 5$ for a uniformly rotating, $0.5 M_{\odot}$ WD
\cite{ShapiroTeukolsky}.
We do not know the spin of any stellar-mass BH, but the
NSs and WDs we find in nature are typically spinning at well
below these maximum possible rates. Most known pulsars have
$\chi_2 \sim 10^{-3}-10^{-2}$ (recycled pulsars being the
major exception), and it appears that WDs typically have
$\chi_2 \sim 10^{-2}$ \cite{WDspin}.

Our estimates here are based on low-order, post-Newtonian
equations for circular-orbit binaries.
Of course, the capture orbits of interest to us can be moderately
eccentric in the LISA band, and they are sufficiently close to the MBH that
higher-order post-Newtonian terms can
be comparable in size to the lower-order terms that we do include.
We expect that these higher-order PN terms, and modifications for $e \ne 0$,
would modify our estimates on the effects of $\vec S_2$ by
factors of order unity, but we anticipate that our PN estimates
 at least give the order of
magnitude.

Two other recent works that consider the effect
of the CO's spin on the capture waveform are by Hartl~\cite{Hartl_03}
and Burko~\cite{Burko_03}.

\subsection{Effect of $\vec L \cdot \vec S_2$ and $\vec S_1 \cdot \vec S_2$
terms on the waveform phase}

For circular orbits, let  $\omega \equiv 2\pi \nu + \dot{\tilde\gamma}$
be the angular frequency of the CO around the MBH. Defining $\Psi$ to be
the accumulated orbital phase from time $t_i$ to $t_f$,
\be\label{Psi}
\Psi = \int_{t_i}^{t_f} \omega\, dt = \int_{\omega_i}^{\omega_f}
\frac{\omega}{\dot\omega} d\omega \, ,
\ee
Kidder~\cite{Kidder_95} finds that the change in the orbital phase $\Psi$ due
to non-zero $\vec S_2$ is
\begin{eqnarray}\label{delpsi}
\Delta\Psi &=& \chi_2 \frac{125}{256} (\hat L \cdot \hat S_2)
[(M\omega_i)^{-2/3} -(M\omega_f)^{-2/3}] \nonumber\\
&+&\chi_1\chi_2 \bigl(\frac{1235}{1536} \hat S_1 \cdot \hat S_2
- \frac{3605}{1536}  \hat L \cdot \hat S_1 \, \hat L \cdot \hat S_2\bigr)
[(M\omega_i)^{-1/3} -(M\omega_f)^{-1/3}]+O(\mu/M).
\end{eqnarray}
[Actually, Eq.~(\ref{delpsi}) refers to the case where the dot
products $\hat L \cdot \hat S_2$,
$\hat L \cdot \hat S_1$, and $\hat S_1 \cdot \hat S_2$ are constants
(e.g., the case where these three angular momenta are
all perfectly aligned or anti-aligned, so there is no
Lense-Thirring precession).
For the realistic case, where $\hat L$ and $\hat S_2$ undergo significant
Lense-Thirring precession, these dot products would be replaced by
appropriately weighted time-averages.]
For our case of inspiral orbits near plunge and $(t_f - t_i)$ of
order a year, we have $[(M\omega_i)^{-1/3} -(M\omega_f)^{-1/3}] \sim
1-2$ and $[(M\omega_i)^{-2/3} -(M\omega_f)^{-2/3}] \sim 3-10$.
Therefore, when $\chi_2$ and $\chi_1$ are of order one,
$\Delta\Psi$ is at most a few radians, while $\Psi$ itself
is of order $10^6$.

\subsection{Effect of $S_2$ on pericenter precession, $d\gamma/dt$}

The $\vec S_2$ contribution to $d\gamma/dt$ is given (to lowest nontrivial
PN order)
by Brumberg~\cite{Brumberg}, in his Eqs.~(4.4.41) and ~(4.4.45).
We do not reproduce those equations here, but just note that the
relevant terms are (for low eccentricity) of order $\chi_2 (\mu/M)
(M/r)^{3/2} \omega$ and $\chi_1 \chi_2 (\mu/M) (M/r)^{2} \omega$, respectively.
Therefore, the integrated effect of $\vec S_2$ on $\gamma$ is
$\Delta\gamma \sim \chi_2 (\mu/M) N_{orb}$ radians, where $N_{orb}$ is
the number of orbits during the integration time.
Hence, $\Delta\gamma$ is of order one radian for $\chi_2 \sim 1$
and typical values $\mu/M \sim 10^{-5}$
and $N_{orb} \sim 10^5$.

\subsection{Effect of $\vec S_2$ on $\hat S_1$ and $\hat L$}

We next estimate how $\vec S_2$ affects the Lense-Thirring precession.
Our estimates are based on the following
post-Newtonian precessional equations
given by Apostolatos {\it et al.}~\cite{haris}, which were obtained by averaging
the spin-orbit, spin-spin, and radiation reaction torques
(all calculated to lowest nontrivial post-Newtonian order)
over one complete orbit:
\begin{eqnarray}
{\dot {\vec L}} &=&
{1\over r^3} \left[2\vec S_1 +
{{3M}\over{2\mu}}\vec S_2 \right ] \times \vec L
- {3\over 2}{1\over r^3} \bigl[(\vec S_2\cdot \hat L)\vec S_1 + (\vec
S_1 \cdot \hat L)\vec S_2 \bigr]\times \hat L
-{32\over5}{\mu^2\over r}\left ( {M\over r}\right )^{5/2} \hat L \;,
\label{spin1}\\
{\dot {\vec S}_1} &=&
{2\over r^3}\vec L \times \vec S_1
+ {1\over r^3}\left[ {1\over 2} \vec S_2
-{3\over 2}(\vec S_2\cdot \hat L)\hat L\right]\times \vec S_1\;,
\label{spin2}\\
{\dot {\vec S}_2} &=&
{{3}\over {2r^3}} {M\over {\mu}}\vec L \times \vec S_2
+ {1\over r^3}\left[ {1\over 2} \vec S_1
-{3\over 2} (\vec S_1\cdot \hat L)\hat L\right]\times \vec S_2\;.
\label{spin3}
\end{eqnarray}
[Actually, Eqs.~(\ref{spin1})--(\ref{spin3})
are a simplified version of Eqs.~(11a)--(11c) in
Apostolatos {\it et al.}, differing by fractional corrections of order
$\mu/M$.]

Two results that follow from Eqs.~(\ref{spin1})--(\ref{spin3}) are:
(i) the two spin magnitudes,
$|\vec S_1|$ and $|\vec S_2|$, are constants of the motion; and (ii)
the total angular momentum,
$\vec J =\vec S_1+\vec L+\vec S_2$,  is also constant, {\it except} for the
orbital momentum that is radiated away:
\be\label{Jdot}
\dot {\vec J} =  -{32\over5}{\mu^2\over r}\left ( {M\over r}\right )^{5/2} \hat L \; .
\ee
Therefore, $\Delta \vec J$---the total change in $\vec J$ over the observed
inspiral---has magnitude of order $|\Delta \vec L| \sim (\mu M)$.

Combining this with the fact that the magnitudes of
$\vec S_1$, $\vec L$, and $\vec S_2$ are roughly in the ratio
\be\label{ratio}
|\vec S_1|:|\vec L|:|\vec S_2|\sim \chi_1: \mu/M: \chi_2 (\mu/M)^2
\ee
(i.e., $|\vec S_1| >> |\vec L| >> |\vec S_2|$),
one easily sees that the yearly change in the direction of $\vec S_1$
can be of order $\chi_1^{-1}(\mu/M)$ radians at most, no matter how
$\hat S_2$ and $\hat L$ evolve. Thus our approximation of treating
$\hat S_1$ as fixed, used throughout, remains
valid when we include the effects of $\vec S_2$.

Finally, we consider the motion of $\hat L$.
For $\vec S_2 = 0$, we have seen that $\vec L$
simply precesses around $\vec S_1$ at rate $\dot \alpha$ given by
Eq.~(\ref{alphadot}). Because the terms in Eq.~(\ref{spin1})
involving $\vec S_2$
are all smaller than the dominant $(2/r^3)\vec S_1 \times \vec L$
term by an amount of order $\chi_2 (\mu/M)$ or $(\chi_2/\chi_1) (\mu/M)$,
they represent a small perturbation on the simple precession picture, which
can clearly be absorbed into a time-varying $\lambda$ (the angle
between $\hat S_1$ and $\hat L$) and a perturbed precession rate $\dot\alpha$.
These extra terms in $\dot{\hat L}$ are of order $\chi_2 (\mu/M) (M/r)^{3/2}\omega$
and $\chi_1 \chi_2 (\mu/M) (M/r)^{2}\omega$, respectively---the same as
for the $\vec S_2$ terms in $d\gamma/dt$---and so again integrate up
to a yearly difference of order one radian at most.

\subsection{Conclusion}

If the CO's angular momentum is close to maximal (i.e., within a factor
of a few), then the CO's spin is marginally relevant for the dynamics over
timescales of order a year.  However, if the CO is not rapidly
rotating, or if one is just searching for short stretches (lasting
$\sim 2$ weeks) of the waveforms in the data (e.g., as the first
step of a hierarchical search), then the CO's spin can be safely neglected.
(For such short stretches, in addition to the fact that the waveform
phase errors from neglecting $\vec S_2$ are much less than one radian,
it seems likely that these errors can be partially compensated for
by errors in the other physical parameters.)


\newpage

\begin{figure}[htb]
\input{epsf}
\centerline{\epsfysize 6cm \epsfbox{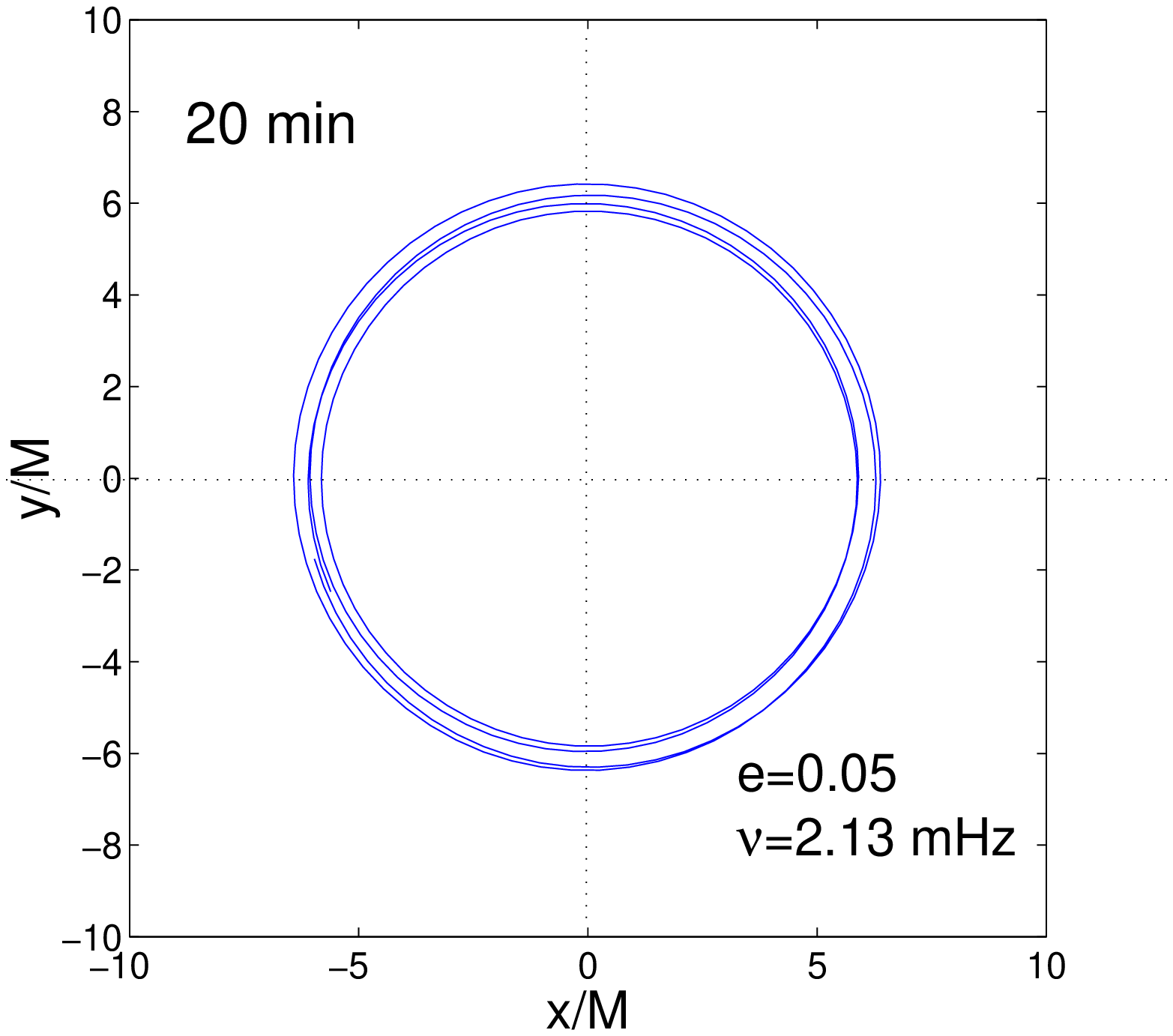}
            \epsfysize 6cm \epsfbox{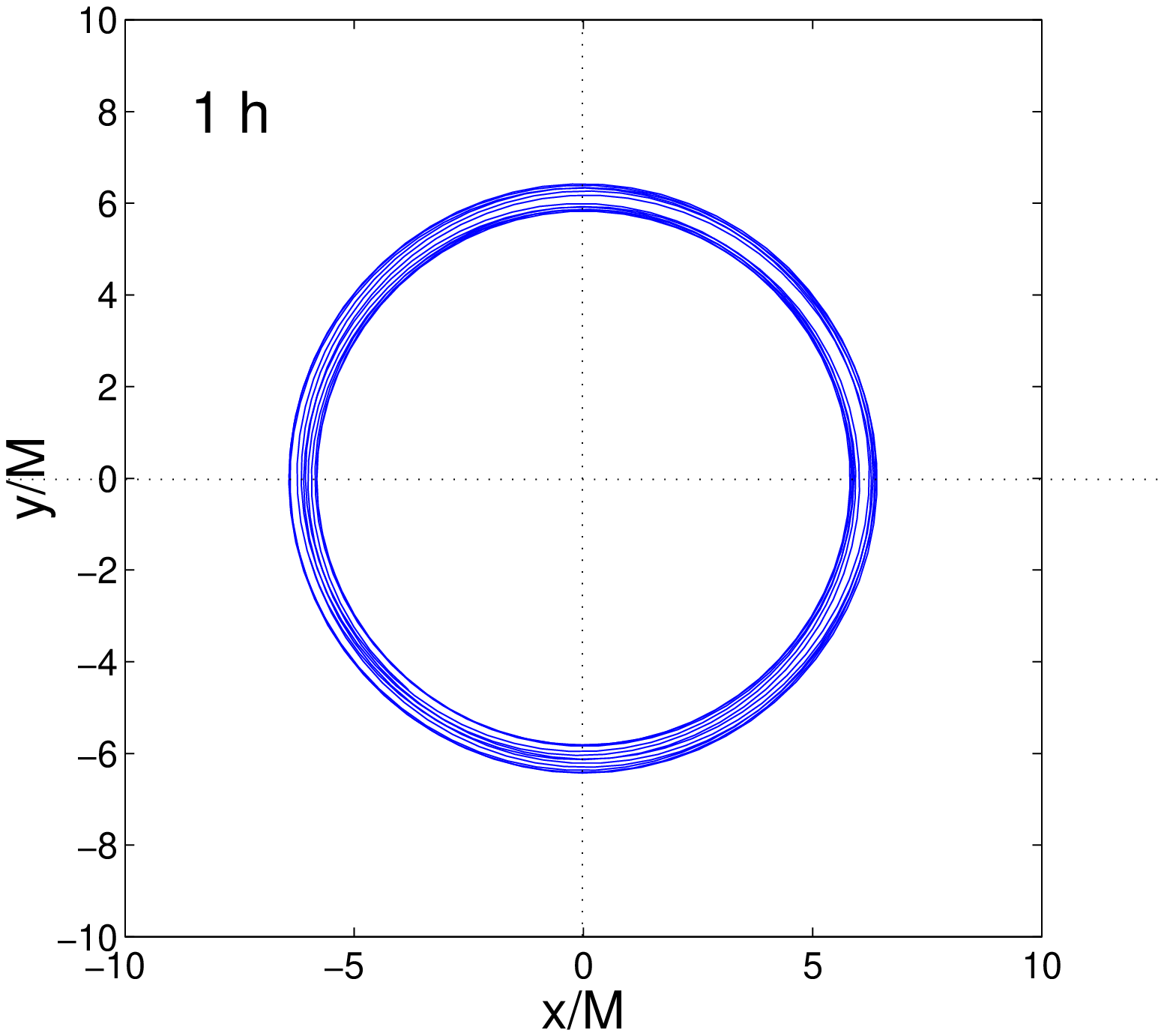}
            \epsfysize 6cm \epsfbox{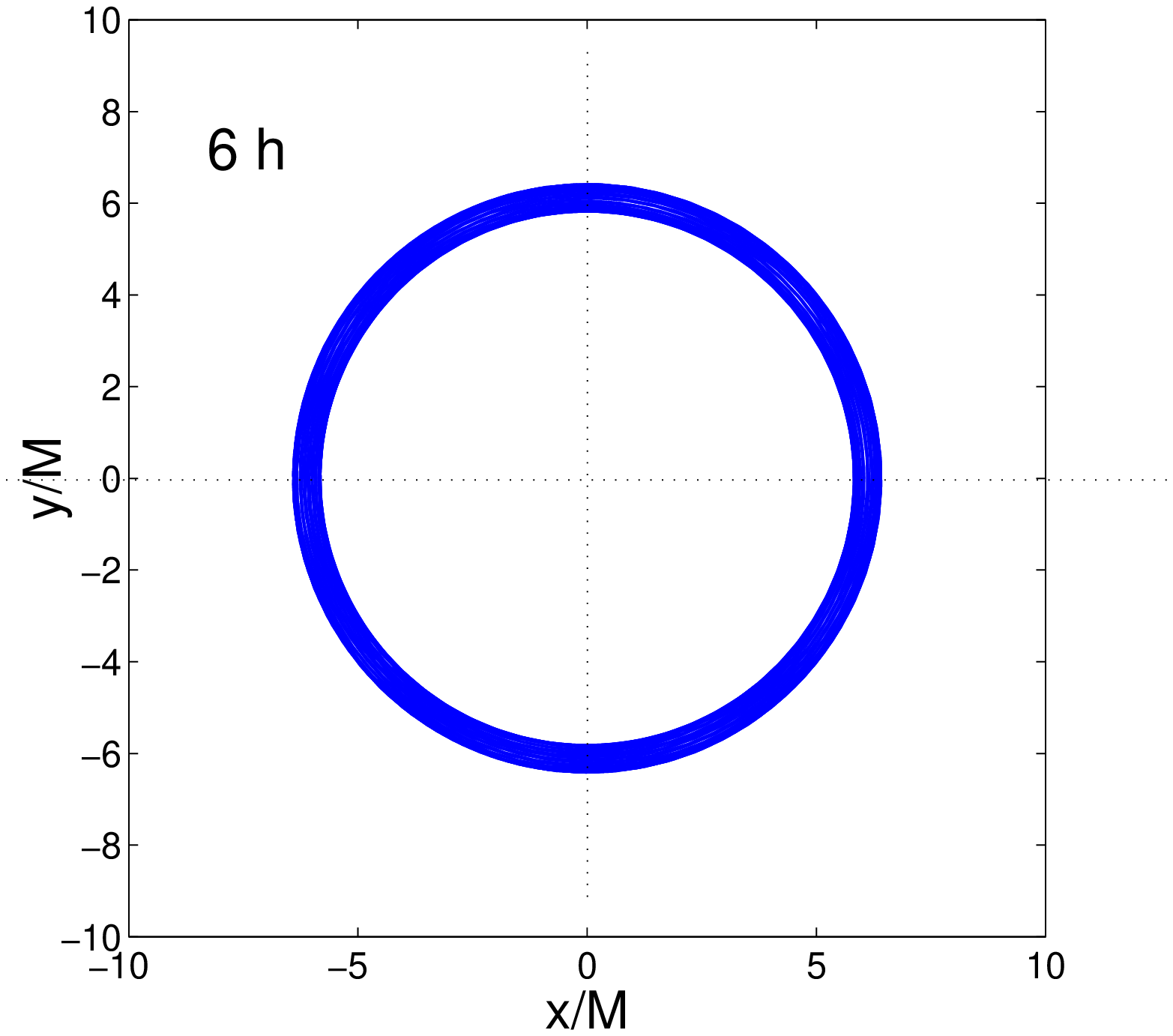}}
\caption{\label{fig:orb1}
Sample equatorial orbits:
Shown (left to right, respectively) are
20 minutes, 1 hour, and 6 hours-long samples of the CO's trajectory
just before approaching the LSO.
$(x,y)$ is a Cartesian coordinate system in the orbital (equatorial)
plane, centered at the MBH.
The axes give the distance in units of the MBH's mass $M$.
This sequence of plots shows a case with a very small LSO eccentricity,
$e=0.05$. The other physical parameters are set as follows:
CO's mass: $\mu=10 M_{\odot}$;
MBH's mass: $M=10^6 M_{\odot}$;
MBH's spin magnitude: $S=M^2$;
}
\end{figure}

\begin{figure}[htb]
\input{epsf}
\centerline{\epsfysize 6cm \epsfbox{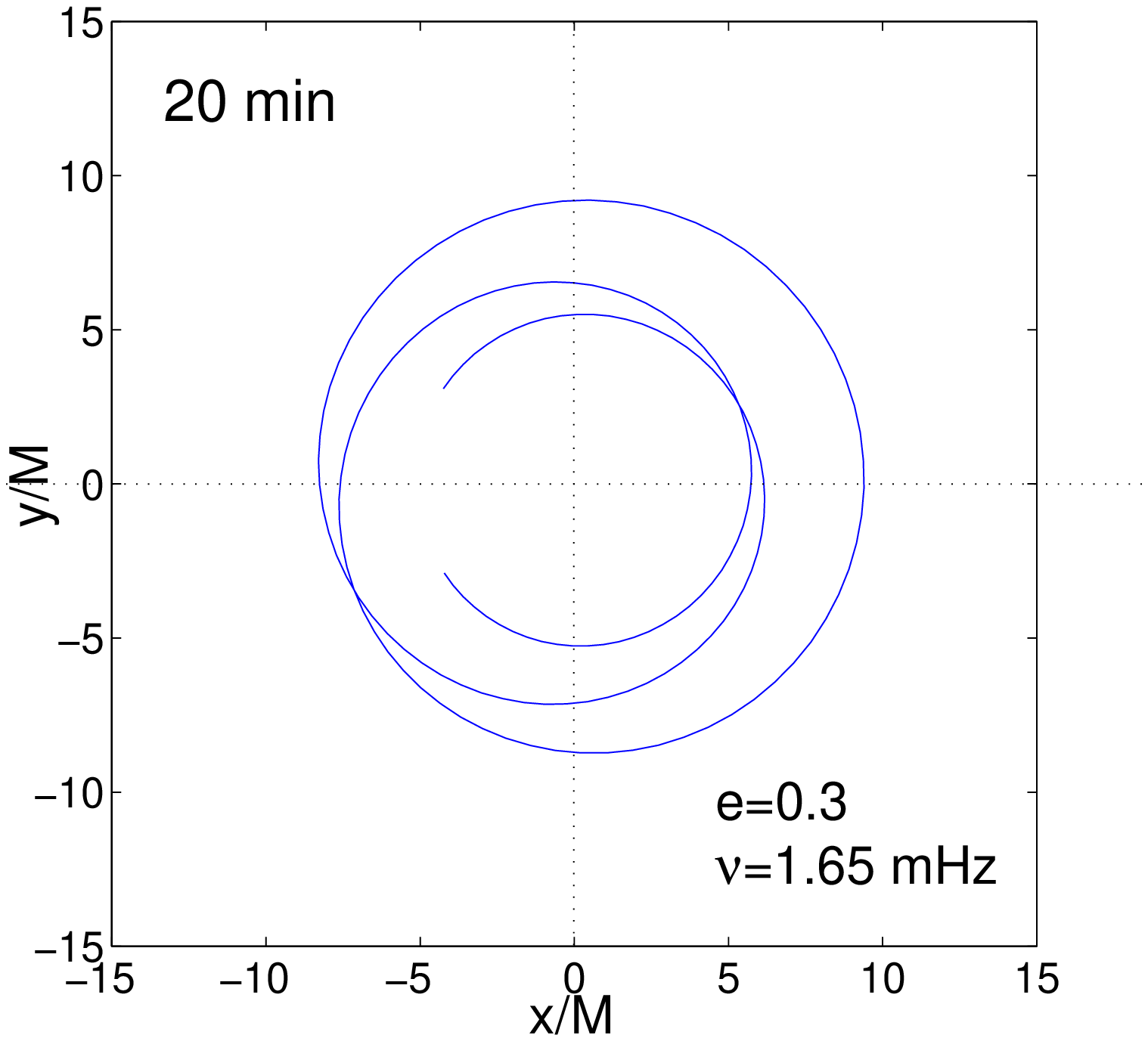}
            \epsfysize 6cm \epsfbox{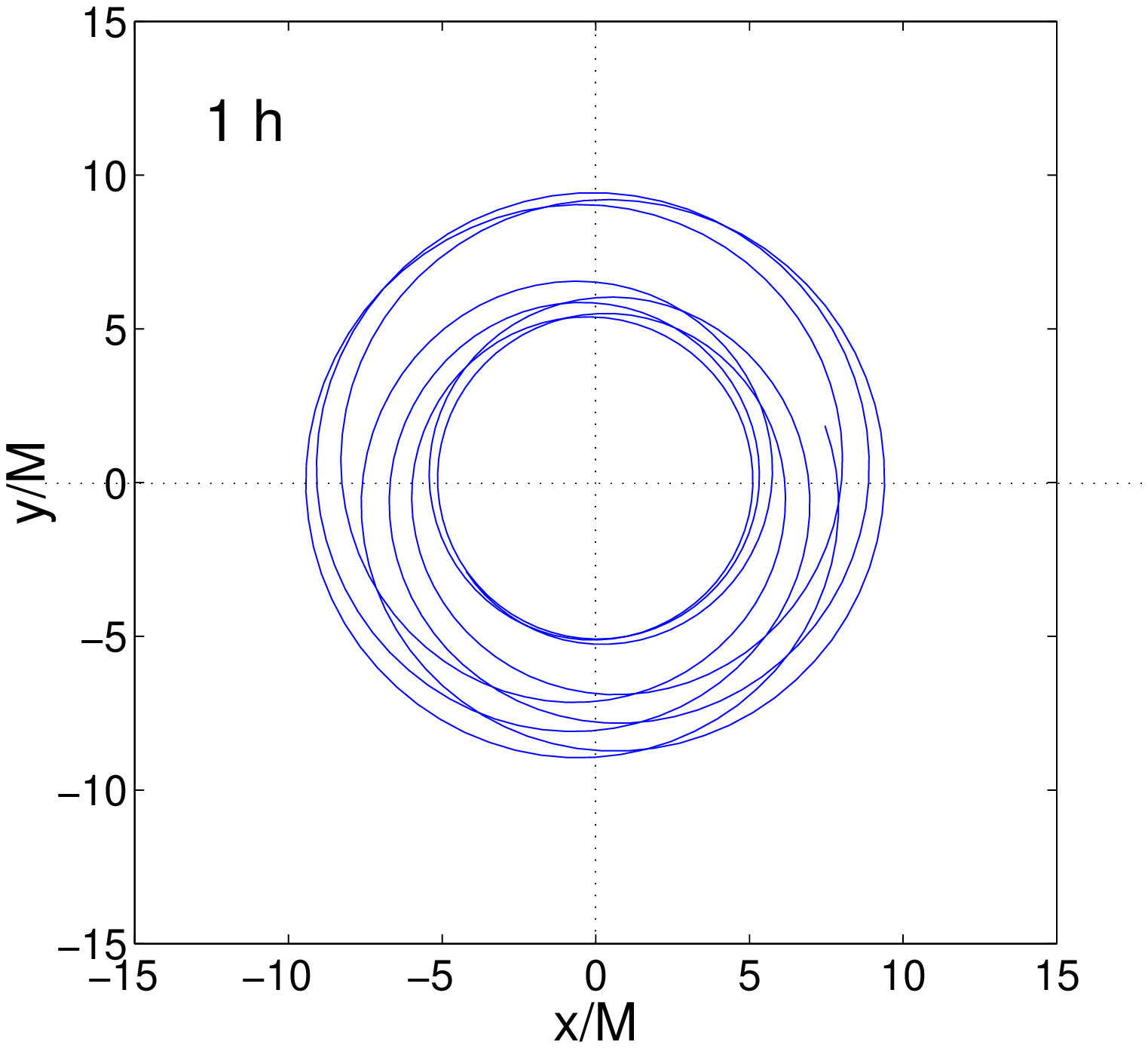}
            \epsfysize 6cm \epsfbox{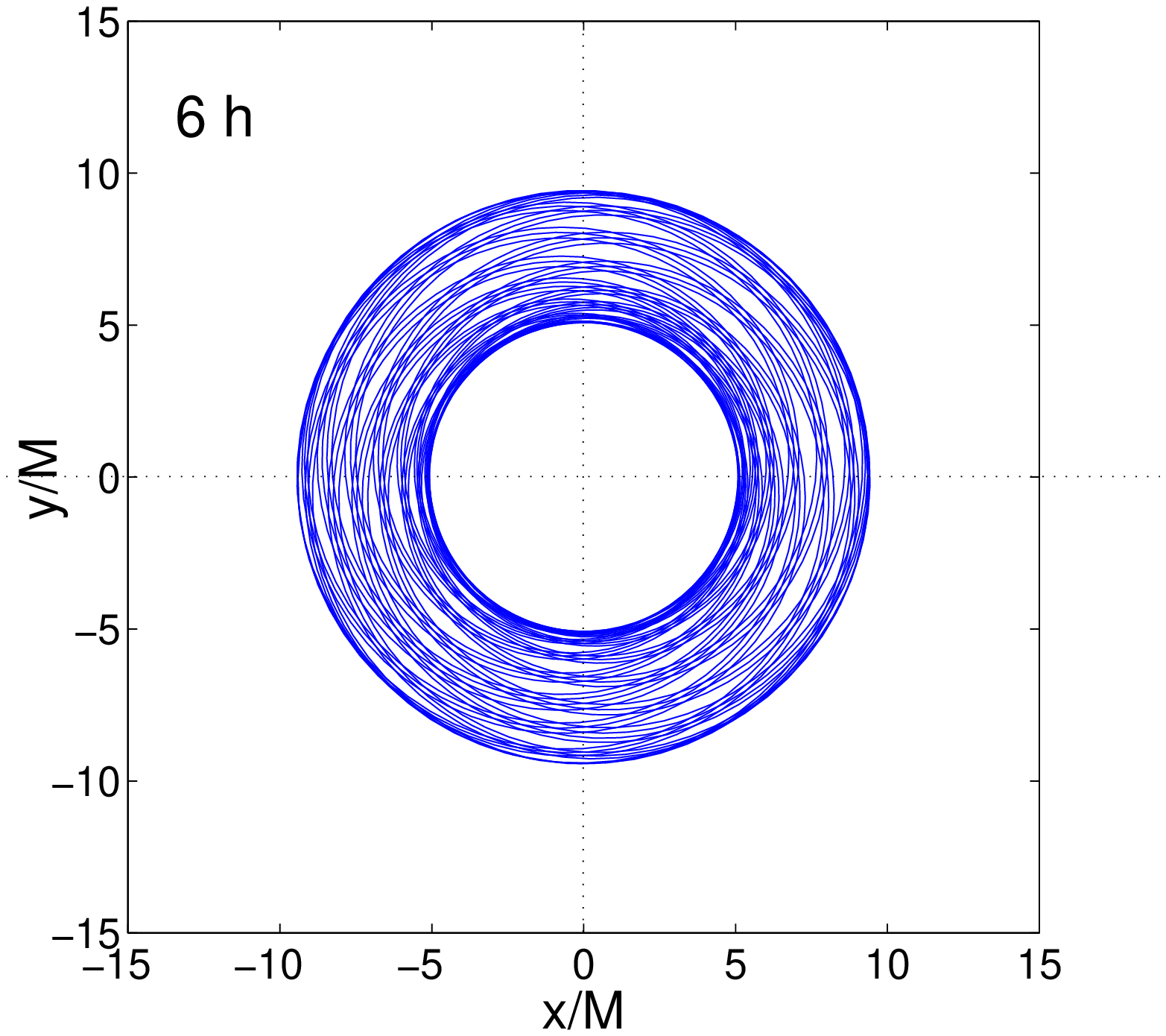}}
\caption{\label{fig:orb2}
Same as in Fig.\ (\ref{fig:orb1}), for a trajectory with LSO eccentricity
$e=0.3$.
}
\end{figure}

\begin{figure}[htb]
\input{epsf}
\centerline{\epsfysize 6cm \epsfbox{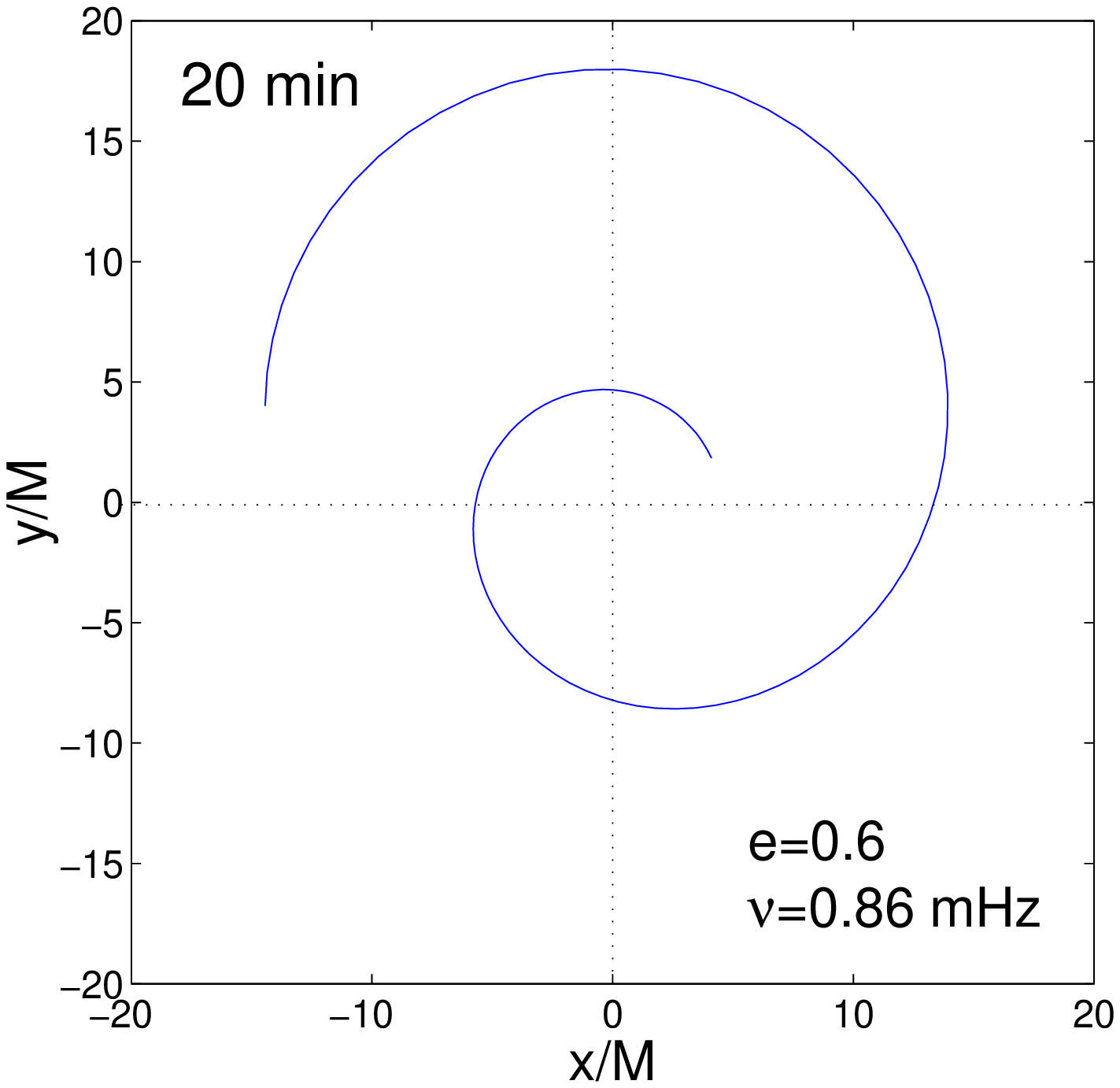}
            \epsfysize 6cm \epsfbox{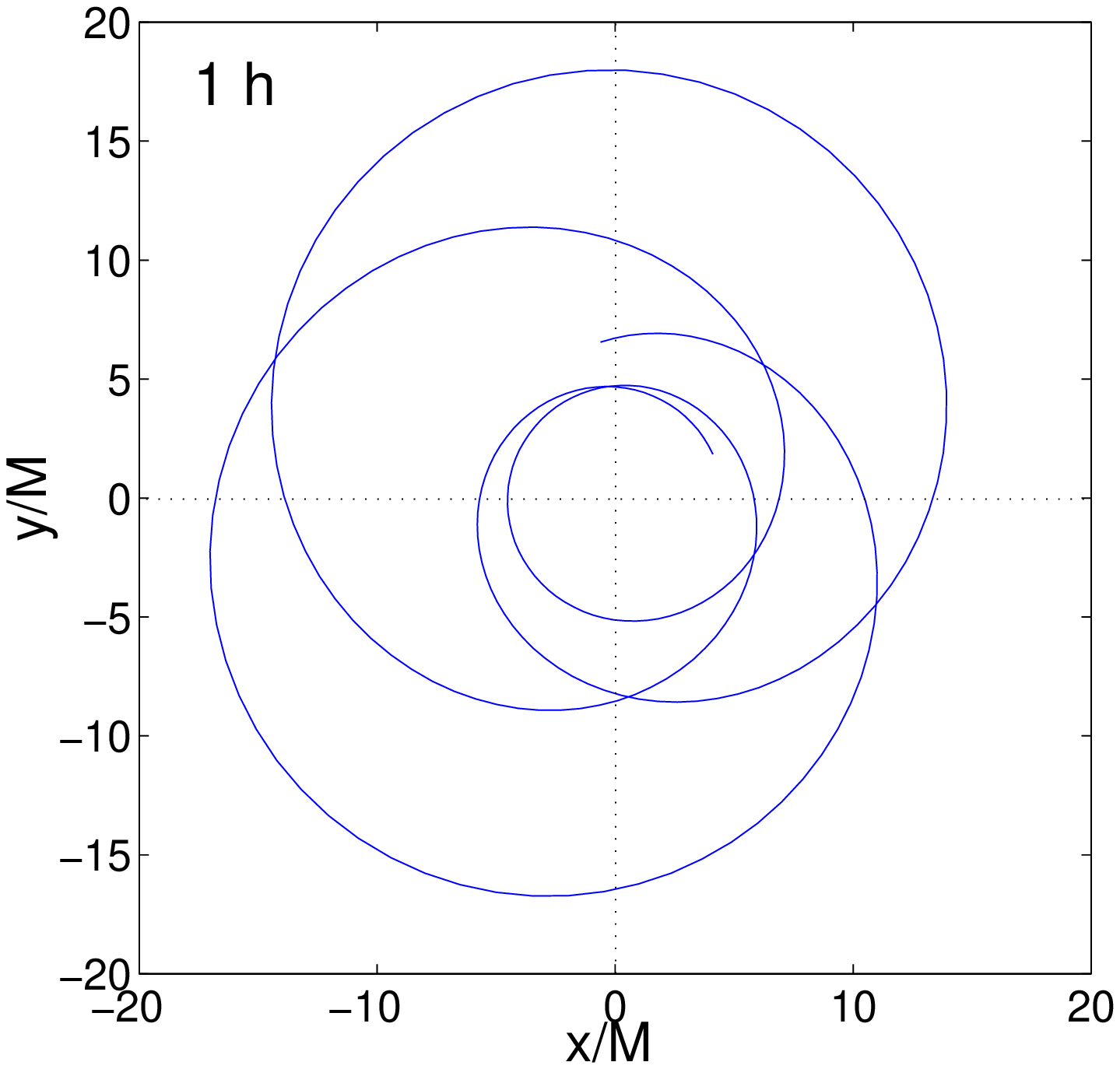}
            \epsfysize 6cm \epsfbox{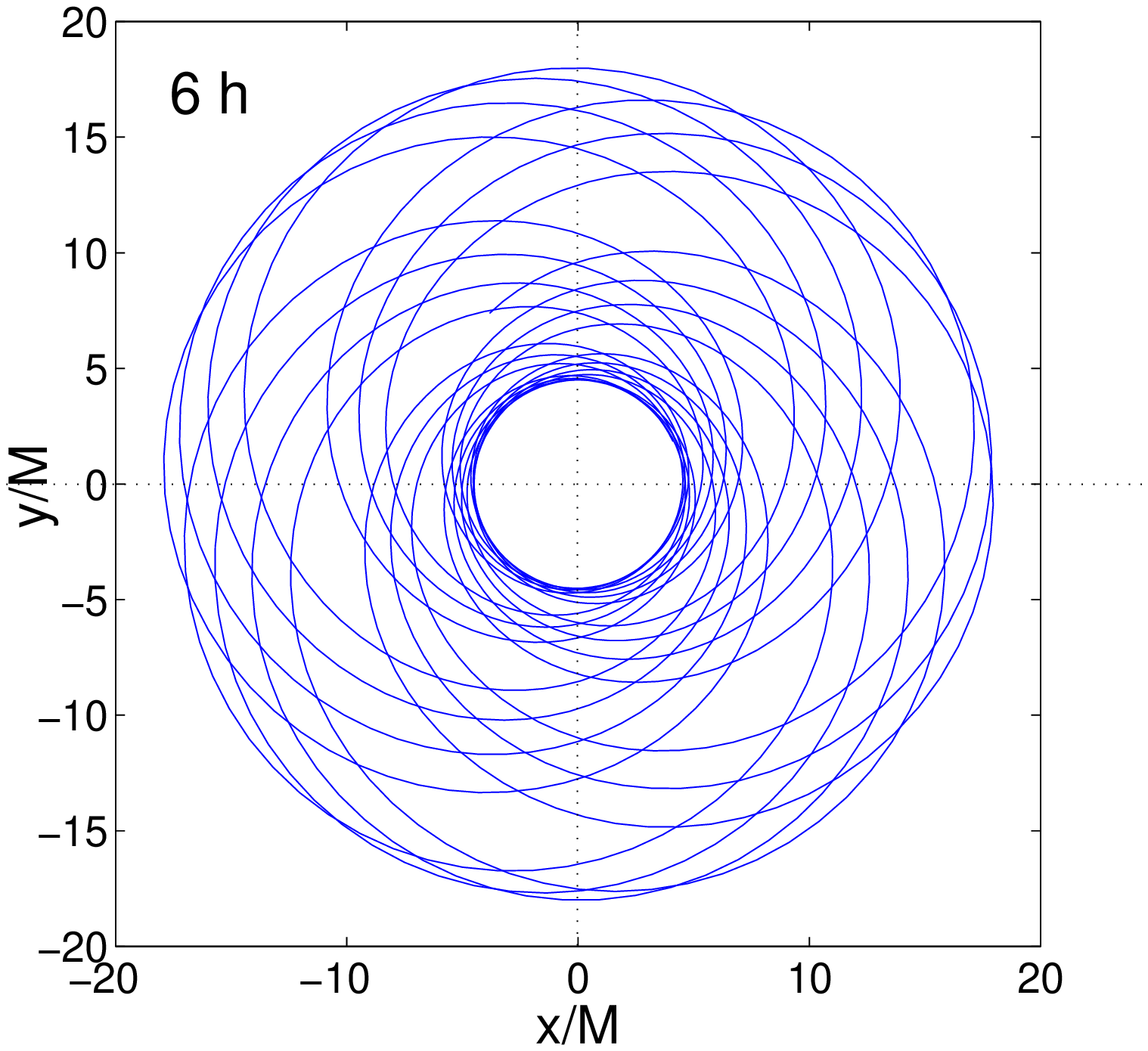}}
\caption{\label{fig:orb3}
Same as in Fig.\ (\ref{fig:orb1}), for a trajectory with LSO eccentricity
$e=0.6$.
}
\end{figure}

\begin{figure}[htb]
\input{epsf}
\centerline{\epsfysize 7cm \epsfbox{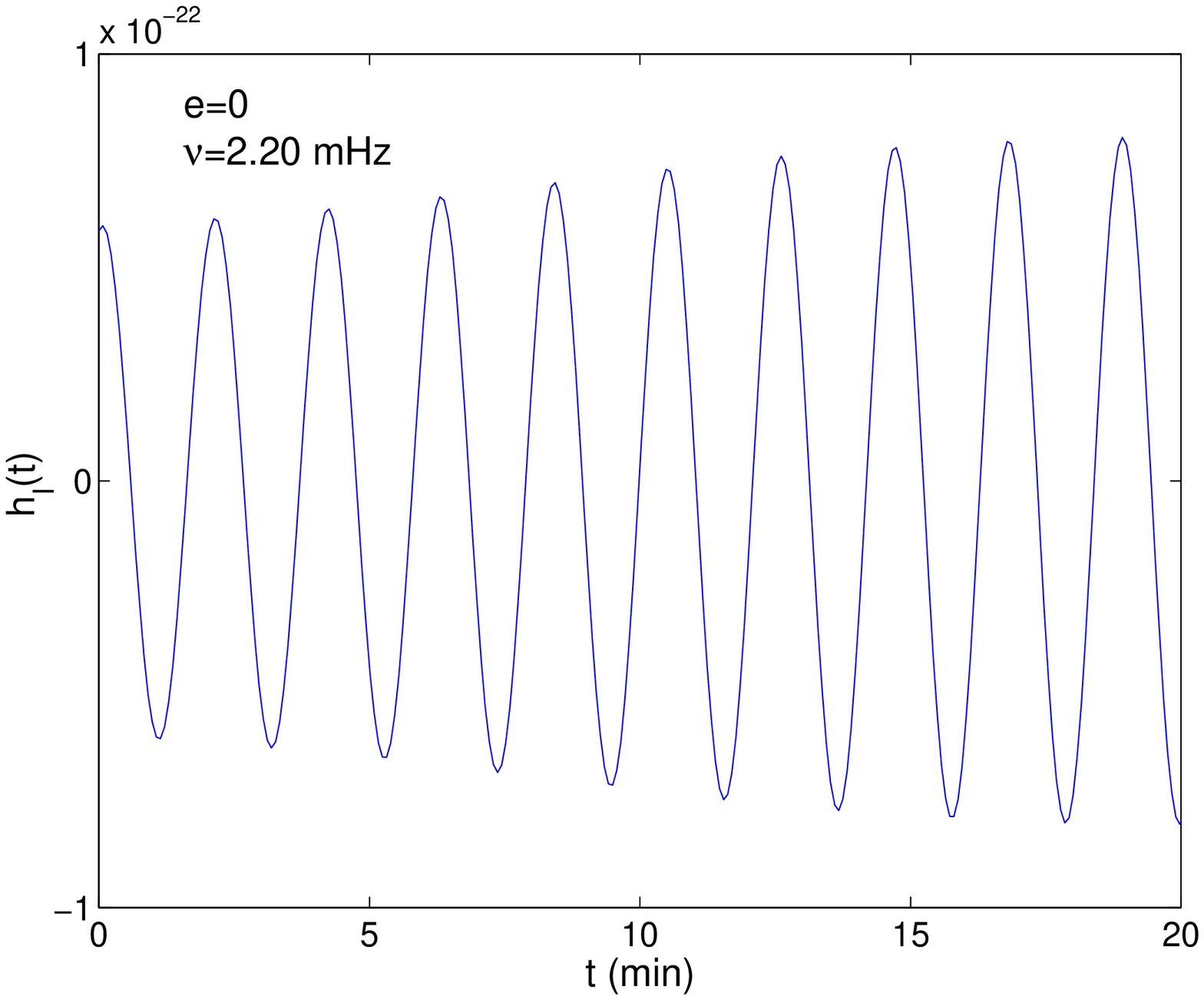}}
\centerline{\epsfysize 7cm \epsfbox{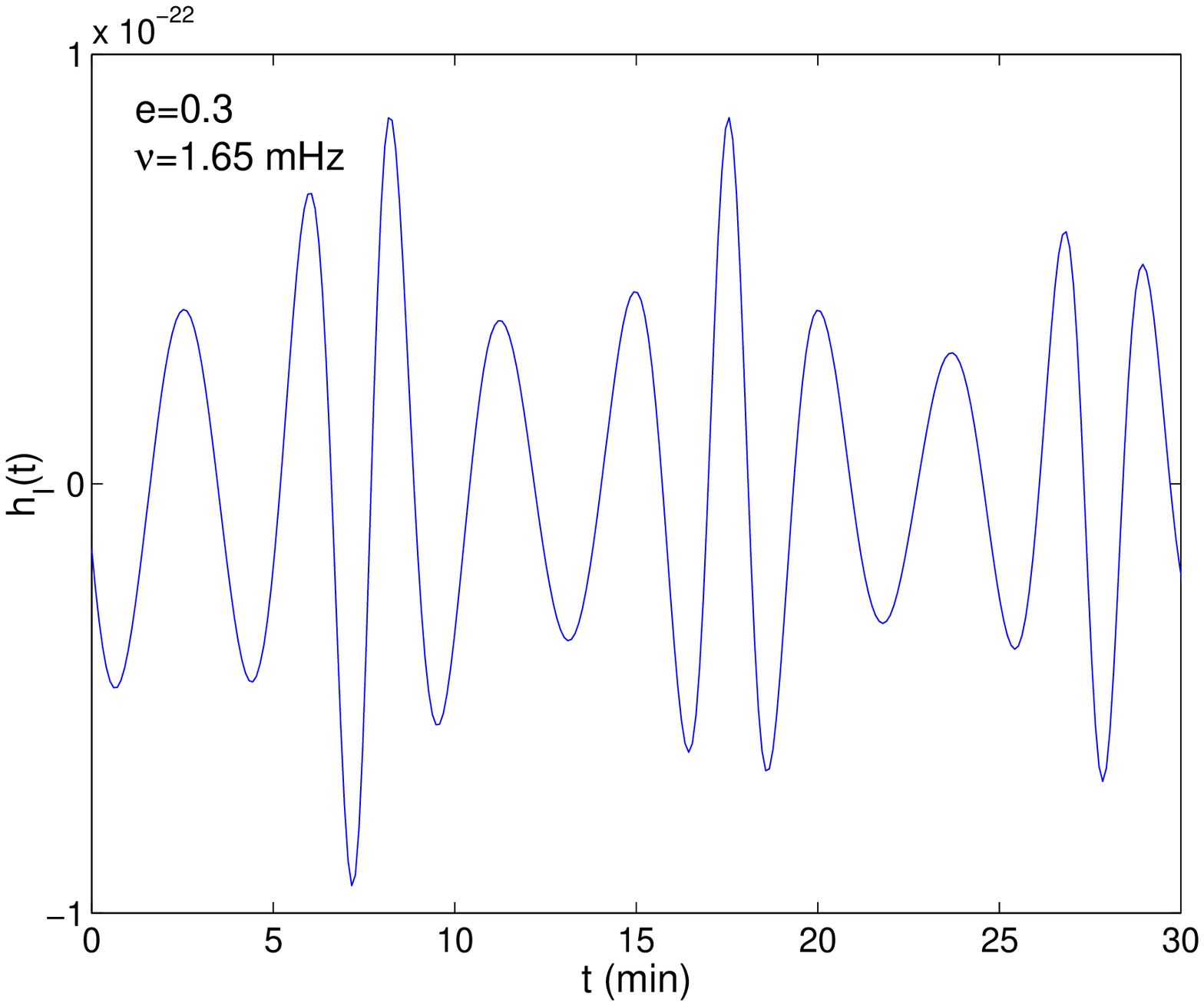}}
\centerline{\epsfysize 7cm \epsfbox{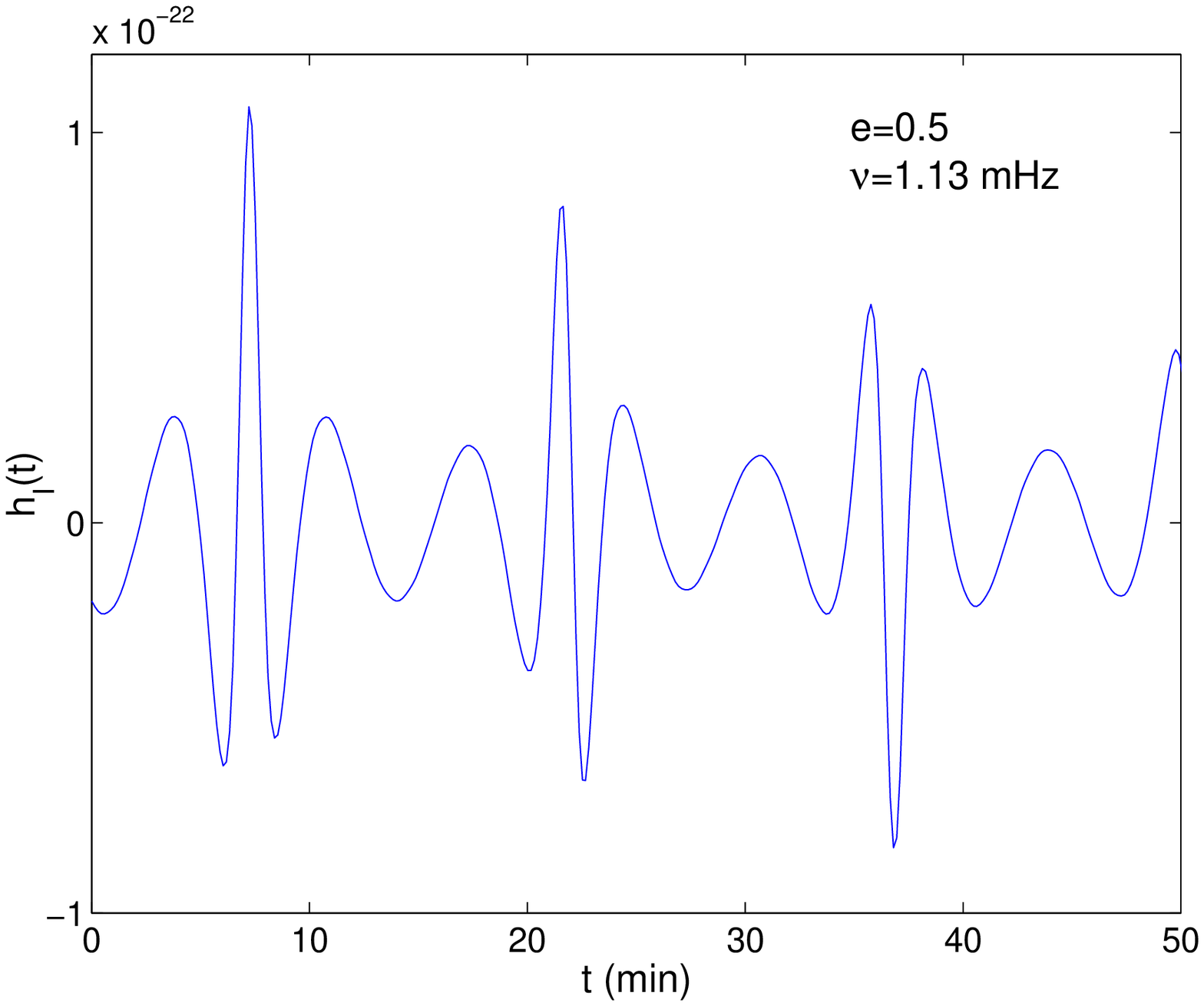}}
\caption{\label{fig:wf1}
Sample waveforms.
Shown is the responce function $h_I(t)$ (as defined in the text)
during the last minutes before the final plunge. The three panels show
cases with LSO eccentricity 0, 0.3, and 0.5 (top to bottom, respectively).
The other physical parameters are set as follows:
CO's mass: $\mu=10 M_{\odot}$;
MBH's mass: $M=10^6 M_{\odot}$;
MBH's spin magnitude: $S=M^2$;
Angle between MBH's spin and orbital angular momentum:
$\lambda=30^{\circ}$.
}
\end{figure}

\begin{figure}[htb]
\input{epsf}
\centerline{\epsfysize 9cm \epsfbox{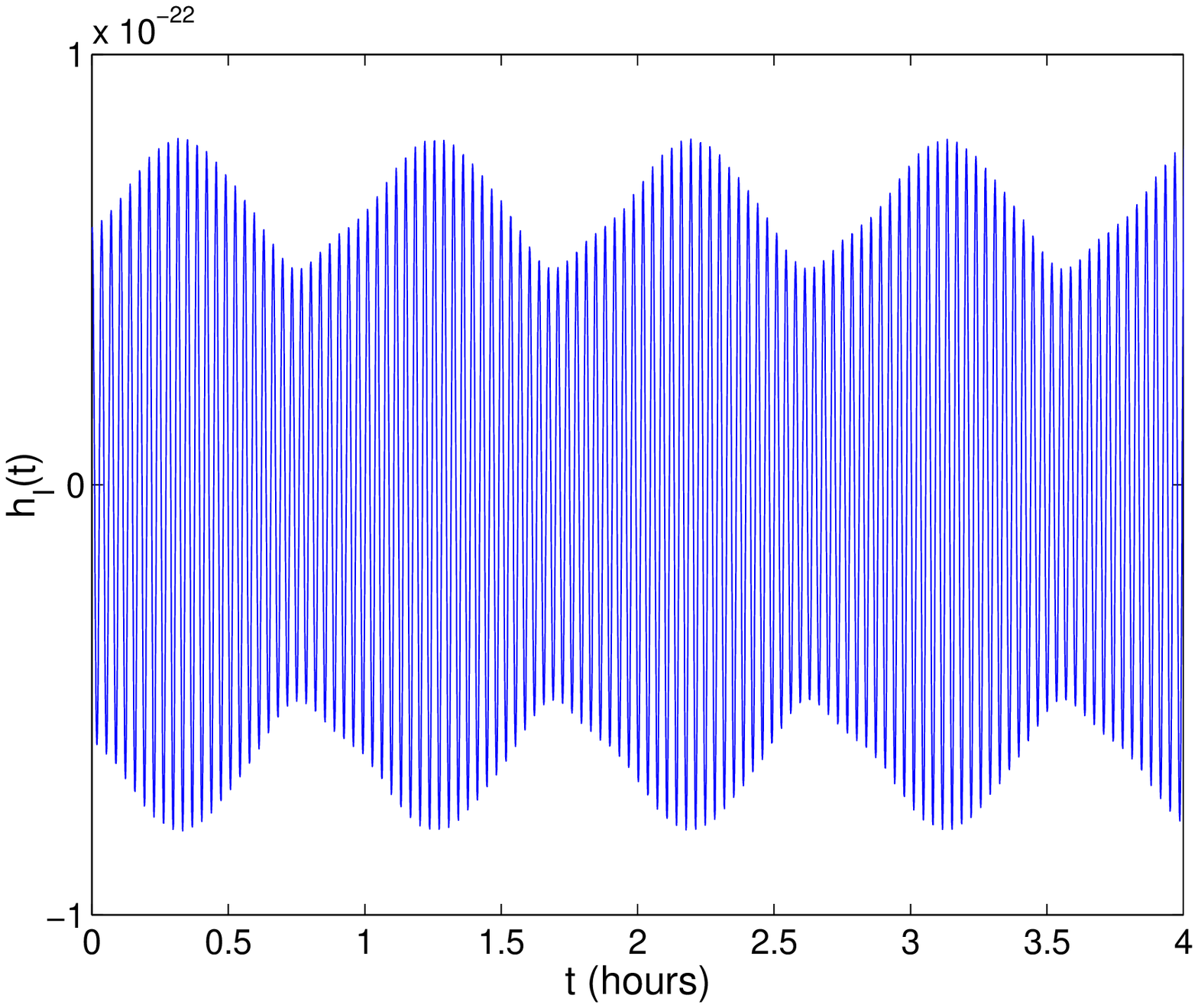}}
\caption{\label{fig:wf2}
A longer waveform segment shows amplitude modulations
due to precession of the orbital plane.
(The physical parametes are set here as in the upper panel of
Fig.\ \ref{fig:wf1}.)
}
\end{figure}


\begin{thebibliography}{99}

\bibitem{Pre} K. Danzmann et al., {\it LISA- Laser Interferometer
    Space Antenna, Pre-Phase A Report}, Max-Planck-Institut
    f\"{u}r Quantenoptic, Report MPQ 233 (1998).

\bibitem{ryan_multipoles}
    F.\ D.\ Ryan, Phys.\ Rev.\ D {\bf 56}, 1845 (1997).

\bibitem{rates}
        S.\ Sigurdsson and M.\ J.\ Rees, Mon.\ Not.\ R.\ Astron.\ Soc.\
        {\bf 284}, 318 (1997); S.\ Sigurdsson, Class.\ Quantum Grav.\
        {\bf 14}, 1425 (1997); M.\ Freitag, Class.\
         Quantum Grav.\ {\bf 18}, 4033 (2001).

\bibitem{Poisson_03}
E.\ Poisson, Living Reviews in Relativity, submitted (gr-qc/0306052).

\bibitem{wg1_website}  http://www.tapir.caltech.edu/listwg1/


\bibitem{pm}
P.\ C.\ Peters and J.\ Mathews, Phys.\ Rev.\ {\bf 131}, 435 (1963);
P.\ C.\ Peters, Phys.\ Rev.\ {\bf 136}, B1224 (1964).

\bibitem{Cutler-Kennefick-Poisson}
C.\ Cutler, D.\ Kennefick and E.\ Poisson, Phys. Rev. D
{\bf 50}, 3816 (1994).

\bibitem{Glampedakis-Kennefick}
K.\ Glampedakis and D.\ Kennefick, Phys.\ Rev.\ D {\bf 66}, 044002 (2002).

\bibitem{scott1}
        S.\ A.\ Hughes, Phys. Rev. D {\bf 61}, 084004 (2000).

\bibitem{cutler98}
    C. Cutler, Phys. Rev. D. {\bf 57}, 7089 (1998).

\bibitem{Cornish}
    N.~J. Cornish and L.~J. Rubbo, gr-qc/0209011.

\bibitem{Richstone_98}
    D. Richstone {\it et al.}, Nature {\bf 395}, A14 (1998).

\bibitem{Kormendy_02} J. Kormendy and K. Gebhardt, in Proceedings
of 20th Texas Symposium on Relativistic Astrophysics, eds. H. Martel
and J. C. Wheeler, AIP (2002); astro-ph/0105230.

\bibitem{Hils_Bender_95}
    D. Hils and P.~L. Bender, ApJ {\bf 445}, L7 (1995).

\bibitem{sterl_notes}
E.\ S.\ Phinney, unpublished notes.

\bibitem{finnthorne} L.\ S.\ Finn and K.\ S.\ Thorne, Phys.\ Rev.\ D
    {\bf 62}, 124021 (2000).

\bibitem{GHK} K. Glampedakis, S. A. Hughes, and D. Kennefick,
 Phys.\ Rev.\ D {\bf 66}, 064005 (2002).

\bibitem{Freitag_03a}
    M. Freitag, ApJ {\bf 583} L21 (2003).


\bibitem{wg1_members}
    Members of WG1 communicate through regular telecons.
    Progress reports are posted on the WG1 website, Ref.\ \cite{wg1_website}.

\bibitem{peterseim96}
        M. Peterseim, O. Jennrich and K. Danzmann,  Class.\ Quantum Grav.\
        {\bf 13}, 279 (1996).

\bibitem{vecchio03}
    A. Vecchio; astro-ph/0304051.

\bibitem{Poisson96}
    E.\ Poisson, Phys. Rev. D {\bf 54}, 5939 (1996).

\bibitem{haris}
    T. Apostolatos, C. Cutler, G.~J. Sussman, and  K.~S. Thorne,
    Phys. Rev. D {\bf 49} 6274 (1994).

\bibitem{BCV}
    A. Buonanno, Y. Chen, and M. Vallisneri, Phys.\ Rev.\ D {\bf 67},  024016 (2003).

\bibitem{MTW}
    C.\ W.\ Misner, K.\ S.\ Thorne, and J.\ A.\ Wheeler, {\it
    Gravitation} (Freeman, San Francisco, 1973), chapter 33.

\bibitem{AET} M. Tinto, F. B. Estabrook, and J. W. Armstrong,
Phys.\ Rev. D {\bf 65}  082003 (2002).

\bibitem{300years} K.S. Thorne, in {\it 300 Years of Gravitation},
            ed. S.W. Hawking and W. Israel (Cambridge University
            Press, Cambridge, 1987), pp. 330-458.

\bibitem{markovic}
    Markovi\'c, D.\ M.\ 1993, Phys.\ Rev.\ D 48, 4738.

\bibitem{JunkerSchaefer}
    W. Junker and G. Sch\"afer, Mon.\ Not.\ R.\ astr.\ Soc.\
    {\bf 254}, 146 (1992).


\bibitem{Brumberg}
    V. A. Brumberg, {\it Essential Relativistic Celestial Mechanics}
(IOP Publishing, Bristol, 1991).

\bibitem{ryan96}
    F.\ D.\ Ryan, Phys.\ Rev.\ D {\bf 53}, 3064 (1996).

\bibitem{Barker}
    B. M. Barker and R. F. O'Connell, Phys.\ Rev.\ D {\bf 12}, 329 (1975).

\bibitem{Blanchet02} L. Blanchet, G. Faye, B. R. Iyer, and B. Joguet,
Phys.\ Rev.\ D {\bf 65}, 061501 (2002).

\bibitem{cutler_flanagan} C. Cutler, and E.~E. Flanagan, Phys. Rev. D {\bf 49}  2658 (1994).

\bibitem{Hughes02}
        S.\ A.\ Hughes, Mon.\ Not.\ R.\ Astron.\ Soc.\ {\bf 331}, 805 (2002).

\bibitem{stochUL}
    B. Abbott {\it et al.}; to be submitted.

\bibitem{FarmerPhinney}
    A. J. Farmer and E. S. Phinney; astro-ph/0304393.

\bibitem{Nelemans_2001c}
    G. Nelemans, L. R. Yungelson, and S. F. Portegies Zwart, Astron.\
    and Astrophys.\ {\bf 375}, 890 (2001).

\bibitem{Webbink_Han_98}
    R. F. Webbink and Z. Han, in Proceedings
2nd International LISA Symposium, ed. W. Folkner, AIP (1998);
astro-ph/0105230.
 Nelemans, L. R. Yungelson, and S. F. Portegies Zwart, Astron.\
  and Astrophys.\ {\bf 375}, 890 (2001).

\bibitem{Cornish_confusion}
 N. J. Cornish, gr-qc/0304020.

\bibitem{FreitagNew} M.\ Freitag, to appear in the proceedings of ``The Astrophysics
    of Gravitational Wave Sources'', a workshop held at the University of Maryland,
    2003 (astro-ph/0306064).

\bibitem{NR} W. H. Press, S. A. Teukolsky, W. T. Vetterling, and B. P.
Flannery, {\it Numerical Recipes} (Cambridge University Press, Cambridge,
1992).

\bibitem{cutler_vecchio}
    C. Cutler and A. Vecchio, in Proceedings
    2nd International LISA Symposium, ed. W. Folkner, AIP (1998).

\bibitem{D} D. H. McNamara, J. B. Madsen, J. Barnes, and F. B. Ericksen,
Publications of the Astronomical Society of the Pacific, {\bf 112},
202 (2000).

\bibitem{ShapiroTeukolsky}
    S. L. Shapiro and S. A. Teukolsky, \emph{Black Holes, White Dwarfs,
    and Neutron Stars} (Wiley-Interscience, 1983).

\bibitem{WDspin} S. D. Kawaler, astro-ph/0301539.

\bibitem{Hartl_03}
    M.\ D.\ Hartl; gr-qc/0302103.

\bibitem{Burko_03}
    L.\ Burko; gr-qc/0308003.

\bibitem{Kidder_95}
    L. E. Kidder, Phys.\ Rev.\ D {\bf 52}, 821 (1995).

\end{thebibliography}
\end{document}